\documentclass[a4paper,11pt]{article}
\usepackage{jheppub}
\usepackage{float} 
\usepackage{graphicx} 
\usepackage{slashed}
\usepackage{mathrsfs}
\usepackage{comment}
\usepackage{braket}
\usepackage{tikz}
\usetikzlibrary{calc}
\usetikzlibrary{decorations.markings}
\usetikzlibrary{decorations.text}
\usetikzlibrary{decorations.pathreplacing}
\usetikzlibrary{decorations.pathmorphing}

\usepackage{tikzit}

\tikzstyle{new style 0}=[fill=white, draw=black, shape=rectangle]
\tikzstyle{new edge style 0}=[snake=coil, draw=black]

\tikzstyle{interm}=[-, line width=1 pt, dashed, draw={rgb,255: red,126; green,126; blue,126}]
\tikzstyle{gluon}=[-, draw=black, line width=1 pt, decorate, decoration={{coil,amplitude=1mm,segment length=1.3mm}}]
\tikzstyle{quark}=[-, draw=black, line width=1 pt, postaction={decorate, decoration={{markings,mark=at position 0.5 with {\arrow[#1]{stealth}}}}}]
\tikzstyle{basic}=[-, line width=1 pt]

\def\v#1{\mathbf{#1}}
\def\mf#1{\mathbb{#1}}
\usepackage{orcidlink}
\tikzstyle{new style 0}=[fill=white, draw=black, shape=rectangle]
\tikzstyle{new edge style 0}=[snake=coil, draw=black]

\tikzstyle{interm}=[-, line width=1 pt, dashed, draw={rgb,255: red,126; green,126; blue,126}]
\tikzstyle{gluon}=[-, draw=black, line width=1 pt, decorate, decoration={{coil,amplitude=1mm,segment length=1.3mm}}]
\tikzstyle{quark}=[-, draw=black, line width=1 pt, postaction={decorate, decoration={{markings,mark=at position 0.5 with {\arrow[#1]{stealth}}}}}]
\tikzstyle{basic}=[-, line width=1 pt]

\title{\boldmath Soft Gluon Wave Function and Evolution Operator in the CGC at Next-to-Leading Order}

\author{Ramkumar~Radhakrishnan~\orcidlink{0000-0001-6838-9153}}
\affiliation{Department of Physics and Astronomy,\\ North Carolina State University, Raleigh, NC 27695, USA.}
\emailAdd{rradhak2@ncsu.edu}

\abstract{We construct the soft gluon light-cone wave function of a fast-moving hadron and the associated unitary evolution operator $\Omega$ up to $\mathcal{O}(g^{2})$ in pure Yang–Mills theory $(N_{f} = 0)$, within the Color Glass Condensate (CGC) framework. Working in light-cone gauge, we perform a Born–Oppenheimer separation between fast valence and soft modes and implement the eikonal approximation, which allows $\Omega$ to be written as a fully normal ordered series in soft gluon creation and annihilation operators. The expansion coefficients are functionals of the non-commuting valence color charge density operators $\rho$. We fix the coefficients by using the unitarity and explicit diagrammatic calculations within light-cone perturbation theory (LCPT). As an application, we diagonalize the pure Yang–Mills soft Hamiltonian through orders $g$ and $g^{2}$ and show that the off diagonal mixing elements between Fock sectors cancel leading to the coherent background field energy proportional to $\rho^{2}$.}

\begin{document}
\maketitle
\flushbottom
\section{Introduction} \label{sec:intro}

The dynamics of gluon saturation \cite{morreale2021mining,albacete2014gluon,blaizot2017high,mueller1986gluon} at small-$x$ can be systematically studied within the framework of Color Glass Condensate (CGC) effective theory~\cite{gelis2010color,weigert2005evolution,McLerran:1993ka,McLerran:1993ni}. In this regime, the rapidity evolution of gluon distributions is governed by the nonlinear BK and JIMWLK equations ~\cite{iancu2001renormalization,iancu2001nonlinear,weigert2002unitarity,kovner2000relating,jalilian1998wilson,jalilian19981wilson,jalilian1997bfkl,jalilian1997intrinsic,kovchegov2000unitarization,kovchegov1999small,lublinsky2017high}. While these equations describe the rapidity evolution of Wilson line correlators, physical scattering observables — such as single and double inclusive gluon production ~\cite{kovner2011particle,altinoluk2018double} and more generally, particle production ~\cite{altinoluk2011particle,iancu2021forward,dominguez2013universality,iancu2023dihadron,levin2010gluon,dumitru2010two,iancu2019forward,iancu2013jimwlk,levin2019j,gotsman2015cgc,bartels2008inclusive,altinoluk2009inclusive,levin2010gluon,altinoluk2012particle,gribov1983semihard,salzaar1} are obtained from matrix elements of operators evaluated between boosted hadronic states. Single inclusive gluon production, in particular, is the expectation value of the gluon number operator in the boosted projectile wave function, convoluted with the scattering amplitude off the target color field~\cite{venugopalan2007introduction}. Consequently, an explicit and controlled construction of the light-cone wave function (LCWF) of a fast-moving hadron is not only of theoretical but phenomenological interest. Morevoer, it provides a systematic framework for computing saturation corrections by separating the projectile evolution from its multiple eikonal interactions with the dense target~\cite{Li:2021zmf,Li:2021yiv,Li:2021ntt,Kovchegov:2024aus}.

From the light-cone perspective, boosting a hadron by a rapidity interval gives us an additional longitudinal phase space in which new soft gluon modes are generated. The effect of the boost can be represented by a unitary operator $\Omega$ acting on the initial valence state, describing the emission, absorption and recombination of soft gluons. Soft gluons with longitudinal momenta below the cutoff $(\Lambda)$ do not contribute to the scattering. As the energy of the hadron increases, their longitudinal momenta are boosted above the cutoff and they become dynamically relevant, contributing to physical observables such as scattering amplitudes and particle production. At leading order, $\Omega$ generates the Weizsäcker–Williams gluon field ~\cite{kovner2005pursuit,kovner2017exploring}, whereas beyond leading order it develops multi-gluon and color correlated structures responsible for fluctuations, correlations and higher-order effects.

It is precisely these structures that motivate the present construction. At leading order, $\Omega$ reduces to a coherent operator, i.e., an exponential of a single-gluon emission amplitude proportional to the valence color charge density $\rho$ and the state it produces is built from independently emitted Weizs\"acker--Williams gluons. Multi-gluon correlations are already present at this order: they are generated when observables are averaged over the valence color charge configurations, and double inclusive gluon correlations have indeed been computed within the leading order coherent operator framework ~\cite{kovner2011particle,altinoluk2018double,kovner2017exploring}. However, for a fixed configuration of the valence charge density, the leading order state carries no connected correlations among the emitted soft gluons: every multi-gluon observable factorizes into products of single-gluon quantities. The coefficients that first appear beyond leading order in the normal ordered expansion of $\Omega$ encode genuinely connected multi-gluon emission amplitudes, present even prior to the averaging over $\rho$. These coefficients are built from the non-Abelian color charge density and carry nontrivial color and momentum structure. They provide additional quantum contributions to fluctuations, double inclusive azimuthal correlations~\cite{McLerran:2014uka,Dumitru:2014yza,McLerran:2015sva} and the near side ridge, beyond those generated by the leading order coherent operator. We require the explicit construction of the higher coefficients of $\Omega$ to compute these observables.

The purpose of this work is to construct the soft gluon wave function and the corresponding evolution operator $\Omega$ explicitly up to order $g^{2}$ within the CGC framework, using light-cone perturbation theory (LCPT) ~\cite{kovchegov2013quantum,munier2025unitary}. Some contributions at orders $g^{2}$ and $g^{3}$ were obtained in Ref.~\cite{lublinsky2017high}, although that work did not contain all contributions at order $g^{2}$. There are also several related works in which light-cone wave functions were computed using LCPT (see Refs.~\cite{munier2025extracting,lappi2026two,munier2026unitary,hanninen2018one,lappi2017one}). We restrict ourselves to pure Yang–Mills theory and neglect quark degrees of freedom in order to isolate the gluonic dynamics relevant at high energies. We perform a systematic Born–Oppenheimer separation between fast valence and soft modes, then implement the eikonal approximation, and compute the perturbative expansion of $\Omega$ by matching its action on vacuum, one, two and three-gluon incoming states to the corresponding LCWF components~\cite{McLerran:1993ka,McLerran:1993ni,lublinsky2017high}. We obtain explicit expressions for all coefficients in the normal ordered expansion of $\Omega$ up to order $g^{2}$, constrained by unitarity and fixed by direct LCPT calculations. The logic underlying our strategy is that $\Omega$, acting on the free Fock states, must reproduce exactly the multi-particle structure generated by the pure Yang--Mills light-cone Hamiltonian. Operationally, we parameterize $\Omega$ as a fully normal ordered series in soft gluon creation and annihilation operators with operator valued coefficients, act with it on elementary Fock states and equate the result, sector by sector in Fock space, with the corresponding LCWF components computed in LCPT. This one-to-one matching fixes the coefficients associated with real transitions, while the unitarity condition $\Omega^{\dagger}\Omega = 1$, imposed order by order in the coupling, supplies the relations that determine the remaining (diagonal and vacuum) coefficients.

As an application of the formalism developed here, we use the evolution operator to perform a systematic diagonalization of the pure Yang–Mills light-cone Hamiltonian governing the soft sector of a fast hadron up to orders $g$ and $g^{2}$. The operator $\Omega$ is constructed perturbatively and chosen such that, order by order, it cancels the off-diagonal matrix elements between different Fock sectors and makes the transformed Hamiltonian diagonal to the desired accuracy. At order $g$ the Hamiltonian is fully diagonalized and at order $g^{2}$ the diagonalization leaves only the coherent background field energy proportional to $\rho^{2}$. Beyond this application, the wave function constructed here is the input required for the computation of physical observables at next-to-leading order. In particular, it sets the stage for the study of single inclusive gluon production at NLO at mid-rapidity~\cite{rr}. Along with the single inclusive cross section we will also obtain the moments of the gluon multiplicity (number) operator and the  particle number fluctuations at order $g^{4}$ in an upcoming paper.

Our article is structured as follows: In section \ref{sec:bgandsetup} we review the wave functional formalism, introduce the light-cone QCD Hamiltonian and discuss the Born–Oppenheimer separation together with the eikonal approximation within the CGC framework. In section \ref{sec:LCWFLCPT} we introduce the light-cone wave function formalism and review light-cone perturbation theory used to construct the soft gluon wave functions. In section \ref{sec:Omega} we construct the perturbative evolution operator and derive its action on the relevant Fock states. In section \ref{sec:coeff} we determine the coefficients of the evolution operator by matching to the LCWFs and imposing unitarity constraints. In section \ref{sec:diagonal} we diagonalize the QCD Hamiltonian at orders $g$ and $g^{2}$.
\section{Background and the setup} \label{sec:bgandsetup}
\subsection{Light-cone coordinates and conventions}
We will review the basic setup of QCD on the light cone~\cite{harindranath1996introduction,burkardt1996light,pritchard1980qcd}. In light-cone coordinates a four vector $x^{\mu}$ is given by the components
\begin{align}
    x^{\mu} & = (x^{+},x^{-},\v{x}),
\end{align}
where $\v{x} = (x^{1},x^{2})$. The longitudinal light-cone coordinates are defined by
\begin{align}
x^{\pm} = \frac{1}{\sqrt{2}}(x^{0}\pm x^{3}).
\end{align}
Throughout this work, quantization is carried out on hypersurfaces of fixed $x^{+}$. Therefore $x^{+}$ plays the role of light-cone time along which quantum states evolve, while $x^{-}$ is the corresponding longitudinal spatial coordinate. Similarly, the momentum four vector is written as
\begin{align}
p^{\mu} & = (p^{+},p^{-},\v{p}),
\end{align}
with
\begin{align}
p^{\pm} = \frac{1}{\sqrt{2}}(p^{0}\pm p^{3}).
\end{align}
The quantity $p^{+}$ is the longitudinal momentum and is positive for physical particles. The light-cone energy is denoted by $p^{-}$, which generates translations in light-cone time $x^{+}$. The scalar product of two four vectors is given by
\begin{align}
p\cdot x & = p^{+}x^{-}+p^{-}x^{+}-\v{p}\cdot\v{x}.
\end{align}
For an on-shell massless particle, $p^{2} = 0$ and therefore the light-cone energy is
\begin{align} \label{eq:disper}
p^{-} & = \frac{\v{p}^{2}}{2p^{+}}.
\end{align}
This relation plays an important role in light-cone perturbation theory, since all energy denominators are expressed in terms of differences of light-cone energies. The integration measure is defined as
\begin{align}
d^{3}p \equiv dp^{+}\,d^{2}\v{p},
\end{align}
and we use the shorthand notation
\begin{align}
\delta^{3}(k-p) \equiv \delta(k^{+}-p^{+})\,\delta^{(2)}(\v{k}-\v{p}).
\end{align}
Throughout this work we employ the light-cone gauge $A^{+} = 0$. In this gauge, only the transverse components $A^{i}$ with $i=1,2$ remain as dynamical degrees of freedom, while $A^{-}$ becomes a constrained non-dynamical field determined through the equations of motion. This considerably simplifies the structure of the Yang–Mills Hamiltonian and makes the physical gluonic degrees of freedom manifest. The gluon creation and annihilation operators satisfy the canonical commutation relation
\begin{align}
\left[a^{a}_{i}(k), a^{\dagger\, b}_{j}(p) \right] & = (2\pi)^{3}
\delta^{ab}\,\delta_{ij}\,\delta^{3}(k-p),
\end{align}
where $a,b$ are color indices in the adjoint representation and $i,j$ denote transverse polarization indices.
\subsection{Light-cone QCD Hamiltonian}
The derivation of the QCD Hamiltonian in light-cone gauge is standard and has been discussed extensively in the literature ~\cite{lublinsky2017high,kovner2005high,kovner2005remarks}.Our approximation $(N_{f} = 0)$ is sufficient for describing the soft gluon sector relevant for high energy evolution in the Color Glass Condensate framework, where the dynamics at small-$x$ is controlled by the gluonic degrees of freedom. The resulting light-cone Yang–Mills Hamiltonian can be written entirely in terms of these transverse fields and naturally separates into electric and magnetic contributions,
\begin{align}\label{eq:LCQCD}
H_{LC\; YM}
& =\int dx^{-}d^{2}\mathbf{x}\left(\frac{1}{2}\Pi^{a}(x^{-},\, \mathbf{x})\,\Pi^{a}(x^{-},\, \mathbf{x})\,+\,\frac{1}{4}F_{ij}^{a}(x^{-},\, \mathbf{x})\, F_{ij}^{a}(x^{-},\, \mathbf{x})\right),
\end{align}
where  the electric and magnetic pieces have the form:
 \begin{equation}\begin{split}
&\Pi^{a}(x^{-},\,\mathbf{x})\,\equiv\,-\,\frac{1}{\partial^{+}}(D_{i}^{ab}\partial^{+}A_{i}^{b}),\\
&F_{ij}^{a}(x^{-},\,\mathbf{x})\,\equiv\,\partial_{i}A_{j}^{a}\,-\,\partial_{j}A_{i}^{a}\,-\, gf^{abc}A_{i}^{b}A_{j}^{c}.\\
\end{split}\end{equation}
After substituting the explicit expressions for the gauge fields and their conjugate momenta into the light-cone Yang–Mills Hamiltonian, the Hamiltonian naturally separates into a free part and an interaction part, $H_{LC\; YM} = H_{0} + H_{\text{int.}}$. The free Hamiltonian defines the unperturbed Fock states and the interaction Hamiltonian generates transitions between different gluonic sectors order by order in the coupling. The free part of the Hamiltonian is given by
\begin{align}\label{eq:free}
    H_{0}\, & = \int dx^{-}\, d^{2}\mathbf{x}\,\left(\frac{1}{2}(\partial_{i}A_{j}^{a})^{2}\right).
\end{align}
The eigenstates of $H_{0}$ are the free gluon Fock states constructed by acting with gluon creation operators on the perturbative vacuum. All nonlinear dynamics of the theory are described in the interaction Hamiltonian. In the full QCD theory, the interaction part contains cubic and quartic gluon self interactions together with quark–gluon and quark–quark interactions. The interacting Hamiltonian is given by
\begin{align} \label{eq:Hint}
    H_{\text{int.}}\, & = \int dx^{-}\, d^{2}\v{x}\,\bigg(-gf^{abc}A_{i}^{b}A_{j}^{c}\partial_{i}A_{j}^{a}\,+\,\frac{g^{2}}{4}f^{abc}f^{ade}A_{i}^{b}A_{j}^{c}A_{i}^{d}A_{j}^{e}\,\notag \\
&-\, gf^{abc}(\partial_{i}A_{i}^{a})\frac{1}{\partial^{+}}(A_{j}^{b}\partial^{+}A_{j}^{c})\,+\,\frac{g^{2}}{2}f^{abc}f^{ade}\frac{1}{\partial^{+}}(A_{i}^{b}\partial^{+}A_{i}^{c})\frac{1}{\partial^{+}}(A_{j}^{d}\partial^{+}A_{j}^{e})\bigg).
\end{align}
The first term corresponds to the standard three-gluon interaction vertex and the second term corresponds to the four gluon interaction vertex. The remaining terms are instantaneous interaction contributions characteristic of light-cone quantization. Once the Hamiltonian has been formulated in light-cone gauge, the theory can be quantized canonically in the standard way. The transverse gauge fields are expanded in terms of creation and annihilation operators. The gauge field expansion is written as
\begin{align} \label{eq:gaugequant}
    A_{i}^{a}(x) & =\int_{0}^{\infty}\frac{dk^{+}}{2\pi}\int\frac{d^{2}\mathbf{k}}{(2\pi)^{2}}\frac{1}{\sqrt{2k^{+}}}\left(a_{i}^{a}(k^{+},\mathbf{k})e^{-ik\cdot x}+a_{i}^{a\dagger}(k^{+},\mathbf{k})\,e^{ik\cdot x}\right).
\end{align}
Substituting the field expansions i.e., Eq. \eqref{eq:gaugequant} into the free Hamiltonian Eq. \eqref{eq:free} yields
\begin{align} \label{eq:freecan}
    H_{0} & =\int_{0}^{\infty}\frac{dk^{+}}{2\pi}\, \int\frac{d^{2}\mathbf{k}}{(2\pi)^{2}}\, \frac{\mathbf{k}^{2}}{2k^{+}}\bigg(a_{i}^{a\dagger}(k^{+},\mathbf{k})\, a_{i}^{a}(k^{+},\mathbf{k}) \bigg),
\end{align} 
The interaction Hamiltonian is quantized in sec. \ref{sec:Eikonalapprox}.
\subsection{Born-Oppenheimer formalism and the eikonal approximation}\label{sec:Eikonalapprox}
In CGC, gluon modes carrying different longitudinal momenta play distinct dynamical roles. The degrees of freedom with large longitudinal momentum are referred to as the valence degrees of freedom, while modes carrying comparatively smaller longitudinal momentum are identified as the soft degrees of freedom ~\cite{McLerran:1993ni,McLerran:1993ka}. The valence modes represent the energetic constituents of the hadron and act as sources, whereas the soft modes correspond to low longitudinal momentum gluons act as a field. Within the Color Glass Condensate framework, this separation is implemented by introducing longitudinal momentum cutoffs that divide the Hilbert space into two momentum regions. The soft gluons are restricted to the interval $\Lambda<k^{+}<\vee$, while the valence modes reside at $k^{+}\geq \vee$. Here $\Lambda$ is the IR longitudinal momentum cutoff of the soft sector and $\vee$ defines the boundary separating the soft and valence degrees of freedom. The precise separation between the soft and valence sectors is not universal, but instead depends on the physical process and the kinematic regime under consideration. The choice of these cutoffs is decided by the rapidity interval over which one wishes to evolve the wave function. In the CGC framework, physical observables should ultimately be independent of the arbitrary separation scale $\vee$; changing this scale simply redistributes gluon modes between the valence and soft sectors. Since the valence modes carry much larger longitudinal momenta than the soft gluons, they evolve on much shorter longitudinal time scales. Consequently, from the perspective of the soft sector, the valence fields behave approximately as static background color sources.
Accordingly, the gluon field is decomposed into valence and soft components,
\begin{align} \label{eq:decompose}
    A_{i}^{a}(x)=\underline{A}_{i}^{a}(x)+\overline{A}_{i}^{a}(x).
\end{align}
Substituting this decomposition into the interaction Hamiltonian $H_{\text{int.}}$, defined in Eq.~\eqref{eq:Hint}, generates a large number of terms involving soft–soft, soft–valence and valence–valence interactions. The explicit mode decomposition and the resulting operator structures are presented in Appendix~\ref{sec:App2}. The separation between valence and soft modes naturally leads to the eikonal approximation. Since the longitudinal momenta carried by the valence fields are parametrically larger than those of the soft gluons, contributions suppressed by powers of $\frac{k^{+}_{\text{soft}}}{k^{+}_{\text{valence}}}$ can consistently be neglected. Keeping such terms would introduce non-eikonal power corrections that are beyond the perturbative accuracy considered in the present work. Operationally, this approximation implies that inverse longitudinal derivatives acting on valence fields are suppressed. For example, contributions of the form $\frac{1}{\partial^{+}}\left(\underline{A}_{j}^{b}\partial^{+}\overline{A}_{j}^{c}\right)$ are neglected. After performing the mode decomposition and applying the eikonal approximation, the interaction Hamiltonian can be organized as in Ref.~\cite{lublinsky2017high}
\begin{align} \label{eq: expandH}
H_{\text{int}}& = H_{g}+H_{ggg}+H_{gggg}+H_{gg-\text{inst.}}+H_{gggg-\text{inst.}}+H_{V}.
\end{align}
The first five terms, describe interactions involving at least one soft field. These terms govern the evolution of the soft sector in the presence of the valence background. The last term, $H_{V}$, contains only valence–valence interactions and does not affect the soft dynamics. Within the Born-Oppenheimer approximation, the effect of the fast valence sector on the soft fields is fully observed in its color charge density. We define the total valence charge density as
\begin{align}
   \rho^{a}(-\mathbf{p}) =  & -if^{abc}\int_{\vee}^{\infty}\frac{dk^{+}}{2\pi}\,\int\,\frac{d^{2}\mathbf{k}}{(2\pi)^{2}}\, a_{j}^{\dagger b}(k^{+},\, \mathbf{k})\, a_{j}^{c}(k^{+},\, \mathbf{k}+\mathbf{p}).
\end{align}
This charge density operator represents the fast valence degrees of freedom and provides the background field that drives the soft evolution, consistent with both the Born-Oppenheimer and eikonal approximations. From the definition and commutation relations it follows that the valence current $\rho^{a}(\mathbf{p})$ satisfies the $SU(N_{c})$ algebra:
 \begin{align}\label{rhoalge}
\left[\rho^{a}(\mathbf{k})\,,\,\,\rho^{b}(\mathbf{p})\right]=if^{abc}\rho^{c}(\mathbf{k}+\mathbf{p})\:.
  \end{align}
Now, one can express all the components of interacting Hamiltonian in terms of creation and annihilation operators. This is carried out by substituting the mode expansions of gauge field (Eq. \eqref{eq:gaugequant}). The final result is (see \ref{sec:App2} for details)
 \begin{align}\label{eikohg}
\mathsf{H}_{g} & =\int_{\Lambda}^{\vee}\frac{dk^{+}}{2\pi}\int\frac{d^{2}\mathbf{k}}{(2\pi)^{2}}\frac{g\mathbf{k}^{i}}{\sqrt{2}|k^{+}|^{3/2}}\left[a_{i}^{a\dagger}(k^{+},\, \mathbf{k})\rho^{a}(-\mathbf{k})+a_{i}^{a}(k^{+},\, \mathbf{k})\rho^{a}(\mathbf{k})\right],
 \end{align}
 \begin{align} \label{eq:H_g-gg}
H_{g\: gg}= & -\int_{\Lambda}^{\vee}\frac{dk^{+}}{2\pi}\,\frac{dp^{+}}{2\pi}\, dq^{+}\,\int\frac{d^{2}\mathbf{k}}{(2\pi)^{2}}\,\frac{d^{2}\mathbf{p}}{(2\pi)^{2}}\, d^{2}\mathbf{q}\,\frac{igf^{abc}}{2\sqrt{2k^{+}p^{+}q^{+}}}\notag \\ 
&\times\left[\left(\mathbf{q}^{i}-\frac{q^{+}}{p^{+}+q^{+}}\mathbf{k}^{i}\right)a_{i}^{a}(k)a_{j}^{b\dagger}(p)a_{j}^{c\dagger}(q)\delta^{(3)}(-k+p+q)\right.\notag \\ 
&+\left.\left(\mathbf{p}^{i}+\mathbf{q}^{i}+\frac{p^{+}+q^{+}}{q^{+}-p^{+}}\mathbf{k}^{i}\right)a_{i}^{a\dagger}(k)a_{j}^{b\dagger}(p)a_{j}^{c}(q)\delta^{(3)}(k+p-q),\,+\,h.c.\right],
 \end{align}
    \begin{align} \label{eq:H_gg-inst.}
H_{gg-\text{inst.}} & =\int_{\Lambda}^{\vee}\frac{dp^{+}}{2\pi}\frac{dq^{+}}{2\pi}\int\frac{d^{2}\mathbf{p}}{(2\pi)^{2}}\frac{d^{2}\mathbf{q}}{(2\pi)^{2}}\frac{ig^{2}f^{abc}}{\sqrt{p^{+}q^{+}}}\,\left(\frac{q^{+}\rho^{a}(-\mathbf{p}-\mathbf{q})}{2(p^{+}+q^{+})^{2}}\, a_{i}^{b\dagger}(p^{+},\,\mathbf{p})\, a_{i}^{c\dagger}(q^{+},\,\mathbf{q})\right.\notag \\ 
&\left.-\, h.c.\,-\,\frac{(p^{+}+q^{+})\rho^{a}(-\mathbf{p}+\mathbf{q})}{2(p^{+}-q^{+})^{2}}\, a_{i}^{b\dagger}(p^{+},\,\mathbf{p})\, a_{i}^{c}(q^{+},\,\mathbf{q})\right),
\end{align}
\begin{align} \label{eq:H_g-ggg}
    H_{gggg} & = \frac{g^{2}}{8} f^{abc}f^{ade}\, \int\, \frac{dp^{+}}{2\pi}\, \frac{dk^{+}}{2\pi}\, \frac{dr^{+}}{2\pi}\, dq^{+}\, \int\, \frac{d^{2}p}{(2\pi)^{2}}\, \frac{d^{2}k}{(2\pi)^{2}}\, \frac{d^{2}r}{(2\pi)^{2}}\, d^{2}q\,\frac{1}{\sqrt{p^{+}k^{+}r^{+}q^{+}}} \notag \\ & \Bigg[\frac{(p^{+}+k^{+})q^{+}}{(r^{+}+q^{+})(k^{+}-p^{+})}a^{b}_{i}(k)a^{c\dagger}_{i}(p)a^{d\dagger}_{j}(r)a^{e\dagger}_{j}(q) \delta^{(3)}(r+q-k+p) + \rm{h.c.,} \notag \\ & - \frac{p^{+}(r^{+}+q^{+})}{(q^{+}-r^{+})(k^{+}+p^{+})}a^{b\dagger}_{i}(k)a^{c\dagger}_{i}(p)a^{d\dagger}_{j}(r)a^{e}_{j}(q)\delta^{(3)}(k+p+r-q)+ \rm{h.c.,}\notag \\ & +\frac{g^{2}}{8} f^{abc}f^{ade}\, \int\, \frac{dp^{+}}{2\pi}\, \frac{dk^{+}}{2\pi}\, \frac{dr^{+}}{2\pi}\, dq^{+}\, \int\, \frac{d^{2}p}{(2\pi)^{2}}\, \frac{d^{2}k}{(2\pi)^{2}}\, \frac{d^{2}r}{(2\pi)^{2}}\, d^{2}q\,\frac{1}{\sqrt{p^{+}k^{+}r^{+}q^{+}}}\notag \\ & \Bigg[a^{b}_{i}(k)a^{c\dagger}_{j}(p)a^{d\dagger}_{i}(r)a^{e\dagger}_{j}(q) \delta^{(3)}(-k+p+r+q)+ \rm{h.c.,}+a^{b\dagger}_{i}(k)a^{c}_{j}(p)a^{d\dagger}_{i}(r)a^{e\dagger}_{j}(q)\notag \\ & \delta^{(3)}(k+r+q-p)+ \rm{h.c.,}\Bigg],
\end{align}
\begin{align} \label{eq:H_gg-gg}
    H_{gggg}+H_{gggg\,\text{inst.}} & = \frac{g^{2}}{8} f^{abc}f^{ade}\, \int\, \frac{dp^{+}}{2\pi}\, \frac{dk^{+}}{2\pi}\, \frac{dr^{+}}{2\pi}\, dq^{+}\, \int\, \frac{d^{2}p}{(2\pi)^{2}}\, \frac{d^{2}k}{(2\pi)^{2}}\, \frac{d^{2}r}{(2\pi)^{2}}\, d^{2}q\,\frac{1}{\sqrt{p^{+}k^{+}r^{+}q^{+}}} \notag \\ & \Bigg[-\frac{(k^{+}q^{+}+p^{+}r^{+})}{(p^{+}-k^{+})(r^{+}-q^{+})}a^{b}_{i}(p)a^{c\,\dagger}_{i}(k)a^{d}_{j}(r)a^{e\dagger}_{j}(q)\delta^{(3)}(-p-r+k+q) \notag \\ & -\frac{(k^{+}q^{+}+p^{+}r^{+})}{(p^{+}-k^{+})(r^{+}-q^{+})} a^{b}_{i}(p)a^{c\,\dagger}_{i}(k)a^{d\dagger}_{j}(r)a^{e}_{j}(q) \delta^{(3)}(r-q+k-p)\notag \\ & -\frac{(k^{+}q^{+}+r^{+}p^{+})}{(p^{+}+k^{+})^{2}(r^{+}+q^{+})^{2}}a^{b}_{i}(p)a^{c}_{i}(k)a^{d\,\dagger}_{j}(r)a^{e\,\dagger}_{j}(q)\delta^{(3)}(-p+r-k+q)\Bigg]\notag \\ & 
    + \frac{g^{2}}{8}\, f^{abc}f^{ade}\,\int\, \frac{dp^{+}}{2\pi}\, \frac{dk^{+}}{2\pi}\, \frac{dr^{+}}{2\pi}\, dq^{+}\, \int\, \frac{d^{2}p}{(2\pi)^{2}}\, \frac{d^{2}k}{(2\pi)^{2}}\, \frac{d^{2}r}{(2\pi)^{2}}\, d^{2}q\,\frac{1}{\sqrt{p^{+}k^{+}r^{+}q^{+}}} \notag \\ & \Bigg[a^{b}_{i}(p)a^{c}_{j}(k)a^{d\dagger}_{i}(r)a^{e\dagger}_{j}(q)\,\delta^{(3)}(r+q-p-k) + a^{b}_{i}(p)a^{c\dagger}_{j}(k)a^{d}_{i}(r)a^{e\dagger}_{j}(q)\notag \\ &  \delta^{(3)}(-p+k-r+q) +a^{b}_{i}(p)a^{c\dagger}_{j}(k)a^{d\dagger}_{i}(r)a^{e}_{j}(q)\, \delta^{(3)}(-p+k+r-q)\Bigg].
\end{align}
\section{Light-cone Wave Functions and Light-cone Perturbation Theory} \label{sec:LCWFLCPT}
\subsection{Light-cone quantization and Fock space expansion} \label{sec:LCquant}
In light-cone quantization, quantum states evolve along the light-cone time direction $x^{+}$ , while the generator of evolution is the light-cone Hamiltonian. The eigenstates of the free Hamiltonian form a natural Fock basis constructed from gluon creation operators acting on the perturbative vacuum. A generic hadronic state $(\ket{\Psi})$ can therefore be expanded as a superposition of multiparton Fock states,
\begin{align} \label{eq: LCQuant}
\ket{\Psi} & = \psi_{0}\ket{0} + \psi_{g}\ket{g} +
\psi_{gg}\ket{gg} + \psi_{ggg}\ket{ggg} +\cdots,
\end{align}
where the coefficients $\psi_{n}$ are the light-cone wave function amplitudes corresponding to different gluonic components. Each Fock component carries definite longitudinal momentum, transverse momentum, polarization and color quantum numbers. Physically, the LCWF describes the probability amplitude for finding a particular gluonic configuration inside the fast-moving hadron. At high energies, where gluon densities become large, the multi gluonic components become increasingly important and it describes the nonlinear dynamics characteristic of the Color Glass Condensate regime. One of the main advantages of light-cone quantization is that the vacuum structure remains comparatively simple. The nontrivial dynamics of the theory are therefore transferred directly into the wave function and this makes the LCWF particularly suitable for describing high energy scattering processes, particle production and small-$x$ evolution. 

Within the CGC framework, the hadron is separated into fast valence degrees of freedom and softer gluonic modes. Consequently, the light-cone wave function naturally organizes itself in powers of the valence color charge density operator $\rho$. It is important to stress that the number of soft gluons in a given Fock component and the number of $\rho$ insertions multiplying it are two independent labels: at a fixed order in the coupling, a component with a fixed number of soft gluons in general receives contributions carrying different powers of $\rho$. We therefore label each component of the wave function by the pair $(n,m)$, where $n$ counts the soft gluons in the Fock state and $m$ counts the insertions of the valence color charge density,
\begin{align} \label{sec:Wfunc}
\ket{\Psi} & = \mf{N}\ket{0} + \underbrace{\ket{\Psi_{g\rho}}}_{\mathcal{O}(g)} + \underbrace{\Big(\ket{\Psi_{gg\rho}} + \ket{\Psi_{gg\rho\rho}}\Big)}_{\mathcal{O}(g^{2})} +\cdots.
\end{align}
The single soft gluon component appears at order $g$ with a single insertion of $\rho$, while at order $g^{2}$ the two-gluon component contains both a single-$\rho$ piece, arising from a gluon splitting after a single emission off the valence source, and a double-$\rho$ piece, arising from two independent emissions off the valence source. This also has Fock components with more soft gluons and/or more powers of $\rho$, which first arise beyond the order considered in this work (as well as the order $g^{2}$ corrections to the lower components). The coefficients appearing in these sectors are not ordinary $c$-number functions but operator valued quantities acting in the valence Hilbert space. This is due to the non-Abelian nature of the color charge density (see Eq. \eqref{rhoalge}), which also implies that different color structures generally do not commute. As a result, the ordering of operators becomes important throughout the perturbative construction of the wave function. The explicit computation of these LCWF components is carried out using light-cone perturbation theory.
\subsection{Light-cone perturbation theory (LCPT)} \label{sec:LCPT1}
Light-cone perturbation theory is the Hamiltonian formulation of old fashioned time ordered perturbation theory which is used with light-cone quantization. In this approach, the interaction Hamiltonian acts perturbatively on the free Fock states and generates transitions between components containing different numbers of gluons. Starting from the decomposition
\begin{align}
H & = H_{0} + H_{\text{int.}},
\end{align}
the interacting wave function is obtained by solving the eigenvalue equation perturbatively,
\begin{align}
H\ket{\Psi} = E\ket{\Psi}.
\end{align}
To first order in perturbation theory~\cite{Landau:1991wop,shankar2012principles}, the correction to an incoming state $\ket{n}$ is given by
\begin{align}
\ket{\Psi^{(1)}_{n}} & =  \sum_{m\neq n}\,\frac{\ket{m}\bra{m}H_\text{int.}\ket{n}}{E_{n}-E_{m}},
\end{align}
where the sum runs over all intermediate Fock states generated by the interaction Hamiltonian. The quantities $E_{n}$ and $E_{m}$ are the light-cone energies of the corresponding states. For a multiparticle state, the total light-cone energy is simply the sum of the individual gluon energies, i.e., $\sum_{i}\frac{\mathbf{k}_{i}^{2}}{2k_{i}^{+}}$. Unlike covariant perturbation theory, LCPT is formulated directly in terms of physical intermediate states ordered in light-cone time. Each perturbative contribution therefore admits a transparent physical interpretation in terms of sequential gluon emission and absorption processes. The interaction vertices appearing in the Hamiltonian generate transitions between different Fock sectors, while the energy denominators describe the propagation of intermediate off shell states. Higher order wave functions arise from repeated insertions of the interaction Hamiltonian and therefore contain contributions generated through successive transitions between intermediate Fock sectors. Because the soft gluons evolve in the background of the valence sector, the resulting wave functions contain multiple insertions of the valence color charge density operator $\rho$. The details of these calculations are presented in Appendix~\ref{sec:App5}.
\subsection{Soft gluon wave function in the CGC} \label{sec:LCWF1}
We now introduce the evolution operator from a quantum field theoretic perspective. The construction closely follows the operator based approach developed in the work of Ref.~\cite{kovner2019entanglement}, where the evolution operator is interpreted as a unitary transformation that both generates the soft gluon wave function and diagonalizes the interacting Yang–Mills Hamiltonian order by order in perturbation theory. In perturbative quantum field theory, the interacting eigenstates of the Hamiltonian are not identical to the free Fock states. The interaction Hamiltonian mixes components with different particle numbers between states containing different numbers of gluons. Consequently, the physical states of the theory are dressed states consisting of superpositions of multiparticle Fock components.

This dressing can be implemented systematically through a unitary evolution operator. Let $H = H_{0} + H_{\text{int.}}$ denote the light-cone Hamiltonian, where $H_{0}$ is the free Hamiltonian and $H_{\text{int.}}$ contains the interaction terms. The interacting eigenstates are related to the free Fock states by a unitary transformation,
\begin{align}\label{eq:dressing}
\ket{\Psi_{n}} = \Omega\,\ket{n},
\end{align}
where $\ket{n}$ denotes a free Fock state, $\ket{\Psi_{n}}$ is the corresponding dressed eigenstate of the full Hamiltonian, and $\Omega$ is the evolution operator. In perturbation theory, $\Omega$ can be given an explicit representation as an adiabatic time evolution operator. As discussed in Appendix A of Ref.~\cite{kovner2019entanglement}, the interaction is adiabatically switched on, so that $H(t \rightarrow -\infty)\rightarrow H_{0}$, and one considers the time evolution operator from the far past to $t=0$,
\begin{align}
U(-\infty,0)=\mathcal{T}\left[\exp\left[-i\int_{-\infty}^{0}dt\,H(t)\right]\right],
\end{align}
By the adiabatic theorem, $U(-\infty,0)$ maps an eigenstate of $H_{0}$, prepared in the far past before the interaction is switched on, into an eigenstate of the full interacting Hamiltonian at $t=0$. Up to an overall phase, this is exactly the defining property of the operator $\Omega$ in Eq.~\eqref{eq:dressing} and we therefore identify $\Omega = U(-\infty,0)$. In particular, the dressed wave function of the soft vacuum, $\ket{\Psi_{0}} = \Omega\ket{0}$, is obtained by adiabatically evolving the free (soft) vacuum in the presence of the interaction with the valence sources.

At leading order in the coupling and the highest order of $\rho$, the only interaction term that survives is the linear coupling of the soft gluon field to the valence color charge density, $H_{g}$ in Eq.~\eqref{eikohg} and the evolution operator reduces to a coherent operator,
\begin{align}\label{eq:coherentC}
\mathcal{C} = \exp\left\{\int \frac{d^{3}p}{(2\pi)^{3}}\,\Big[\mf{A}^{b}_{j}(p)\,a^{\dagger\, b}_{j}(p) - \mf{A}^{\dagger\, b}_{j}(p)\,a^{b}_{j}(p)\Big]\right\},
\end{align}
where the single-gluon emission amplitude $\mf{A}^{b}_{j}(p) \propto g\, \mathbf{p}^{j}\rho^{b}(-\mathbf{p})/\mathbf{p}^{2}$ (given explicitly in Eq.~\eqref{A1coeffM}) is the Weizs\"acker--Williams field of the valence charge density. Acting on the soft vacuum, $\mathcal{C}$ dresses the valence state with the classical Weizs\"acker--Williams gluon cloud ~\cite{kovner2005pursuit,kovner2019entanglement}. Beyond leading order, $\Omega$ contains multi-gluon operator structures that go beyond the coherent operator; their construction is the subject of the present work.

The evolution operator can equivalently be viewed as the operator that diagonalizes the interacting Hamiltonian. Indeed, acting with $H$ on Eq.~\eqref{eq:dressing} and using $H\ket{\Psi_{n}} = E_{n}\ket{\Psi_{n}}$, one finds $\Omega^{\dagger}H\Omega\ket{n} = E_{n}\ket{n}$. The transformed Hamiltonian $H_{\text{diag}}\equiv \Omega^{\dagger} H \Omega$ is therefore diagonal in the free Fock basis,
\begin{align}
H_{\text{diag}} = \sum_{n}E_{n}\ket{n}\bra{n},
\end{align}
where $\ket{n}$ are the free Fock states and $E_{n}$ are the eigenvalues of the full interacting Hamiltonian. The purpose of the perturbative construction is thus to choose $\Omega$ such that the off diagonal matrix elements of $\Omega^{\dagger}\,H\,\Omega$ between different Fock sectors are eliminated order by order in the coupling. For a quadratic Hamiltonian, the diagonalization can be performed exactly through a canonical transformation. In Yang–Mills theory, however, the Hamiltonian contains genuinely non-Abelian interaction vertices that couple infinitely many gluonic Fock sectors. Consequently, the diagonalization is much more intricate and can only be carried out perturbatively.

\section{Perturbative expansion of the evolution operator ($\Omega$)} \label{sec:Omega}
In the previous section, the evolution operator $\Omega$ was defined implicitly through Eq.~\eqref{eq:dressing}. It is the unitary operator that maps each free Fock state onto the corresponding dressed eigenstate of the full Hamiltonian. Beyond leading order, no closed form for $\Omega$ is available and it must be constructed perturbatively. Our strategy is the following. We first write down the most general ansatz for $\Omega$ as a fully normal ordered series in the soft gluon creation and annihilation operators, $a^{\dagger}$ and $a$, with unknown operator valued coefficients, retaining only those operator structures that can contribute up to $\mathcal{O}(g^{2})$. The unknown coefficients are then fixed by two complementary sets of conditions: 
\begin{itemize}
    \item The unitarity requirement $\Omega^{\dagger}\Omega=1$, imposed order by order in the coupling. 
    \item The requirement that $\Omega$, acting on the free Fock states, reproduces the dressed states computed directly in LCPT, i.e., $\Omega\ket{n} = \ket{\Psi_{n}}$ for the vacuum, one-, two- and three-gluon incoming states.
\end{itemize}    
The reason for acting on several different incoming states is that each coefficient in the normal ordered expansion, containing a definite number of annihilation and creation operators, is probed only by the transition between the corresponding incoming and outgoing Fock sectors; acting on states of increasing gluon number therefore fixes the coefficients sector by sector.

To this end, it is convenient to express $\Omega$ in a fully normal ordered form with respect to the creation and annihilation operators of the soft sector. In our $N_{f} = 0$ approximation, the evolution operator can be written entirely in terms of normal ordered products of soft gluon creation and annihilation operators, $a$ and $a^{\dagger}$ with operator-valued coefficients multiplying each normal ordered structure.

    These coefficients are functionals of the valence color charge density $\rho$. In CGC formalism, the color charge densities satisfy a non-Abelian algebra i.e., Eq. \eqref{rhoalge}. As a consequence, the coefficients appearing in the normal ordered expansion are themselves non commuting operators. Their ordering therefore carries physical significance and must be treated carefully throughout the construction of the evolution operator. The ansatz, truncated at the operator structures relevant through $\mathcal{O}(g^{2})$, reads as follows. The coefficients $\mf{A}$ and $\mf{C}$ below are of order $g$, while $\mf{N}-1$, $\mf{B}$, $\mf{D}$ and $\mf{E}$ are of order $g^{2}$; normal ordered structures with more creation or annihilation operators than those displayed first contribute beyond $\mathcal{O}(g^{2})$ and are omitted.
 \begin{align}\label{Eq:OmegaExp}
    \Omega  & = \mf{N} \notag \\ & +
                \int \frac{d^3p}{(2\pi)^3} 
                \bigg[ 
                   \mf{A}^{b}_{1j}(p)a^{b}_{j}(p) 
                +  \mf{A}^{b}_{2j} (p) a^{\dagger b}_{j}(p)\bigg] \notag \\ & 
               +\int \frac{d^3p}{(2\pi)^3} \frac{d^3q}{(2\pi)^3} 
               \bigg[\mf{B}^{cb}_{1kj}(p,q) a^{b}_{j}(p)a^{c}_{k}(q) 
                   + \mf{B}^{cb}_{2kj}(p,q) a^{\dagger b}_{j}(p)a^{\dagger c}_{k}(q)
               + \mf{B}^{cb}_{3kj} (p,q) a^{\dagger c}_{k}(q)a^{b}_{j}(p)
               \bigg] \notag \\ & 
               + \int 
               \frac{d^3p}{(2\pi)^3} \frac{d^3q}{(2\pi)^3} \frac{d^3r}{(2\pi)^3}
               \bigg[
                \mf{C}^{ dcb}_{1lkj}(p,q,r) a^{\dagger d}_{l}(r)  a^{\dagger c}_{k}(q)  a^{b}_{j}(p) 
               + \mf{C}^{dcb}_{2lkj}(p,q,r) a^{\dagger d}_{l}(r) a^{c}_{k}(q) a^{b}_{j}(p)\bigg]\notag \\ &  +\int \frac{d^3p}{(2\pi)^3} \frac{d^3q}{(2\pi)^3} \frac{d^3r}{(2\pi)^3} \frac{d^3s}{(2\pi)^3} \bigg[\mf{D}^{edcb}_{1mlkj}(p,q,r,s)a^{\dagger e}_{m}(s)a^{\dagger d}_{l}(r)a^{\dagger c}_{k}(q)a^{b}_{j}(p)\notag \\ & + \mf{D}^{edcb}_{2mlkj}(p,q,r,s)a^{\dagger e}_{m}(s)a^{\dagger d}_{l}(r)a^{c}_{k}(q)a^{b}_{j}(p)  + \mf{D}^{edcb}_{3mlkj}(p,q,r,s) a^{\dagger e}_{m}(s)a^{d}_{l}(r)a^{c}_{k}(q)a^{b}_{j}(p) \bigg] \notag \\ &
               + \int \frac{d^3p}{(2\pi)^3} \frac{d^3q}{(2\pi)^3} \frac{d^3r}{(2\pi)^3} \frac{d^3s}{(2\pi)^3} \frac{d^3u}{(2\pi)^3} \frac{d^3v}{(2\pi)^3}\bigg[\notag \\ & \hspace{1.25em}\mf{E}^{hfedcb}_{1\,onmlkj}(p,q,r,s,u,v)a^{\dagger h}_{o}(v)a^{\dagger f}_{n}(u)a^{\dagger e}_{m}(s)a^{\dagger d}_{l}(r)a^{c}_{k}(q)a^{b}_{j}(p) \notag \\ & + \mf{E}^{hfedcb}_{2\,onmlkj}(p,q,r,s,u,v)a^{\dagger h}_{o}(v)a^{\dagger f}_{n}(u)a^{\dagger e}_{m}(s)a^{d}_{l}(r)a^{c}_{k}(q)a^{b}_{j}(p) \notag \\ & + \mf{E}^{hfedcb}_{3\,onmlkj}(p,q,r,s,u,v)a^{\dagger h}_{o}(v)a^{\dagger f}_{n}(u)a^{e}_{m}(s)a^{d}_{l}(r)a^{c}_{k}(q)a^{b}_{j}(p) \bigg] 
\end{align} 
Before presenting the explicit action on the Fock states, it is useful to exploit this unitarity condition in order to constrain the operator coefficients appearing in the expansion of $\Omega$. The perturbative coefficients are not ordinary $c$-number functions but non commuting operators acting in the valence color space, as we discussed above. To illustrate this with less complexity, we expand the evolution operator perturbatively,
\begin{align}
    \Omega & = 1 + g\,\Omega^{(1)} + g^{2}\, \Omega^{(2)} + \mathcal{O}(g^{3}),
\end{align}
where $g$ is the coupling constant, $\Omega^{(1)}$ and $\Omega^{(2)}$ contain both the soft and valence degrees of freedom. The unitarity condition implies
\begin{align}
    \Omega^{\dagger}\Omega = 1 + g\,\left(\Omega^{\dagger\, (1)} + \Omega^{(1)}\right) + g^{2}\, \left(\Omega^{(2)} + \Omega^{\dagger\, (2)}+ \Omega^{\dagger\,(1)}\Omega^{(1)}\right) + \mathcal{O}(g^{3}) = 1
\end{align}
Requiring this relation to hold order by order in the coupling produces a hierarchy of nontrivial consistency conditions among the coefficients. At leading order one obtains the anti-Hermiticity condition
\begin{align} \label{eq:unitarityfirst}
    \Omega^{\dagger\, (1)} & = -\Omega^{(1)},
\end{align}
while at second order the coefficients satisfy
\begin{align}\label{eq:unitaritysecond}
    \Omega^{(2)} + \Omega^{\dagger\, (2)} & = - \Omega^{\dagger\,(1)}\Omega^{(1)}. 
\end{align}
These relations are essential because the action of $\Omega$ on individual Fock states determines only part of the perturbative structure. The remaining components, particularly diagonal and vacuum contributions, are fixed uniquely by demanding exact unitarity order by order in $g$. Physically, these constraints guarantee that $\Omega$ is unitary and hence that the norm of any evolved state, i.e., the total probability, is conserved order by order in perturbation theory. Implementing the first order condition, Eq.~\eqref{eq:unitarityfirst}, together with the terms of the second order condition, Eq.~\eqref{eq:unitaritysecond}, that involve $\mf{N}$ and $\mf{A}$, yields the following relations among the perturbative coefficients appearing in the Fock space expansion of $\Omega$ in Eq. \eqref{Eq:OmegaExp}:
\begin{align}
 \mf{N}^{2}  &= 1-  \int \frac{d^{3}p}{(2\pi)^{3}} \mf{A}^{\dagger b}_{2j}(p) \mf{A}^{b}_{2j}(p) \quad
 \implies \quad \mf{N} = \mf{N}_{0} + \mf{N}_{2}   = 1-  \frac{1}{2}\int \frac{d^{3}p}{(2\pi)^{3}} \mf{A}^{\dagger b}_{j}(p) \mf{A}^{b}_{j}(p),
 \end{align}
 \begin{align}
 \mf{A}^{\dagger\,  b}_{1j}(p) & =  -\mf{A}^{b}_{2j}(p),
 \end{align}
 \begin{align}
  \mf{C}^{dcb}_{2lkj}(p,q,r) & =  -\mf{C}^{\dagger\, bcd}_{1jkl}(r,q,p).
\end{align}
From here onward, we will denote $\mf{A}_{2}$ as $\mf{A}$\footnote{$\mf{A}$ is at least of order $g$.} and $\mf{A}^{\dagger}_{2}$ as $\mf{A}^{\dagger}$ and $\mf{C}_{1}$ as $\mf{C}$ and $\mf{C}^{\dagger}_{1}$ as $\mf{C}^{\dagger}$. The remaining constraints follow from the $\mathcal{O}(g^{2})$ term of the unitarity condition, Eq.~\eqref{eq:unitaritysecond}, projected onto the independent normal ordered operator structures of Eq.~\eqref{Eq:OmegaExp}. They read
\begin{align}
     \mf{B}^{cb}_{1kj}(p,q) & = \frac{1}{2}\,\bigg[-2\,\mf{B}^{\dagger\, cb}_{2kj}(p,q)+\mf{A}^{\dagger\, c}_{k}(q)\mf{A}^{\dagger\, b}_{j}(p)+ \mf{A}^{\dagger\, b}_{j}(p)\mf{A}^{\dagger\, c}_{k}(q)\notag \\ & +2\,\int \frac{d^{3}r}{(2\pi)^{3}} \mf{A}^{\dagger\, d}_{l}(r)\mf{C}^{\dagger\, bcd}_{jkl}(r,q,p)\bigg],
\end{align}
\begin{align}
    \mf{B}^{\dagger\, cb}_{3kj}(p,q) & = -\mf{B}^{bc}_{3jk}(q,p) - \mf{A}^{b}_{j}(p)\mf{A}^{\dagger\, c}_{k}(q) - \mf{A}^{\dagger\, c}_{k}(q)\mf{A}^{b}_{j}(p) -\int\, \frac{d^{3}r}{(2\pi)^{3}}\bigg[ \mf{A}^{\dagger\, d}_{l}(r)\mf{C}^{bdc}_{jlk}(q,r,p)\notag \\ &  + \mf{A}^{\dagger\, d}_{l}(r)\mf{C}^{dbc}_{ljk}(q,p,r)+ \mf{C}^{\dagger\, dcb}_{lkj}(p,q,r)\mf{A}^{d}_{l}(r) +  \mf{C}^{\dagger\, cdb }_{klj}(p,r,q)\mf{A}^{d}_{l}(r)\notag \\ & -\int\, \frac{d^{3}s}{(2\pi)^{3}}\mf{C}^{\dagger\,deb}_{lmj}(p,s,r)\mf{C}^{edc}_{mlk}(q,r,s) -\int \frac{d^{3}s}{(2\pi)^{3}} \mf{C}^{\dagger\,deb}_{lmj}(p,s,r)\mf{C}^{dec}_{lmk}(q,s,r)\bigg],
\end{align}
\begin{align}
    \mf{D}^{\dagger\,edcb }_{2mlkj}(p,q,r,s) = & - \mf{D}^{cbed}_{2kjml}(r,s,p,q)-\mf{A}^{b}_{j}(p)\mf{C}^{\dagger\,edc}_{mlk}(q,r,s) -\mf{A}^{\dagger\, e}_{m}(s)\mf{C}^{bcd}_{jkl}(r,q,p)\notag \\ & -\mf{C}^{\dagger\,deb}_{lmj}(p,s,r)\mf{A}^{c}_{k}(q) -\mf{C}^{bcd}_{jkl}(r,q,p)\mf{A}^{\dagger\, e}_{m}(s) -\int\, \frac{d^{3}r^{\prime}}{(2\pi)^{3}}\bigg[\mf{C}^{\dagger\, d^{\prime}eb}_{l^{\prime}mj}(p,s,r^{\prime})\notag \\ & \mf{C}^{d^{\prime}cd}_{l^{\prime}kl}(r,q,r^{\prime}) + \mf{C}^{\dagger\, dd^{\prime}b}_{ll^{\prime}j}(p,r^{\prime},r)\mf{C}^{d^{\prime}ce}_{l^{\prime}km}(s,q,r^{\prime}) +\mf{C}^{bcd^{\prime}}_{jkl^{\prime}}(r^{\prime},q,p)\mf{C}^{\dagger\, d^{\prime}ed}_{l^{\prime}ml}(r^{\prime},s,r) \notag \\ & +\mf{C}^{\dagger\, d^{\prime}eb}_{l^{\prime}mj}(p,s,r^{\prime})\mf{C}^{cd^{\prime}d}_{kl^{\prime}l}(r,r^{\prime},q)+\mf{C}^{\dagger\,dd^{\prime}b}_{ll^{\prime}j}(p,r^{\prime},r)\mf{C}^{cd^{\prime}b}_{kl^{\prime}j}(s,r^{\prime},q)\bigg],
\end{align}
\begin{align}
     \mf{D}^{edcb}_{3mlkj}(p,q,r,s) & = -\mf{D}^{\dagger\, bdce}_{1jlkm}(s,q,r,p)+\int\, \frac{d^{3}r^{\prime}}{(2\pi)^{3}}\mf{C}^{\dagger\, d^{\prime}ce}_{l^{\prime}km}(s,q,r^{\prime})\mf{C}^{\dagger\, bdd^{\prime}}_{jll^{\prime}}(r^{\prime},r,p) \notag  \\ & + \mf{A}^{\dagger\, b}_{j}(p)\mf{C}^{\dagger\, dce}_{lkm}(s,q,r)  +\int \frac{d^{3}r^{\prime}}{(2\pi)^{3}} \mf{C}^{\dagger\, dd^{\prime}e}_{ll^{\prime}m}(s,r^{\prime},r)\mf{C}^{\dagger\, bcd^{\prime}}_{jkl^{\prime}}(r^{\prime},q,p)\notag\\ & +\mf{C}^{\dagger\, dce}_{lkm}(s,q,r)\mf{A}^{\dagger\, b}_{j}(p),
\end{align}
\begin{align}
    \mf{E}^{\dagger hfedcb}_{2\,onmlkj}(p,q,r,s,u,v) & =    - \mf{E}^{dcbhfe}_{2\,lkjonm}(s,u,v,p,q,r)-\mf{C}^{\dagger\,hfd}_{onl}(r,u,v)\mf{C}^{cbe}_{kjm}(s,p,q)\notag \\ &  - \mf{C}^{bce}_{jkm}(s,q,p)\mf{C}^{\dagger\,hfd}_{onl}(r,u,v),
\end{align}
\begin{align}
    \mf{E}^{hfedcb}_{3 onmlkj}(p,q,r,s,u,v) = &  -\mf{E}^{\dagger bcedfh}_{1jkmlno}(v,u,r,s,q,p)+\mf{C}^{\dagger bcf}_{jkn}(u,q,p)\mf{C}^{\dagger edh}_{mlo}(v,r,s).
\end{align}
Now that all unitarity constraints have been established, we can proceed to determine the explicit action of the evolution operator $\Omega$ on the relevant soft gluon Fock states. In particular, we will construct its action on the vacuum, one gluon, two-gluon and three-gluon states, retaining terms up to the required perturbative order in the coupling.

We begin with the action of $\Omega$ on the vacuum state. This is the simplest sector, but it already contains the essential structure of soft gluon emission from the valence color sources. The resulting expression takes the following form:

\begin{align} \label{eq:Omegaonvacmom1}
    \Omega\ket{0} & = \mf{N}\ket{0}
    +(2\pi)^{3/2}\int \frac{d^3p}{(2\pi)^3}\mf{A}^{b}_{j}(p)
    \ket{g^{b}_{j}(p)} \notag \\ & 
    +(2\pi)^{3}\int \frac{d^3p}{(2\pi)^3} \frac{d^3q}{(2\pi)^3}  \mf{B}^{cb}_{2kj}(p,q)\ket{g^{b}_{j}(p)g^{c}_{k}(q)}.
\end{align}
We next act with $\Omega$ on a one gluon state. Since $\Omega$ is normal ordered, each structure in Eq.~\eqref{Eq:OmegaExp} contributes in a simple way: every annihilation operator either contracts with the incoming gluon or annihilates the vacuum, while the creation operators produce the gluons of the outgoing Fock component. Thus the term $\mf{A}_{1}\,a$ absorbs the incoming gluon and produces the vacuum component; the terms $\mf{N}$ and $\mf{B}_{3}\,a^{\dagger}a$ preserve the gluon number, the latter describing the interaction of the gluon with the valence background; $\mf{A}_{2}\,a^{\dagger}$ and $\mf{C}_{1}\,a^{\dagger}a^{\dagger}a$ raise the gluon number by one, producing two-gluon components; and $\mf{D}_{1}\,a^{\dagger}a^{\dagger}a^{\dagger}a$ produces the three-gluon component. Structures whose annihilation operators cannot all be saturated by the single incoming gluon (those with two or more annihilation operators) do not contribute in this sector. In writing the result below we keep only the terms that carry new information through $\mathcal{O}(g^{2})$; contributions whose coefficients are already fixed in the vacuum sector (such as the two-gluon component generated by $\mf{A}_{2}\,a^{\dagger}$) are omitted. The explicit action of $\Omega$ on the single-gluon state is then given as
\begin{align}\label{eq:Omegaongmom}
    \Omega\ket{g^{a}_{i}(k)} & =  -\frac{1}{(2\pi)^{3/2}} \mf{A}^{\dagger\, a}_{i}(k)\ket{0}
   + \mf{N}\ket{g^{a}_{i}(k)} 
    +\int  \frac{d^3p}{(2\pi)^3} \mf{B} ^{ba}_{3ji}(k,p)\ket{g^{b}_{j}(p)} \notag \\ &  
    + (2\pi)^{3/2}\int \frac{d^3p}{(2\pi)^3}\frac{d^3q}{(2\pi)^3} \mf{C}^{cba}_{kji}(k,p,q)\ket{g^{c}_{k}(q)g^{b}_{j}(p)}  \notag \\ & + (2\pi)^{3} \int \frac{d^3 q}{(2\pi)^3} \frac{d^3r}{(2\pi)^3} \frac{d^3p}{(2\pi)^3}  \mf{D}^{dcba}_{1lkji}(k,p,q,r) \ket{g^{b}_{j}(p)g^{d}_{l}(r)g^{c}_{k}(q)}.
\end{align}
The coefficients $\mf{A}_{1}, \mf{A}_{2}, \mf{B}_{2}, \mf{C}_{1}, \mf{D}_{1}, \mf{D}_{2},\mf{E}_{1},\mf{E}_{2}$ are determined by matching the perturbative expansion of the evolution operator to the corresponding LCWF. The remaining coefficients are not fixed independently from perturbation theory alone. Instead, they are determined by imposing the unitarity condition. The coefficients $\mf{D}_{2}, \mf{E}_{1}$ require special treatment and are fixed specifically by the action of $\Omega$ on incoming two-gluon states, as the corresponding constraints cannot be extracted solely from the vacuum or one gluon sectors. Throughout the construction, we retain only those operator structures that contribute to the evolution operator up to $\mathcal{O}(g^{2})$. Consequently, terms whose coefficients are already fully constrained by lower order relations or whose contributions enter only at higher perturbative order, are systematically omitted.\\
\raggedbottom
\begin{align}\label{eq:Omegaongg}
    \Omega\ket{g^{a}_{i}(k)g^{b}_{j}(p)} & = \mf{N}\ket{g^{a}_{i}(k)g^{b}_{j}(p)}
    -\frac{1}{(2\pi)^{3/2}}\, \mf{A}^{\dagger\, a}_{i}(k) \ket{g^{b}_{j}(p)} - \frac{1}{(2\pi)^{3/2}}\, \mf{A}^{\dagger\, b}_{j}(p) \ket{g^{c}_{k}(q)} \notag \\ & 
    + \int \frac{d^{3}q}{(2\pi)^{3}}\, \mf{A}^{c}_{k}(q)\ket{g^{c}_{k}(q)g^{a}_{i}(k)g^{b}_{j}(p)}
    + \frac{1}{(2\pi)^{3}}\, \bigg[\mf{B}^{ba}_{1ji}(k,p)+\mf{B}^{ab}_{1ij}(p,k)\bigg] \ket{0} \notag \\ & + \int \frac{d^{3}q}{(2\pi)^{3}}\, \bigg[\mf{B}^{ca}_{3ki}(k,q)\ket{g^{c}_{k}(q)g^{b}_{j}(p)}+\mf{B}^{cb}_{3kj}(p,q)\ket{g^{a}_{i}(k)g^{c}_{k}(q)}\bigg]\notag \\ &+ (2\pi)^{3/2}\, \int \frac{d^{3}r}{(2\pi)^{3}}\, \frac{d^{3}q}{(2\pi)^{3}}\, \bigg[\mf{C}^{edb}_{mlj}(p,q,r)\ket{g^{a}_{i}(k)g^{d}_{l}(r)g^{c}_{k}(q)} \notag \\ & + \mf{C}^{eda}_{mli}(k,q,r)\ket{g^{d}_{l}(r)g^{c}_{k}(q)g^{b}_{j}(p)}\bigg]\notag \\ & 
    - \frac{1}{(2\pi)^{3/2}} \int \frac{d^{3}q}{(2\pi)^{3}}\,\bigg[\mf{C}^{\dagger\, abc}_{ijk}(q,p,k)+\mf{C}^{\dagger\, bac}_{jik}(q,k,p)\bigg]\ket{g^{c}_{k}(q)}  \notag \\ & 
    +  \int \frac{d^{3}q}{(2\pi)^{3}}\frac{d^{3}r}{(2\pi)^{3}} \,\bigg[ \mf{D}^{dcba}_{2lkji}(k,p,q,r)+\mf{D}^{dcab}_{2lkij}(p,k,q,r)\bigg]\ket{g^{c}_{k}(q)g^{d}_{l}(r)}\notag \\ & 
    +   (2\pi)^{3}\, \int  \frac{d^3r}{(2\pi)^3} \frac{d^3s}{(2\pi)^3} \frac{d^3u}{(2\pi)^3} \frac{d^3q}{(2\pi)^3}\,  \bigg[\mf{E}^{fedcba}_{1\,nmlkji}(k,p,q,r,s,u)\notag \\ & +\mf{E}^{fedcab}_{1\,nmlkij}(p,k,q,r,s,u)\bigg] \ket{g^{d}_{l}(r)g^{c}_{k}(q)g^{e}_{m}(s)g^{f}_{n}(u)}.
\end{align}
The only remaining unfixed coefficient is  $\mf{E}_{2}$, which is determined by examining three-gluon incoming states. The action of $\Omega$ on three-gluon state is given as follows
\begin{align}\label{eq:Omegaonggg}
    \Omega\ket{g^{a}_{i}(k)g^{b}_{j}(p)g^{c}_{k}(q)} & = \mf{N}\ket{g^{a}_{i}(k)g^{b}_{j}(p)g^{c}_{k}(q)}  -\frac{1}{(2\pi)^{3/2}}\, \mf{A}^{\dagger\, a}_{i}(k) \ket{g^{b}_{j}(p)g^{c}_{k}(q)} \notag \\ & - \frac{1}{(2\pi)^{3/2}}\, \mf{A}^{\dagger\, b}_{j}(p) \ket{g^{a}_{i}(k)g^{c}_{k}(q)} - \frac{1}{(2\pi)^{3/2}}\, \mf{A}^{\dagger\, c}_{k}(q) \ket{g^{a}_{i}(k)g^{b}_{j}(p)}\notag \\ & +  \frac{1}{(2\pi)^{3}}\bigg[\mf{B}^{ba}_{1ji}(k,p) \ket{g^{c}_{k}(q)} + \mf{B}^{bc}_{1jk}(q,p) \ket{g^{a}_{i}(k)}+\mf{B}^{ca}_{1ki}(k,q) \ket{g^{b}_{j}(p)}\notag \\ & +\mf{B}^{ab}_{1ij}(p,k) \ket{g^{c}_{k}(q)} + \mf{B}^{cb}_{1kj}(p,q) \ket{g^{a}_{i}(k)}+\mf{B}^{ac}_{1ik}(q,k) \ket{g^{b}_{j}(p)}  \bigg]\notag \\ & + \int \frac{d^{3}r}{(2\pi)^{3}}\, \bigg[\mf{B}^{da}_{3li}(k,r)\ket{g^{d}_{l}(r)g^{b}_{j}(p)g^{c}_{k}(q)}+\mf{B}^{db}_{3lj}(p,r)\ket{g^{a}_{i}(k)g^{d}_{l}(r)g^{c}_{k}(q)}\notag \\ & +\mf{B}^{dc}_{3lk}(q,r)\ket{g^{a}_{i}(k)g^{d}_{l}(r)g^{b}_{j}(p)}\bigg] \notag \\ & - \frac{1}{(2\pi)^{3/2}}\int \frac{d^{3}r}{(2\pi)^{3}}\,\bigg[\mf{C}^{\dagger\,bcd}_{jkl}(r,q,p)\ket{g^{a}_{i}(k)g^{d}_{l}(r)}+\mf{C}^{\dagger\,bad}_{jil}(r,k,p)\ket{g^{c}_{k}(q)g^{d}_{l}(r)}\notag \\ & +\mf{C}^{\dagger\,cad}_{kil}(r,k,q)\ket{g^{b}_{j}(p)g^{d}_{l}(r)}+\mf{C}^{\dagger\,cbd}_{kjl}(r,p,q)\ket{g^{a}_{i}(k)g^{d}_{l}(r)}\notag \\ & +\mf{C}^{\dagger\,abd}_{ijl}(r,p,k)\ket{g^{c}_{k}(q)g^{d}_{l}(r)}+\mf{C}^{\dagger\,acd}_{ikl}(r,q,k)\ket{g^{b}_{j}(p)g^{d}_{l}(r)}\bigg]\notag \\ & +  \frac{1}{(2\pi)^{3}}\int \frac{d^{3}r}{(2\pi)^{3}}\bigg[\mf{D}^{dcba}_{3lkji}(k,p,q,r)+\mf{D}^{dcab}_{3lkij}(p,k,q,r)+\mf{D}^{dabc}_{3lijk}(q,p,k,r)\notag \\ & + \mf{D}^{dbca}_{3ljki}(k,q,p,r)+\mf{D}^{dacb}_{3likj}(p,q,k,r)+\mf{D}^{dbac}_{3ljik}(q,k,p,r)\bigg]\ket{g^{d}_{l}(r)}\notag \\ & + \int \frac{d^{3}r}{(2\pi)^{3}}\frac{d^{3}s}{(2\pi)^{3}}\Bigg[\bigg[\mf{D}^{edcb}_{2mlkj}(p,q,r,s)+\mf{D}^{edbc}_{2mljk}(q,p,r,s)\bigg]\ket{g^{a}_{i}(k)g^{d}_{l}(r)g^{e}_{m}(s)}\notag \\ & + \bigg[\mf{D}^{edca}_{2mlki}(k,q,r,s)+\mf{D}^{edac}_{2mlik}(q,k,r,s)\bigg]\ket{g^{b}_{j}(p)g^{d}_{l}(r)g^{e}_{m}(s)}\notag \\ & +\bigg[\mf{D}^{edab}_{2mlij}(p,k,r,s)+\mf{D}^{edba}_{2mlji}(k,p,r,s)\bigg]\ket{g^{c}_{k}(q)g^{d}_{l}(r)g^{e}_{m}(s)}\Bigg]\notag \\ & +\int \frac{d^{3}r}{(2\pi)^{3}}\frac{d^{3}s}{(2\pi)^{3}}\frac{d^{3}u}{(2\pi)^{3}}\,\bigg[\mf{E}^{fedcba}_{2nmlkji}(k,p,q,r,s,u)+\mf{E}^{fedbca}_{2nmljki}(k,q,p,r,s,u)\notag \\ & +\mf{E}^{fedabc}_{2nmlijk}(q,p,k,r,s,u)+\mf{E}^{fedbac}_{2nmljik}(q,k,p,r,s,u)\notag \\ & +\mf{E}^{fedcab}_{2nmlkij}(p,k,q,r,s,u) +\mf{E}^{fedacb}_{2nmlikj}(p,q,k,r,s,u)\bigg]\ket{g^{d}_{l}(r)g^{e}_{m}(s)g^{f}_{n}(u)}.
\end{align}
The coefficients obtained from $\Omega\ket{0}$ were computed in Ref.~\cite{lublinsky2017high}, using LCPT about the vacuum. The remaining coefficients are fixed by matching the states generated by $\Omega$ to the light-cone wave functions obtained from LCPT. In other words, we compare the explicit expressions for $\Omega\ket{g}, \Omega\ket{gg}$, etc., to the corresponding LCWF components and this one-to-one matching determines the coefficients at the order we are working.

\subsection{Review of the vacuum incoming state}\label{sec:vac}
\subsubsection{One outgoing gluon}
We begin by reviewing the results obtained in Ref. ~\cite{lublinsky2017high}. Not all of the wave functions computed in this work are needed for our purposes. Only the LCWFs at $\mathcal{O}(g)$ and $\mathcal{O}(g^{2})$ enter the construction of $\Omega$ operator, and we will present these explicitly. For clarity, we will organize the relevant wave functions according to the number of outgoing gluons and the color charge densities $(\rho)$. There is one diagram at order $g$, it represents a single outgoing gluon emission. 

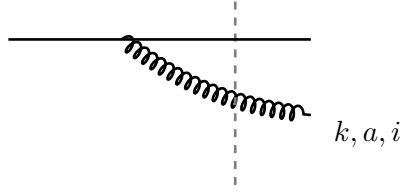
\begin{figure}[htbp]
\centering
    \begin{tikzpicture}[baseline={([yshift=-.8ex] current bounding box.center)}]
	\begin{pgfonlayer}{nodelayer}
		\node [style=none] (4) at (-12, 12) {};
		\node [style=none] (5) at (-7, 10) {};
		\node [style=none] (15) at (-9, 13) {};
		\node [style=none] (16) at (-9, 8) {};
        \node [style=none] (11) at (-5.5, 9.5) {$k, a, i$};
		\node [style=none] (18) at (-15, 12) {};
		\node [style=none] (19) at (-7, 12) {};
	\end{pgfonlayer}
	\begin{pgfonlayer}{edgelayer}
		\draw [style=gluon, bend right=15] (4.center) to (5.center);
		\draw [style=interm] (15.center) to (16.center);
		\draw [style=basic] (18.center) to (19.center);
	\end{pgfonlayer}
\end{tikzpicture}
\end{figure}
Here the dashed vertical lines denote the intermediate states. Applying the LCPT rules we have
\begin{align}
    \left|\Psi_{g\rho}^{LO}\right\rangle \, & =\- -\int_{\Lambda}^{\vee} dk^{+} \int d^{2}\mathbf{k}\, \ket{g^{a}_{i}(k)}\frac{\left\langle g_{i}^{a}(k)\left|\, H_{g}\,\right|0\right\rangle}{E_{g}(k)}.
\end{align}
Using the matrix element from Eq. \eqref{m1} and $E_{g}(k) = \frac{\v{k}^{2}}{2k^{+}}$
\begin{equation}
\left|\Psi_{g\rho}^{LO}\right\rangle \,=\- -\,\int_{\Lambda}^{\vee}\frac{dk^{+}}{\sqrt{k^{+}}}\,\int d^{2}\mathbf{k}\,\left|g_{i}^{a}(k)\right\rangle\frac{g\, \mathbf{k}^{i}}{2\pi^{3/2}\mathbf{k}^{2}}\,\rho^{a}(-\mathbf{k}).
\end{equation}

\subsubsection{Two outgoing gluons}
\subsubsection*{Single $\rho$ part}
To order $g^2$, there are three  diagrams contributing to the production of two-gluons in the final state We first consider,

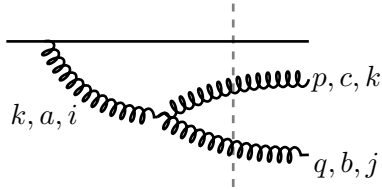
\begin{figure}[htbp]
\centering
    \begin{tikzpicture}[baseline={([yshift=-.8ex] current bounding box.center)}]
	\begin{pgfonlayer}{nodelayer}
		\node [style=none] (4) at (-14, 12) {};
		\node [style=none] (5) at (-11, 10) {};
		\node [style=none] (6) at (-7, 11) {};
		\node [style=none] (7) at (-11, 10) {};
		\node [style=none] (15) at (-9, 13) {};
		\node [style=none] (16) at (-9, 8) {};
		\node [style=none] (17) at (-7, 9) {};
		\node [style=none] (18) at (-15, 12) {};
		\node [style=none] (19) at (-7, 12) {};
        \node [style=none] (11) at (-14,10) {$k, a, i$};
        \node [style=none] (12) at (-6.0, 11) {$p,c,k$};
        \node [style=none] (13) at (-6.0, 8.75) {$q,b,j$};
	\end{pgfonlayer}
	\begin{pgfonlayer}{edgelayer}
		\draw [style=gluon, bend right] (4.center) to (5.center);
		\draw [style=gluon, bend right=15] (6.center) to (7.center);
		\draw [style=interm] (15.center) to (16.center);
		\draw [style=gluon, bend right=15] (7.center) to (17.center);
		\draw [style=basic] (18.center) to (19.center);
	\end{pgfonlayer}
\end{tikzpicture}
\end{figure}
\begin{align}\label{ggtwo}
\left|\Psi_{gg\,\rho}^{2}\right\rangle & = \frac{1}{2}\int_{\Lambda}^{\vee}dk^{+}\, dp^{+}\, dq^{+}\,\int d^{2}\mathbf{k}\, d^{2}\mathbf{p}\, d^{2}\mathbf{q}\, \left|g_{j}^{b}(q)\, g_{k}^{c}(p)\right\rangle\notag \\ &  \frac{\left\langle g_{j}^{b}(q)\, g_{k}^{c}(p)\left|H_{ggg}\right|g_{i}^{a}(k)\right\rangle \left\langle g_{i}^{a}(k)\left|H_{g}\right|0\right\rangle }{E_{gg}(p,\, q)\, E_{g}(k)}.
\end{align}
The momentum conservation is imposed via the delta function, it turns the momentum $q$ into either $k$ or $k-p$. By using the relevant matrix elements, Eq. \eqref{m1} and Eq. \eqref{m3}, we get
\begin{align}\label{ggmid}
\left|\Psi_{gg\,\rho}^{2}\right\rangle 
&=-\int_{\Lambda}^{\vee}dk^{+}\int_{\Lambda}^{k^{+}-\Lambda}dp^{+}\int d^{2}\mathbf{k}\,d^{2}\mathbf{p}\, \left|g_{j}^{b}(k-p)\, g_{k}^{c}(p)\right\rangle\,\notag \\ & \frac{igf^{abc}}{16\pi^{3/2}\left(\frac{\mathbf{p}^{2}}{2p^{+}}+\frac{(\mathbf{k}-\mathbf{p})^{2}}{2(k^{+}-p^{+})}\right)\left(\frac{\mathbf{k}^{2}}{2k^{+}}\right)\sqrt{k^{+}p^{+}(k^{+}-p^{+})}}
\Bigg[\left[2\mathbf{p}^{i}-\frac{2p^{+}}{k^{+}}\mathbf{k}^{i}\right]\delta_{jk}\notag \\ & +\left[\frac{k^{+}+p^{+}}{k^{+}-p^{+}}(\mathbf{k}^{j}-\mathbf{p}^{j})-\mathbf{k}^{j}-\mathbf{p}^{j}\right]\delta_{ik}+\left[2\mathbf{k}^{k}-\frac{2k^{+}}{p^{+}}\mathbf{p}^{k}\right]\delta_{ij}\Bigg]
 \left(\frac{g\rho^{a}(-\mathbf{k})\mathbf{k}^{i}}{4\pi^{3/2}|k^{+}|^{3/2}}\right). 
\end{align}
Next, we consider the instantaneous contribution with two outgoing gluons

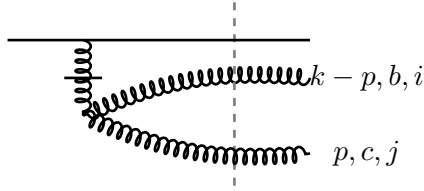
\begin{figure}[htbp]
\centering
    \begin{tikzpicture}[baseline={([yshift=-.8ex] current bounding box.center)}]
	\begin{pgfonlayer}{nodelayer}
		\node [style=none] (4) at (-13, 12) {};
		\node [style=none] (5) at (-13, 10) {};
		\node [style=none] (6) at (-7, 11) {};
		\node [style=none] (7) at (-13, 10) {};
		\node [style=none] (15) at (-9, 13) {};
		\node [style=none] (16) at (-9, 8) {};
		\node [style=none] (17) at (-7, 9) {};
		\node [style=none] (18) at (-15, 12) {};
		\node [style=none] (19) at (-7, 12) {};
		\node [style=none] (20) at (-13.5, 11) {};
		\node [style=none] (21) at (-12.5, 11) {};
        \node [style=none] (100) at (-5.5, 9) {$p,c,j$};
        \node [style=none] (101) at (-5.5, 11) {$k-p,b,i$};
	\end{pgfonlayer}
	\begin{pgfonlayer}{edgelayer}
		\draw [style=gluon] (4.center) to (5.center);
		\draw [style=gluon, bend right=15] (6.center) to (7.center);
		\draw [style=interm] (15.center) to (16.center);
		\draw [style=gluon, bend right=15] (7.center) to (17.center);
		\draw [style=basic] (18.center) to (19.center);
		\draw [style=basic] (20.center) to (21.center);
	\end{pgfonlayer}
\end{tikzpicture}
\end{figure}
\begin{align}\label{ggthree}
\left|\Psi_{gg\,\rho}^{3}\right\rangle & = -\frac{1}{2}\int_{\Lambda}^{\vee}dk^{+}\,\int_{\Lambda}^{k^{+}-\Lambda}dp^{+}\,\int d^{2}\mathbf{k}\, d^{2}\mathbf{p}\,\left|g_{i}^{b}(k-p)\, g_{j}^{c}(p)\right\rangle\notag \\ &  \frac{\left\langle g_{i}^{b}(k-p)\, g_{j}^{c}(p)\left|H_{gg-inst}\right|0\right\rangle }{E_{gg}(k-p,\, p)}.
\end{align}
By using the relevant matrix element (Eq. \eqref{gg1}) and changing the integration variables we get
\begin{align}\label{instgg3}
\left|\Psi_{gg\,\rho}^{3}\right\rangle & =  -\int_{\Lambda}^{\vee}dk^{+}\,\int_{\Lambda}^{k^{+}-\Lambda}dp^{+}\,\int d^{2}\mathbf{k}\, d^{2}\mathbf{p}\, \left|g_{j}^{b}(k-p)\, g_{k}^{c}(p)\right\rangle \, \notag \\ & \frac{ig^{2}f^{abc}(2p^{+}-k^{+})\rho^{a}(-\v{k})\delta_{jk}}{2(2\pi)^{3}\sqrt{p^{+}(k-p)^{+}}k^{+ 2}\bigg(\frac{(\v{k-p})^{2}}{(k-p)^{+}}+\frac{\v{p}^{2}}{p^{+}}\bigg)}.
\end{align}

\subsubsection*{Two $\rho$ part}
The contribution arising from the emission of two-gluons from the valence degrees of freedom is
\pagebreak
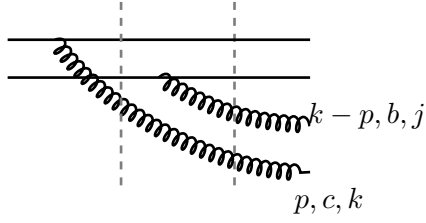
\begin{figure}[htbp]
\centering
   \begin{tikzpicture}[baseline={([yshift=-.8ex] current bounding box.center)}]
	\begin{pgfonlayer}{nodelayer}
		\node [style=none] (4) at (-15, 11) {};
		\node [style=none] (5) at (-7, 11) {};
		\node [style=none] (6) at (-7, 9.75) {};
		\node [style=none] (15) at (-9, 13) {};
		\node [style=none] (16) at (-9, 8) {};
		\node [style=none] (17) at (-7, 8.5) {};
		\node [style=none] (18) at (-15, 12) {};
		\node [style=none] (19) at (-7, 12) {};
		\node [style=none] (20) at (-11, 11) {};
		\node [style=none] (21) at (-13.75, 12) {};
		\node [style=none] (22) at (-12, 13) {};
		\node [style=none] (23) at (-12, 8) {};
        \node [style=none] (11) at (-6.5, 7.75) {$p, c, k$};
        \node [style=none] (12) at (-5.5, 10.0) {$k-p, b, j$};
	\end{pgfonlayer}
	\begin{pgfonlayer}{edgelayer}
		\draw [style=basic] (4.center) to (5.center);
		\draw [style=interm] (15.center) to (16.center);
		\draw [style=basic] (18.center) to (19.center);
		\draw [style=gluon, bend right=15] (20.center) to (6.center);
		\draw [style=gluon, bend right, looseness=0.75] (21.center) to (17.center);
		\draw [style=interm] (22.center) to (23.center);
	\end{pgfonlayer}
\end{tikzpicture}
\end{figure}\\
and its LCPT expression is 
\begin{align}
\left|\Psi_{gg\,\rho\rho}^{1}\right\rangle &\equiv\,\frac{1}{2}\int_{\Lambda}^{\vee}dq^{+}\, dk^{+}\,\int_{\Lambda}^{k^{+}-\Lambda}dp^{+}\,\int\, d^{2}\mathbf{q}\, d^{2}\mathbf{k}\, d^{2}\mathbf{p}\,\left|g_{j}^{b}(k-p)\, g_{k}^{c}(p)\right\rangle \notag \\ & \frac{\left\langle g_{j}^{b}(k-p)\, g_{k}^{c}(p)\left|H_{g}\right|g_{i}^{a}(q)\right\rangle \left\langle g_{i}^{a}(q)\left|H_{g}\right|0\right\rangle }{E_{gg}(p,\, k-p)\, E_{g}(q)}.
\end{align}
Using the appropriate matrix elements (eqs. \eqref{m2}, \eqref{m3}), we obtain
\begin{align}\label{ggrhorho1a}
\left|\Psi_{gg\,\rho\rho}^{1}\right\rangle & = \frac{1}{2}\int_{\Lambda}^{\vee}dk^{+}\, dq^{+}\,\int_{\Lambda}^{k^{+}-\Lambda}dp^{+}\,\int d^{2}\mathbf{k}\, d^{2}\mathbf{p}\, d^{2}\mathbf{q}\,\left|g_{j}^{b}(k-p)\, g_{k}^{c}(p)\right\rangle \notag \\ & \bigg(\delta^{ac}\delta_{ik}\delta^{(3)}(p-q)\, \frac{g\left(\mathbf{k}^{j}-\mathbf{p}^{j}\right)\rho^{b}(-\mathbf{k} +\mathbf{p})}{4\pi^{3/2}|k^{+}-p^{+}|^{3/2}} +\delta^{ab}\delta_{ji}\delta^{(3)}(k-p-q)\,\frac{g\mathbf{p}^{k}\rho^{c}(-\mathbf{p})}{4\pi^{3/2}|p^{+}|^{3/2}}\bigg)\notag \\ &  \times \left(\frac{g\mathbf{q}^{i}\rho^{a}(-\mathbf{q})}{4\pi^{3/2}|q^{+}|^{3/2}}\right)\times \frac{1}{\frac{\mathbf{q}^{2}}{2q^{+}}\left(\frac{\mathbf{p}^{2}}{2p^{+}}+\frac{(\mathbf{k}-\mathbf{p})^{2}}{2(k^{+}-p^{+})}\right)},
\end{align} 
This can be further simplified as
\begin{align}\label{ggrhorho1}
\left|\Psi_{gg\,\rho\rho}^{1}\right\rangle & = \frac{1}{2}\int_{\Lambda}^{\vee}dk^{+}\, dq^{+}\,\int_{\Lambda}^{k^{+}-\Lambda}dp^{+}\,\int d^{2}\mathbf{k}\, d^{2}\mathbf{p}\, d^{2}\mathbf{q}\,\left|g_{j}^{b}(k-p)\, g_{k}^{c}(p)\right\rangle\,\notag \\ &  \Bigg[\frac{1}{\frac{\mathbf{p}^{2}}{2p^{+}}\left(\frac{\mathbf{p}^{2}}{2p^{+}}+\frac{(\mathbf{k}-\mathbf{p})^{2}}{2(k^{+}-p^{+})}\right)} \times \frac{g^{2}(\v{k-p})^{j}p^{k}\rho^{b}(\v{-k+p})\rho^{c}(\v{-p})}{16\pi^{3}|k^{+}-p^{+}|^{3/2}|p^{+}|^{3/2}}\notag \\ & +\frac{1}{\frac{\mathbf{(k-p)}^{2}}{2(k^{+}-p^{+})}\left(\frac{\mathbf{p}^{2}}{2p^{+}}+\frac{(\mathbf{k}-\mathbf{p})^{2}}{2(k^{+}-p^{+})}\right)}\times\frac{g^{2}\v{p}^{k}(\v{k-p})^{j}\rho^{c}(\v{-p})\rho^{b}(\v{p-k})}{16\pi^{3}|k^{+}-p^{+}|^{3/2}|p^{+}|^{3/2}} \Bigg].
\end{align} 
Using the decomposition and the commutation relations
\begin{align}\label{decomposition}
\rho^{b}(-\mathbf{k}+\mathbf{p})\,\rho^{a}(-\mathbf{p}) &= \frac{1}{2}\left[\rho^{b}(-\mathbf{k}+\mathbf{p}),\,\rho^{a}(-\mathbf{p})\right]\,+\,\frac{1}{2}\left\{ \rho^{b}(-\mathbf{k}+\mathbf{p}),\,\rho^{a}(-\mathbf{p})\right\} \\
&= \frac{i}{2}f^{bad}\rho^{d}(-\mathbf{k})\,+\,\frac{1}{2}\left\{ \rho^{b}(-\mathbf{k}+\mathbf{p}),\,\rho^{a}(-\mathbf{p})\right\}, 
\end{align}
We will refer to the replacement in Eq. \eqref{decomposition} as the “decomposition procedure.” Its purpose is to clearly separate contributions coming from two distinct emitters located at different transverse positions from those in which both emitters sit at the same point. In the latter case, the result effectively reduces to emission from a single source, as shown by the color algebra. After applying this decomposition wherever it is relevant, each term in the LCWF is assigned an additional label that reflects the number of $\rho$ operators appearing in that component. So, we can decompose it as 
\begin{align}
\left|\Psi_{gg\,\rho\rho}^{1}\right\rangle \equiv \left|\Psi_{gg\,[\rho,\rho]}^{1a}\right\rangle + \left|\Psi_{gg\,\{\rho,\rho\}}^{1b}\right\rangle.
\end{align}
The first term contains one $\rho$ factor, is given by:
\begin{align}\label{part 1}
\left|\Psi_{gg\,[\rho,\rho]}^{1a}\right\rangle &=\int_{\Lambda}^{\vee}dk^{+}\,\int_{\Lambda}^{k^{+}-\Lambda}dp^{+}\,\int d^{2}\mathbf{k}\, d^{2}\mathbf{p}\,\left|g_{j}^{b}(k-p)\, g_{k}^{c}(p)\right\rangle\,\notag \\ & \frac{-ig^{2}f^{abc}\mathbf{p}^{k}\left(\mathbf{k}^{j}-\mathbf{p}^{j}\right)\rho^{a}(-\mathbf{k})\sqrt{p^{+}(k^{+}-p^{+})}}{16\pi^{3}\left(p^{+}(\mathbf{k}-\mathbf{p})^{2}+(k^{+}-p^{+})\mathbf{p}^{2}\right)}  \left(\frac{1}{(k^{+}-p^{+})\mathbf{p}^{2}}-\frac{1}{p^{+}(\mathbf{k}-\mathbf{p})^{2}}\right).
\end{align}
The second term contains the two $\rho$ factors:
\begin{align}\label{part 2}
\left|\Psi_{gg\,\{\rho,\rho\}}^{1b}\right\rangle &= \int_{\Lambda}^{\vee}dk^{+}\,\int_{\Lambda}^{k^{+}-\Lambda}dp^{+}\,\int d^{2}\mathbf{k}\, d^{2}\mathbf{p}\,\,\left|g_{j}^{b}(k-p)\, g_{k}^{c}(p)\right\rangle\,\notag \\ &  \frac{g^{2}\mathbf{p}^{k}\left(\mathbf{k}^{j}-\mathbf{p}^{j}\right)\left\{ \rho^{c}(-\mathbf{p}),\,\rho^{b}(-\mathbf{k}+\mathbf{p})\right\} }{16\pi^{3}\mathbf{p}^{2}(\mathbf{k}-\mathbf{p})^{2}\sqrt{p^{+}(k^{+}-p^{+})}} . 
\end{align}

\subsubsection{Summary of results}
We now collect all contributions according to their dependence on the valence color charge density operators $\rho$. In the CGC framework, the perturbative light-cone wave function naturally organizes into sectors carrying different powers of the color source. Since the color charge densities are operator valued and obey a non-Abelian algebra, each sector also carries a distinct color structure. Restricting to the accuracy required for the present calculation, the NLO wave function for an incoming vacuum state receives contributions containing one and two insertions of the color charge density. Organizing the result according to these structures, the full NLO wave function can be decomposed as
\begin{align}\label{NLOWF}
\left|\Psi|\right\rangle & = \mf{N}\left|0\right\rangle +\left|\Psi_{g\rho}^{\text{LO}}\right\rangle  
+\left|\Psi_{gg\,\rho}\right\rangle +\left|\Psi_{gg\,\rho\rho}\right\rangle.
\end{align}
The first term represents the vacuum component multiplied by the normalization factor $\mf{N}$, which is fixed by requiring that the full wave function remain normalized to unity order by order in perturbation theory. The second term, $\left|\Psi_{g\rho}^{\text{LO}}\right\rangle $, corresponds to the leading order emission of a single soft gluon from the valence color source. The remaining terms describe two-gluon components generated at higher order. The leading order wave function is given by
\begin{align}\label{lowav1mom}
\left|\Psi_{g\rho}^{\text{LO}}\right\rangle \,& =\- -\,\int_{\Lambda}^{\vee}\frac{dk^{+}}{\sqrt{k^{+}}}\int d^{2}\mathbf{k}\,\left|g_{i}^{a}(k)\right\rangle\,\frac{g\, \mathbf{k}^{i}}{2\pi^{3/2}\mathbf{k}^{2}}\,\rho^{a}(-\mathbf{k}).
\end{align}
$\left|\Psi_{gg\,\rho}\right\rangle$, which is at $\mathcal{O}(g^{2})$, is obtained from eqs. \eqref{part 1}, \eqref{ggmid}, \eqref{instgg3}
\begin{align}
    \left|\Psi_{gg\,\rho}\right\rangle  & = \left|\Psi^{1a}_{gg\,[\rho,\rho]}\right\rangle + \left|\Psi^{2}_{gg\,\rho}\right\rangle + \left|\Psi^{3}_{gg\,\rho}\right\rangle,
\end{align}
\begin{align}\label{psiggrhomom}
    \left|\Psi_{gg\,\rho}\right\rangle  & = \int_{\Lambda}^{\vee}dk^{+}\,\int d^{2}\mathbf{k}\, d^{2}\mathbf{p}\,\int_{\Lambda}^{k^{+}-\Lambda}dp^{+}\,\left|g_{j}^{b}(k-p)\, g_{k}^{c}(p)\right\rangle\,\notag \\ & \Bigg[\frac{-ig^{2}f^{abc}\mathbf{p}^{k}\left(\mathbf{k}^{j}-\mathbf{p}^{j}\right)\rho^{a}(-\mathbf{k})\sqrt{p^{+}(k^{+}-p^{+})}}{16\pi^{3}\left(p^{+}(\mathbf{k}-\mathbf{p})^{2}+(k^{+}-p^{+})\mathbf{p}^{2}\right)}  \left(\frac{1}{(k^{+}-p^{+})\mathbf{p}^{2}}-\frac{1}{p^{+}(\mathbf{k}-\mathbf{p})^{2}}\right) \notag \\ & + \frac{igf^{acb}}{16\pi^{3/2}\left(\frac{\mathbf{p}^{2}}{2p^{+}}+\frac{(\mathbf{k}-\mathbf{p})^{2}}{2(k^{+}-p^{+})}\right)\left(\frac{\mathbf{k}^{2}}{2k^{+}}\right)\sqrt{k^{+}p^{+}(k^{+}-p^{+})}}
\Bigg(\left[2\mathbf{p}^{i}-\frac{2p^{+}}{k^{+}}\mathbf{k}^{i}\right]\delta_{jk}\notag \\ & +\left[\frac{k^{+}+p^{+}}{k^{+}-p^{+}}(\mathbf{k}^{j}-\mathbf{p}^{j})-\mathbf{k}^{j}-\mathbf{p}^{j}\right]\delta_{ik} +\left[2\mathbf{k}^{k}-\frac{2k^{+}}{p^{+}}\mathbf{p}^{k}\right]\delta_{ij}\Bigg)\times
\left(\frac{g\rho^{a}(-\mathbf{k})\mathbf{k}^{i}}{4\pi^{3/2}|k^{+}|^{3/2}}\right) \Bigg]. 
\end{align}
$\left|\Psi_{gg\,\rho\rho}\right\rangle$ is obtained from Eq. \eqref{part 2}
\begin{align}
    \left|\Psi_{gg\, \rho\rho}\right\rangle & \equiv \left|\Psi_{gg\, \{\rho,\rho\}}\right\rangle,
\end{align}
\begin{align}\label{ggrhorhomom}
\left|\Psi_{gg\,\rho\rho}\right\rangle & = \int_{\Lambda}^{\vee}dk^{+}\,\int d^{2}\mathbf{k}\, d^{2}\mathbf{p}\,\int_{\Lambda}^{k^{+}-\Lambda}dp^{+}\,\left|g_{j}^{b}(k-p)\, g_{k}^{c}(p)\right\rangle\, \notag \\ 
& \frac{g^{2}\mathbf{p}^{k}\left(\mathbf{k}^{j}-\mathbf{p}^{j}\right)\left\{ \rho^{c}(-\mathbf{p}),\,\rho^{b}(-\mathbf{k}+\mathbf{p})\right\} }{16\pi^{3}\mathbf{p}^{2}(\mathbf{k}-\mathbf{p})^{2}\sqrt{p^{+}(k^{+}-p^{+})}}. 
\end{align}
The constant $\mf{N}$ is determined by imposing the normalization condition on the full wave function, $\langle\Psi|\Psi\rangle = 1$ to the perturbative order under consideration. The explicit form of normalization constant $(\mf{N})$ is given by
\begin{align}\label{normmom}
    \mf{N} & = 1- \frac{g^2}{8 \pi^3 }\, \int_{\Lambda}^{\vee} \frac{dp^{+}}{p^{+}}\,
  \int d^2p\, \frac{\rho^a(-\v{p}) \rho^a(\v{p})}{\v{p}^2} =   1 - \frac{g^2}{8 \pi^3 }\,\log{\left(\frac{\vee}{\Lambda}\right)}
  \int d^2p \frac{\rho^a(-\v{p}) \rho^a(\v{p})}{\v{p}^2}.
\end{align}

\subsection{Calculation of the single-gluon incoming state}\label{sec:singleglue}
 We will examine all contributions at orders $g$ and $g^{2}$ associated with single-gluon incoming states. The possible final configurations include vacuum, one, two, or three outgoing gluons. Finally, we will normalize the resulting wave function in the similar manner used for the vacuum incoming state and conclude with a summary of the key results.
 
\subsubsection{Vacuum outgoing state}
To order $g$, only one diagram contributes: the absorption of the incoming gluon by the valence $\rho$.

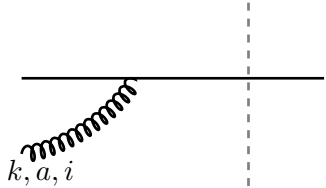
\begin{figure}[htbp]
\centering
   \begin{tikzpicture}[baseline={([yshift=-.8ex] current bounding box.center)}]
	\begin{pgfonlayer}{nodelayer}
		\node [style=none] (2) at (-15, 11) {};
		\node [style=none] (4) at (-15, 9) {};
		\node [style=none] (5) at (-12, 11) {};
		\node [style=none] (9) at (-7, 11) {};
		\node [style=none] (11) at (-14.5, 8.5) {$k, a, i$};
		\node [style=none] (15) at (-9, 13) {};
		\node [style=none] (16) at (-9, 8) {};
	\end{pgfonlayer}
	\begin{pgfonlayer}{edgelayer}
		\draw [style=basic] (2.center) to (9.center);
		\draw [style=gluon, bend right, looseness=0.75] (4.center) to (5.center);
		\draw [style=interm] (15.center) to (16.center);
	\end{pgfonlayer}
\end{tikzpicture}
\end{figure}
\begin{align}
\ket{ \left(\Psi^{a}_{i}(k) \right)_{\rho} }
   & =
   \ket{0}\frac{\bra{0}H_{g}\ket{g^a_{i}(k)}}{E_{g}(k)}
   \\
   &= \frac{g}{ 2\pi^{3/2}}\frac{k^i \rho^{a}(\v{k})}{\sqrt{k^+} \v{k}^{2}} \ket{0}  \,.
\end{align}
Here we used the LCPT transition element of Eq.~\eqref{m1}. The vertical dashed line in the diagram denotes the intermediate state.  
\subsubsection{One outgoing gluon}
\subsubsection*{Single $\rho$ part}
We start with the instantaneous interactions, see Eq.~\eqref{gg2} and   first consider  the process when the plus momentum flows from the valence degrees of freedom. 

\begin{figure}[htbp]
\centering
   \begin{tikzpicture}[baseline={([yshift=-.8ex] current bounding box.center)}]
	\begin{pgfonlayer}{nodelayer}
		\node [style=none] (0) at (-15, 12) {};
		\node [style=none] (4) at (-15, 9) {};
		\node [style=none] (5) at (-7, 9) {};
		\node [style=none] (6) at (-11, 12) {};
		\node [style=none] (7) at (-11, 9) {};
		\node [style=none] (8) at (-7, 12) {};
		\node [style=none] (11) at (-14.5, 8.5) {$k, a, i$};
		\node [style=none] (12) at (-7, 8.5) {$p, b, j$};
		\node [style=none] (15) at (-9, 13) {};
		\node [style=none] (16) at (-9, 8) {};
		\node [style=none] (17) at (-11.5, 10.5) {};
		\node [style=none] (18) at (-10.5, 10.5) {};
	\end{pgfonlayer}
	\begin{pgfonlayer}{edgelayer}
		\draw [style=basic] (0.center) to (8.center);
		\draw [style=gluon] (4.center) to (5.center);
		\draw [style=gluon] (6.center) to (7.center);
		\draw [style=interm] (15.center) to (16.center);
		\draw [style=basic] (17.center) to (18.center);
	\end{pgfonlayer}
\end{tikzpicture}
\end{figure}
\begin{align}\label{oneginst1a}
  \ket{ \left(\Psi^{a}_{i}(k) \right)^{1}_{g \rho\ {\rm inst.} } }
   & =\int_{k^{+}+\Lambda}^{\vee} 
   d p^+\, \int d^2\v{p}\,
   \ket{g_{j}^{b}(p)} 
   \frac{ \bra{g_{j}^{b}(p)}H_{gg\,\text{inst.}}\ket{g_{i}^{a}(k)}}{E_{g}(k)-E_{g}(p)}
\end{align}
Using the matrix element (Eq. \eqref{gg2}), we get
\begin{align}
 \ket{ \left(\Psi^{a}_{i}(k) \right)^{1}_{g \rho\ {\rm inst.} } }   & =  \frac{ig^2 f^{abc}}{(2\pi)^3} 
   \int_{k^+ + \Lambda}^{\vee}
   d p^+\, \int d^2\v{p}\, \ket{g_{j}^{b}(p)}
   \frac{\sqrt{p^{+}k^{+}} (k^++p^+) } {(p^+-k^+)^2} \frac{\rho^c(-\v{p}+\v{k})}{p^{+}\v{k}^2 - k^{+}\v{p}^2} \delta_{ij}
\end{align}

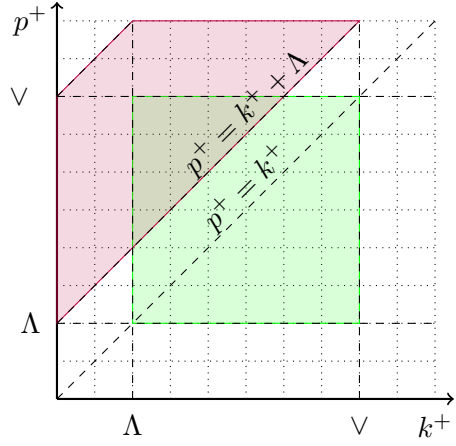
\begin{figure}[htbp]
\centering
\begin{tikzpicture}[baseline={([yshift=-.8ex] current bounding box.center)}]
  \draw[thin, dotted] (0, 0) grid (10,10);
  \draw[thick, ->] (0,0) -- (10.5,0); 
  \draw[thick, ->] (0,0) -- (0,10.5); 
  \draw[draw=green, fill=green, fill opacity = 0.15] (2,2) -- (8,2) -- (8,8) -- (2,8) -- cycle;
  \draw[draw=purple, fill=purple, fill opacity = 0.15] (0,2) -- (8,10) -- (2,10) -- (0,8) -- cycle;
  \draw[dashed] (0,0) -- (10,10);
  \draw[dashed] (0,2) -- (8,10);
  \draw[dashed] (0,8) -- (2,10);
  \draw[dashed] (2,0) -- (2,10);
  \draw[dashed] (0,2) -- (10,2);
  \draw[dashed] (8,0) -- (8,10);
  \draw[dashed] (0,8) -- (10,8);
  \node[] at (10, -0.7) {$k^+$};
  \node[] at (2, -0.7) {$\Lambda$};
  \node[] at (8, -0.7) {$\vee$};
  \node[] at (-0.7, 10) {$p^+$};
  \node[] at (-0.7,2) {$\Lambda$};
  \node[] at (-1.0, 8) {$\vee$};
  \node[rotate=45] at (5, 5.6) {$p^+=k^+$};
  \node[rotate=45] at (5, 7.6) {$p^+=k^++\Lambda$};
\end{tikzpicture}
\caption{Kinematic constraint on the soft gluon plus momentum for the gluon number preserving transitions off the valence background, Eqs.~\eqref{oneginst1a}--\eqref{oneginst2}. The intersection of the shaded regions defines the allowed range of integration with respect to $p^+$.
  The red region is due to the constraint $ \Lambda < p^+ - k^+ < \vee$, the green region --  $ \Lambda < p^+  < \vee $
and  $ \Lambda <  k^+ < \vee$.
}
\label{fig:kinem}
\end{figure}
The integration limits in this and in the subsequent gluon number preserving contributions are obtained by the requirement that both the incoming and outgoing soft gluons, as well as the longitudinal momentum exchanged with the valence sector, lie within the soft window between $\Lambda$ and $\vee$; the resulting allowed region of the $p^{+}$ integration is illustrated in Fig.~\ref{fig:kinem}. Similarly, for the  momentum flowing into the valence degrees of freedom we get identical expression modulo the integration limits:
\begin{align}\label{oneginst2}
  \ket{\left(\Psi^{a}_{i}(k) \right)^{2}_{g \rho\ {\rm inst.} } }
   & =  \frac{ig^2 f^{abc}}{(2\pi)^3}
   \int_{\Lambda}^{k^{+}-\Lambda}
   d p^+\, \int d^2\v{p}\, \ket{g_{j}^{b}(p)}
   \frac{\sqrt{p^{+}k^{+}} (k^++p^+) } {(p^+-k^+)^2} \frac{\rho^c(-\v{p}+\v{k})}{p^{+}\v{k}^2 - k^{+}\v{p}^2} \delta_{ij}
\end{align}

Non-instantaneous interaction also leads to two possible contributions to $g^2$ order. We consider first the diagram when the valence particle transfers plus momentum to the gluon 
(see Eq.~\eqref{m3} for three-gluon interaction vertex)

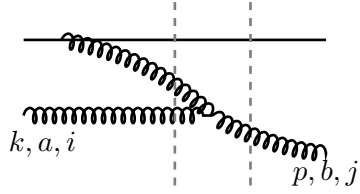
\begin{figure}[htbp]
    \centering
   \begin{tikzpicture}[baseline={([yshift=-.8ex] current bounding box.center)}]
	\begin{pgfonlayer}{nodelayer}
		\node [style=none] (0) at (-15, 12) {};
		\node [style=none] (4) at (-15, 10) {};
		\node [style=none] (5) at (-10, 10) {};
		\node [style=none] (6) at (-14, 12) {};
		\node [style=none] (7) at (-10, 10) {};
		\node [style=none] (8) at (-7, 12) {};
		\node [style=none] (11) at (-14.5, 9.25) {$k, a, i$};
		\node [style=none] (12) at (-7, 8.5) {$p, b, j$};
		\node [style=none] (15) at (-9, 13) {};
		\node [style=none] (16) at (-9, 8) {};
		\node [style=none] (17) at (-7, 9) {};
		\node [style=none] (18) at (-11, 13) {};
		\node [style=none] (19) at (-11, 8) {};
	\end{pgfonlayer}
	\begin{pgfonlayer}{edgelayer}
		\draw [style=basic] (0.center) to (8.center);
		\draw [style=gluon] (4.center) to (5.center);
		\draw [style=gluon, bend left=15] (6.center) to (7.center);
		\draw [style=interm] (15.center) to (16.center);
		\draw [style=gluon, bend right=15, looseness=0.75] (7.center) to (17.center);
		\draw [style=interm] (18.center) to (19.center);
	\end{pgfonlayer}
\end{tikzpicture}
\end{figure}
\begin{align}\label{onegrho1}
  \ket{ \left(\Psi^{a}_{i}(k) \right)^{1}_{g \rho} }
   & = 
  \frac{1}{2}\, \int d^3 r\, d^3p\, d^3q\, 
   \ket{g^{b}_{j}(p)}
   \frac{ \bra{g^{b}_{j}(p)}H_{ggg}\ket{g^{c}_{k}(q)g^{d}_{l}(r)}\bra{g^{c}_{k}(q)g^{d}_{l}(r)}H_{g}\ket{g^{a}_{i}(k)}}
   {(E_{g}(k)-E_{gg}(q,r))(E_{g}(k)-E_{g}(p))}
   \end{align} 
\begin{align}
 \ket{\left(\Psi^{a}_{i}(k) \right)^{1}_{g \rho} }
   & = -\frac{ig^{2}f^{abc}}{8\pi^{3}}\int_{k^{+}+\Lambda}^{\vee}\,dp^{+}\,\int d^{2}\v{p}\,\ket{g^{b}_{j}(p)}\, \frac{\rho^{c}(\v{k-p})(\v{p-k})^{k}}{\sqrt{p^{+}(p-k)^{+}k^{+}}|p-k|^{+\,3/2}}\notag \\ &   \Bigg[\bigg(2\v{k}^{j}-\v{p}^{j}+\frac{p^{+}-2k^{+}}{p^{+}}\v{p}^{j}\bigg) \delta_{ik} +\bigg(2\v{p}^{i}-\v{k}^{i} -\frac{2p^{+}-k^{+}}{k^{+}}\v{k}^{i}\bigg)\delta_{jk} \notag \\ & +\bigg(\frac{p^{+}+k^{+}}{p^{+}-k^{+}}(\v{p-k})^{k} - \v{k}^{k} - \v{p}^{k}\bigg)\delta_{ij}\Bigg]\times \frac{1}{\bigg(\frac{\v{k}^{2}}{k^{+}}-\frac{\v{p}^{2}}{p^{+}}\bigg)\frac{(\v{k-p})^{2}}{(k-p)^{+}}}
\end{align}
Then, we examine the process in which the gluon transfers plus momentum back to the valence degrees of freedom. This process highlights the complementary momentum flow relative to the previous case.

\begin{figure}[htbp]
    \centering
    \begin{tikzpicture}[baseline={([yshift=-.8ex] current bounding box.center)}]
	\begin{pgfonlayer}{nodelayer}
		\node [style=none] (0) at (-15, 12) {};
		\node [style=none] (4) at (-15, 10) {};
		\node [style=none] (5) at (-12, 10) {};
		\node [style=none] (6) at (-9.75, 12) {};
		\node [style=none] (7) at (-12, 10) {};
		\node [style=none] (8) at (-7, 12) {};
		\node [style=none] (11) at (-14.5, 9.25) {$k, a, i$};
		\node [style=none] (12) at (-7, 8.5) {$p, b, j$};
		\node [style=none] (15) at (-9, 13) {};
		\node [style=none] (16) at (-9, 8) {};
		\node [style=none] (17) at (-7, 9) {};
		\node [style=none] (18) at (-11, 13) {};
		\node [style=none] (19) at (-11, 8) {};
	\end{pgfonlayer}
	\begin{pgfonlayer}{edgelayer}
		\draw [style=basic] (0.center) to (8.center);
		\draw [style=gluon] (4.center) to (5.center);
		\draw [style=gluon, bend right, looseness=0.75] (6.center) to (7.center);
		\draw [style=interm] (15.center) to (16.center);
		\draw [style=gluon, bend right=15, looseness=0.75] (7.center) to (17.center);
		\draw [style=interm] (18.center) to (19.center);
	\end{pgfonlayer}
\end{tikzpicture}
\end{figure}
\begin{align}\label{onegrho2}
  \ket{ \left(\Psi^{a}_{i}(k) \right)^{2}_{g \rho} }
   & = 
\frac{1}{2}\,  \int d^{3}r\, d^{3}q\,  d^{3}p\,
  \ket{g^{b}_{j}(p)}\,
  \frac{\bra{g^{b}_{j}(p)}H_{g}\ket{g^{c}_{k}(q)g^{d}_{l}(r)}\bra{g^{c}_{k}(q)g^{d}_{l}(r)}H_{ggg}\ket{g^{a}_{i}(k)}}{(E_{g}(k)-E_{gg}(q,r))(E_{g }(k)-E_{g}(p))}
\end{align}
Using the matrix elements (eqs. \eqref{m2}, \eqref{m3}), we obtain the following
 \begin{align}
 \ket{ \left(\Psi^{a}_{i}(k) \right)^{2}_{g \rho} }
   &  = \int_{\Lambda}^{k^{+}-\Lambda}dp^{+}\, \int d^{2}\v{p}\, \ket{g^{b}_{j}(p)}\,\frac{ig^{2}f^{abc}(\v{k-p})^{k}\rho^{c}(\v{k-p})}{8\pi^{3}\sqrt{k^{+}(k-p)^{+}p^{+}}|k-p|^{+\, 3/2}}\notag \\ &  \Bigg[\bigg(2\v{p}^{i}-\v{k}^{i} +\frac{k^{+}-2p^{+}}{k^{+}}\v{k}^{i}\bigg)\delta_{jk} +\bigg(2\v{k}^{j}-\v{p}^{j}-\frac{2k^{+}-p^{+}}{p^{+}}\v{p}^{j}\bigg)\delta_{ik}\notag \\ & +\bigg(\frac{k^{+}+p^{+}}{k^{+}-p^{+}}(\v{k-p})^{k}-\v{p}^{k}-\v{k}^{k}\bigg)\delta_{ij}\Bigg]\times \frac{1}{\bigg(\frac{\v{k}^{2}}{k^{+}}-\frac{(\v{k-p})^{2}}{(k-p)^{+}}-\frac{\v{p}^{2}}{p^{+}}\bigg) \bigg(\frac{\v{k}^{2}}{k^{+}}-\frac{\v{p}^{2}}{p^{+}}\bigg)}
\end{align} 

\subsubsection*{Two $\rho$ part}
In the expressions below, the wave functions correspond to different time orderings in which an outgoing gluon is first absorbed by the valence parton and the valence subsequently emits another gluon. These orderings must be treated separately, as each leads to a distinct intermediate state and contributes differently to the overall wave function.

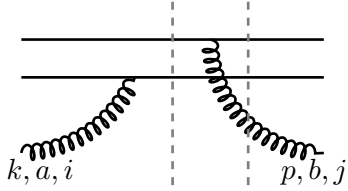
\begin{figure}[htbp]
    \centering
    \begin{tikzpicture}[baseline={([yshift=-.8ex] current bounding box.center)}]
	\begin{pgfonlayer}{nodelayer}
		\node [style=none] (0) at (-15, 12) {};
		\node [style=none] (2) at (-15, 11) {};
		\node [style=none] (4) at (-15, 9) {};
		\node [style=none] (5) at (-12, 11) {};
		\node [style=none] (6) at (-10, 12) {};
		\node [style=none] (7) at (-7, 9) {};
		\node [style=none] (8) at (-7, 12) {};
		\node [style=none] (9) at (-7, 11) {};
		\node [style=none] (11) at (-14.5, 8.5) {$k, a, i$};
		\node [style=none] (12) at (-7.25, 8.5) {$p, b, j$};
		\node [style=none] (13) at (-11, 13) {};
		\node [style=none] (14) at (-11, 8) {};
		\node [style=none] (15) at (-9, 13) {};
		\node [style=none] (16) at (-9, 8) {};
	\end{pgfonlayer}
	\begin{pgfonlayer}{edgelayer}
		\draw [style=basic] (0.center) to (8.center);
		\draw [style=basic] (2.center) to (9.center);
		\draw [style=gluon, bend right] (4.center) to (5.center);
		\draw [style=gluon, bend right=45] (6.center) to (7.center);
		\draw [style=interm] (13.center) to (14.center);
		\draw [style=interm] (15.center) to (16.center);
	\end{pgfonlayer}
\end{tikzpicture}
\end{figure}
\begin{align}
  \ket{ \left(\Psi^{a}_{i}(k) \right)^{1}_{g\, \rho \rho} }
   &=\int_{\Lambda}^{\vee} 
   d p^+\, \int d^2\v{p} \,
   \ket{g_{j}^{b}(p)}\,
   \frac{\bra{g_{j}^{b}(p)}H_{g}\ket{0}\bra{0}H_{g}\ket{g_{i}^{a}(k)}}{E_{g}(k) (E_{g}(k) - E_{g}(p))},
   \end{align}  
 \begin{align}
    \ket{ \left(\Psi^{a}_{i}(k) \right)^{1}_{g\, \rho \rho} }
   & =  
   \frac{g^2}{4\pi^3}
  \int_{\Lambda}^{\vee}\,
   \frac{dp^{+}}{\sqrt{p^{+}}}\, \int d^2\v{p}\,\ket{g_{j}^{b}(p)}\,\frac{\v{k}^{i}\v{p}^{j}\rho^{b}(-\v{p})\rho^{a}(\v{k})}{(p^{+} \v{k}^{2} -k^{+}\v{p}^{2}) \v{k}^{2}}\, \sqrt{k^{+}}.
   \label{m1grr}
\end{align}
Here we used the matrix elements of Eq.~\eqref{m1}. For different time ordering, we obtain the following 

\begin{figure}[htbp]
    \centering
    \begin{tikzpicture}[baseline={([yshift=-.8ex] current bounding box.center)}]
	\begin{pgfonlayer}{nodelayer}
		\node [style=none] (0) at (-15, 12) {};
		\node [style=none] (2) at (-15, 11) {};
		\node [style=none] (4) at (-15, 9) {};
		\node [style=none] (5) at (-9.75, 11) {};
		\node [style=none] (6) at (-13, 12) {};
		\node [style=none] (7) at (-7, 9) {};
		\node [style=none] (8) at (-7, 12) {};
		\node [style=none] (9) at (-7, 11) {};
		\node [style=none] (11) at (-14.5, 8.5) {$k, a, i$};
		\node [style=none] (12) at (-7.25, 8.5) {$p, b, j$};
		\node [style=none] (13) at (-11, 13) {};
		\node [style=none] (14) at (-11, 8) {};
		\node [style=none] (15) at (-9, 13) {};
		\node [style=none] (16) at (-9, 8) {};
	\end{pgfonlayer}
	\begin{pgfonlayer}{edgelayer}
		\draw [style=basic] (0.center) to (8.center);
		\draw [style=basic] (2.center) to (9.center);
		\draw [style=gluon, bend right=15, looseness=0.75] (4.center) to (5.center);
		\draw [style=gluon, bend right] (6.center) to (7.center);
		\draw [style=interm] (13.center) to (14.center);
		\draw [style=interm] (15.center) to (16.center);
	\end{pgfonlayer}
\end{tikzpicture}
\end{figure}
\begin{align}
\ket{\left(\Psi^{a}_{i}(k) \right)^{2}_{g\, \rho \rho}} 
   & =
   \frac{1}{2}\,\int_{\Lambda}^{\vee} d r^+\,  dq^+\,  dp^+\,
   \int d^2 \v{r}\, d^2\v{q}\, d^2\v{p}\,
   \ket{g_{j}^{b}(p)}
   \frac{\bra{g_{j}^{b}(p)}H_{g}\ket{g_{k}^{c}(q)g_{l}^{d}(r)}}{(E_{g}(k)-E_{gg}(q,r))}\notag \\ & \frac{\bra{g_{k}^{c}(q)g_{l}^{d}(r)}H_{g}\ket{g_{i}^{a}(k)}}{(E_{g}(k)-E_{g}(p))},
   \end{align}
With the help of Eq.~\eqref{m2}, we obtain
   \begin{align}
  \ket{\left(\Psi^{a}_{i}(k) \right)^{2}_{g\, \rho \rho}} 
   & = - \frac{g^2}{4\pi^3}
  \int_{\Lambda}^{\vee}\, dp^{+} \,
   \int d^{2}\v{p}\, \ket{g_{j}^{b}(p)}\, 
    \frac{\v{k}^{i}\v{p}^{j}\rho^{a}(\v{k})\rho^{b}(-\v{p}) }{\v{p}^{2}(p^{+}\v{k}^{2}-k^{+}\v{p}^{2})}\sqrt{\frac{p^{+}}{k^{+}}}.
   \label{m2grr}
\end{align}
Therefore, the sum of $\ket{ \left(\Psi^{a}_{i}(k) \right)^{1}_{g \rho \rho} }$ and  $\ket{ \left(\Psi^{a}_{i}(k) \right)^{2}_{g \rho \rho} }$, i.e., eqs. \eqref{m1grr} and \eqref{m2grr}, can be rewritten using the decomposition of valence charges given in Eq.~\eqref{decomposition} as 
\begin{align}
\ket{ \left(\Psi^{a}_{i}(k) \right)^{1}_{g \rho \rho} }
+\ket{ \left(\Psi^{a}_{i}(k) \right)^{2}_{g \rho \rho} }  = 
\ket{ \left(\Psi^{a}_{i}(k) \right)^{1}_{g \{\rho, \rho\} } }
+\ket{ \left(\Psi^{a}_{i}(k) \right)^{2}_{g [\rho, \rho] } }  
\end{align}
with 
\begin{align}\label{onegrho3mom}
\ket{ \left(\Psi^{a}_{i}(k) \right)_{g \{\rho, \rho\} } }
 = &  
 - \frac{g^2}{8\pi^3}
  \int_{\Lambda}^{\vee}\frac{dp^{+}}{\sqrt{k^{+}p^{+}}}\, \int d^2\v{p}\,\ket{g_{j}^{b}(p)}\,\frac{\v{k}^{i}\v{p}^{j}} {\v{k}^{2} \v{p}^2 }\,
  \{\rho^{a}(\v{k}),\rho^{b}(-\v{p}) \}
\end{align}
and 
\begin{align}\label{onegrho4mom}
  \ket{ \left(\Psi^{a}_{i}(k) \right)_{g [\rho, \rho] } }
 & = 
 - \frac{g^2 f^{abc}}{8\pi^3}
 \int_{\Lambda}^{\vee}dp^{+}\, \int d^2\v{p}\, \ket{g_{j}^{b}(p)}\, \frac{\v{k}^{i}\v{p}^{j}} {\v{k}^{2} \v{p}^2 }\,\frac{\v{p}^2 k^{+}+ p^{+} \v{k}^2}{\v{p}^2 k^{+}- p^{+} \v{k}^2}  \rho^{c}(\v{k} -\v{p}) 
\end{align}

\subsubsection{Two outgoing gluons}

\subsubsection*{$\rho$ independent part}
There is only one diagram that contributes to the two-gluon outgoing state. The corresponding wave function arises from a single-gluon splitting into two-gluons via the three-gluon interaction vertex, providing the sole contribution of this type. 

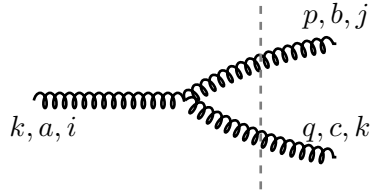
\begin{figure}[htbp]
    \centering
    \begin{tikzpicture}[baseline={([yshift=-.8ex] current bounding box.center)}]
	\begin{pgfonlayer}{nodelayer}
		\node [style=none] (2) at (-15, 10.5) {};
		\node [style=none] (4) at (-11, 10.5) {};
		\node [style=none] (5) at (-7, 12) {};
		\node [style=none] (9) at (-11, 10.5) {};
		\node [style=none] (11) at (-14.75, 9.75) {$k, a, i$};
		\node [style=none] (15) at (-9, 13) {};
		\node [style=none] (16) at (-9, 8) {};
		\node [style=none] (17) at (-7, 9) {};
		\node [style=none] (18) at (-7, 9.75) {$q, c,  k$};
		\node [style=none] (19) at (-7, 12.75) {$p, b, j$};
	\end{pgfonlayer}
	\begin{pgfonlayer}{edgelayer}
		\draw [style=gluon] (2.center) to (9.center);
		\draw [style=gluon, bend left=15, looseness=0.75] (4.center) to (5.center);
		\draw [style=interm] (15.center) to (16.center);
		\draw [style=gluon, bend right=15, looseness=0.75] (9.center) to (17.center);
	\end{pgfonlayer}
\end{tikzpicture}
\end{figure}
\begin{align}
  \ket{ \left(\Psi^{a}_{i}(k) \right)_{gg} }
   & =
   \frac{1}{2}\, \int d^{3}p\, d^{3}q \, 
   \ket{g^{c}_{k}(q)g^{b}_{j}(p)}\, 
   \frac{\bra{g^{c}_{k}(q)g^{b}_{j}(p)} H_{ggg} \ket{g^{a}_{i}(k)}}{(E_{g}(k)-E_{gg}(p,q))},
   \end{align}  
Using the matrix element (Eq. \eqref{m3}), we obtain
   \begin{align}\label{rhoindep}
   \ket{ \left(\Psi^{a}_{i}(k) \right)_{gg} }
   & =  \int d^{3}p\,\int d^{3}q\,\ket{g^{c}_{k}(q)g^{b}_{j}(p)}\,\frac{igf^{abc}\delta^{(3)}(k-p-q)}{8\pi^{3/2}\sqrt{k^{+}p^{+}q^{+}}\bigg(\frac{\v{k}^{2}}{k^{+}}-\frac{\v{p}^{2}}{p^{+}}-\frac{\v{q}^{2}}{q^{+}}\bigg)}\notag \\ & \Bigg[\bigg(\v{p}^{i}-\v{q}^{i}+\frac{q^{+}-p^{+}}{k^{+}}\v{k}^{i}\bigg)\delta_{jk} +\bigg(\v{k}^{j}+\v{q}^{j}-\frac{k^{+}+q^{+}}{p^{+}}\v{p}^{j}\bigg)\delta_{ik}\notag \\ &  +\bigg(\frac{k^{+}+p^{+}}{q^{+}}\v{q}^{k}-\v{p}^{k}-\v{k}^{k}\bigg)\delta_{ij}\Bigg]. 
\end{align}

\subsubsection{Three outgoing gluons}
There are four diagrams that contribute to the three-gluon outgoing state at order $g^{2}$. 

\subsubsection*{$\rho$ independent part}
In this case, the wave function comes from a single-gluon splitting into three-gluons through the four gluon interaction vertex, yielding the unique $g^{2}$ contribution of this type.

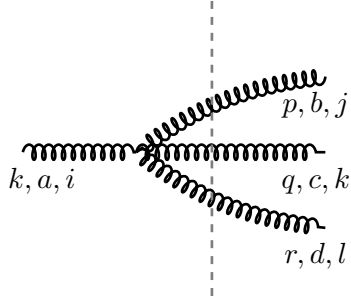
\begin{figure}[htbp]
    \centering
    \begin{tikzpicture}[baseline={([yshift=-.8ex] current bounding box.center)}]
	\begin{pgfonlayer}{nodelayer}
		\node [style=none] (4) at (-15, 10) {};
		\node [style=none] (5) at (-12, 10) {};
		\node [style=none] (6) at (-7, 12) {};
		\node [style=none] (7) at (-12, 10) {};
		\node [style=none] (11) at (-14.5, 9.25) {$k, a, i$};
		\node [style=none] (12) at (-7.25, 11.25) {$p, b, j$};
		\node [style=none] (17) at (-7, 8) {};
		\node [style=none] (18) at (-7, 10) {};
		\node [style=none] (19) at (-7.25, 9.25) {$q, c, k$};
		\node [style=none] (20) at (-7.25, 7.25) {$r, d, l$};
		\node [style=none] (21) at (-10, 14) {};
		\node [style=none] (22) at (-10, 6) {};
	\end{pgfonlayer}
	\begin{pgfonlayer}{edgelayer}
		\draw [style=gluon] (4.center) to (5.center);
		\draw [style=gluon, bend right=15] (6.center) to (7.center);
		\draw [style=gluon, bend right=15] (7.center) to (17.center);
		\draw [style=gluon] (7.center) to (18.center);
		\draw [style=interm] (21.center) to (22.center);
	\end{pgfonlayer}
\end{tikzpicture} 
\end{figure}
\begin{equation}
    \ket{(\Psi^{a}_{i}(k))^{1}_{ggg}}  = \frac{1}{6}\int d^{3}r d^{3}q d^{3}p\ket{g^{d}_{l}(r)g^{c}_{k}(q)g^{b}_{j}(p)}\frac{\bra{g^{d}_{l}(r)g^{c}_{k}(q)g^{b}_{j}(p)}H_{gggg}\ket{g^{a}_{i}(k)}}{E_{g}(k)-E_{ggg}(p,q,r)}
\end{equation}
Using the matrix element in Eq. \eqref{gg2}, we obtain
\begin{align}
     \ket{(\Psi^{a}_{i}(k))^{1}_{ggg}} &  = \frac{g^{2}}{24(2\pi)^{3}}\int\, d^{3}p\, d^{3}q\, d^{3}r\, \ket{g^{d}_{l}(r)g^{c}_{k}(q)g^{b}_{j}(p)}\, \frac{\delta^{(3)}(-k+p+q+r)}{\sqrt{k^{+}p^{+}q^{+}r^{+}}}\Bigg[f^{a'ab}f^{a'cd}\notag \\ & \bigg(\delta_{lj}\delta_{ik}-\delta_{il}\delta_{kj} +\frac{(p^{+}+k^{+})(r^{+}-q^{+})}{(r^{+}+q^{+})(k^{+}-p^{+})}\delta_{lk}\delta_{ij}\bigg) +f^{a'ac}f^{a'bd}\bigg(\delta_{lk}\delta_{ij}-\delta_{jk}\delta_{il} \notag \\ & +\frac{(q^{+}+k^{+})(r^{+}-p^{+})}{(r^{+}+p^{+})(k^{+}-q^{+})}\delta_{ik}\delta_{lj}\bigg) +f^{a'ad}f^{a'bc}\bigg(\delta_{lk}\delta_{ij}-\delta_{lj}\delta_{ik}  \notag \\ & +\frac{(r^{+}+k^{+})(q^{+}-p^{+})}{(p^{+}+q^{+})(k^{+}-r^{+})}\delta_{jk}\delta_{il}\bigg)\Bigg]\times \frac{1}{\left(\frac{\v{k}^{2}}{2k^{+}}-\frac{\v{p}^{2}}{2p^{+}}-\frac{\v{q}^{2}}{2q^{+}}-\frac{\v{r}^{2}}{2r^{+}}\right)}
\end{align}
We consider a contribution in which an incoming gluon splits into two-gluons and one of the resulting gluons subsequently emits another gluon.

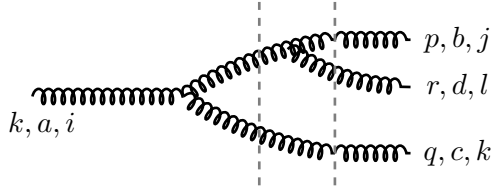
\begin{figure}[htbp]
    \centering
    \begin{tikzpicture}[baseline={([yshift=-.8ex] current bounding box.center)}]
	\begin{pgfonlayer}{nodelayer}
		\node [style=none] (2) at (-15, 10.5) {};
		\node [style=none] (4) at (-11, 10.5) {};
		\node [style=none] (5) at (-7, 12) {};
		\node [style=none] (9) at (-11, 10.5) {};
		\node [style=none] (11) at (-14.75, 9.75) {$k, a, i$};
		\node [style=none] (15) at (-9, 13) {};
		\node [style=none] (16) at (-9, 8) {};
		\node [style=none] (17) at (-7, 9) {};
		\node [style=none] (18) at (-3.75, 9) {$q, c,  k$};
		\node [style=none] (19) at (-3.75, 12) {$p, b, j$};
		\node [style=none] (20) at (-8.25, 11.75) {};
		\node [style=none] (21) at (-5, 9) {};
		\node [style=none] (22) at (-5, 12) {};
		\node [style=none] (23) at (-5, 10.75) {};
		\node [style=none] (24) at (-4, 10.75) {};
		\node [style=none] (25) at (-4, 10.75) {};
		\node [style=none] (26) at (-3.75, 10.75) {$r,d,l$};
		\node [style=none] (27) at (-7, 13) {};
		\node [style=none] (28) at (-7, 8) {};
	\end{pgfonlayer}
	\begin{pgfonlayer}{edgelayer}
		\draw [style=gluon] (2.center) to (9.center);
		\draw [style=gluon, bend left=15, looseness=0.75] (4.center) to (5.center);
		\draw [style=interm] (15.center) to (16.center);
		\draw [style=gluon, bend right=15, looseness=0.75] (9.center) to (17.center);
		\draw [style=gluon, bend right=15] (20.center) to (23.center);
		\draw [style=gluon] (5.center) to (22.center);
		\draw [style=gluon] (17.center) to (21.center);
		\draw [style=interm] (27.center) to (28.center);
	\end{pgfonlayer}
\end{tikzpicture} 
\end{figure}
\begin{align}
    \ket{(\Psi^{a}_{i}(k))^{2}_{ggg}} &  = 2\times\,\frac{1}{12}\, \int\, d^{3}p\,d^{3}q\, d^{3}r\, d^{3}u\, d^{3}v\,\ket{g^{b}_{j}(p)\,g^{c}_{k}(q)\,g^{d}_{l}(r)}\,\notag \\ & \frac{\bra{g^{b}_{j}(p)\,g^{c}_{k}(q)\,g^{d}_{l}(r)}H_{ggg}\ket{g^{f}_{n}(v)\,g^{e}_{m}(u)}}{E_{g}(k)-E_{ggg}(p,q,r)}\,\frac{\bra{g^{f}_{n}(v)\,g^{e}_{m}(u)}H_{ggg}\ket{g^{a}_{i}(k)}}{E_{g}(k)-E_{gg}(u,v)},
\end{align}
There is a factor of 2 because the gluon can also be emitted from the other gluon produced in the splitting. Using the matrix elements in Eqs. \eqref{m2} and \eqref{m3}, we get
\begin{align}
    \ket{(\Psi^{a}_{i}(k))^{2}_{ggg}} & = -\frac{g^{2}\, f^{adf}\,f^{fbc}}{32\pi^{3}}\, \int\, d^{3}p\, d^{3}q\, d^{3}r\, d^{3}v\, \ket{g^{b}_{j}(p)\,g^{c}_{k}(q)\,g^{d}_{l}(r)}\notag \\  & \frac{\delta^{(3)}(k-r-v)\, \delta^{(3)}(v-p-q)}{\sqrt{k^{+}r^{+}v^{+}}\,\sqrt{v^{+}p^{+}q^{+}}\,\left(\frac{\v{k}^{2}}{k^{+}}-\frac{\v{r}^{2}}{r^{+}}-\frac{\v{v}^{2}}{v^{+}}\right)}\,\bigg[\left(\v{r}^{i}-\v{v}^{i}+\frac{v^{+}-r^{+}}{k^{+}}\,\v{k}^{i}\right)\, \delta_{nl} \notag \\ & + \left(\v{k}^{l}+\v{v}^{l}-\frac{k^{+}+v^{+}}{r^{+}}\,\v{r}^{l}\right)\delta_{ni} + \left(\frac{k^{+}+r^{+}}{v^{+}}\,\v{v}^{n} -\v{r}^{n}-\v{k}^{n}\right)\,\delta_{il}\bigg] \times \notag \\ &  \bigg[\left(p^{n}-q^{n}+\frac{q^{+}-p^{+}}{v^{+}}\,v^{n}\right)\delta_{kj} + \left(v^{j}+q^{j}-\frac{v^{+}+q^{+}}{p^{+}}\,p^{j}\right)\,\delta_{nk}\notag \\ & +\left(\frac{v^{+}+p^{+}}{q^{+}}\,k^{k}-p^{k}-v^{k}\right)\,\delta_{jn}\bigg]\times \frac{1}{\left(\frac{k^{2}}{k^{+}}-\frac{p^{2}}{p^{+}}-\frac{q^{2}}{q^{+}}-\frac{r^{2}}{r^{+}}\right)}.
\end{align}
\subsubsection*{One $\rho$ part}
We consider a contribution in which the valence charge emits a gluon, and the incoming gluon subsequently splits into two-gluons through the standard $g\rightarrow gg$ splitting vertex.

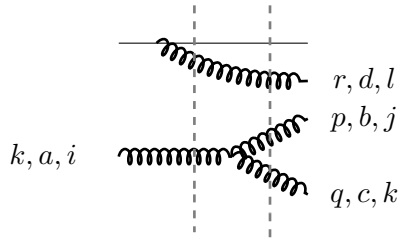
\begin{figure}[htbp]
    \centering
   \begin{tikzpicture}
	\begin{pgfonlayer}{nodelayer}
		\node [style=none] (0) at (-7, 6) {};
		\node [style=none] (1) at (-2, 6) {};
		\node [style=none] (2) at (-7, 3) {};
		\node [style=none] (3) at (-4, 3) {};
		\node [style=none] (4) at (-2, 4) {};
		\node [style=none] (5) at (-2, 2) {};
		\node [style=none] (6) at (-6, 6) {};
		\node [style=none] (7) at (-2, 5) {};
		\node [style=none] (8) at (-5, 7) {};
		\node [style=none] (9) at (-5, 2) {};
		\node [style=none] (10) at (-3, 7) {};
		\node [style=none] (11) at (-3, 1.5) {};
		\node [style=none] (12) at (-9, 3) {};
		\node [style=none] (13) at (-0.5, 5) {};
		\node [style=none] (14) at (-0.5, 4) {};
		\node [style=none] (15) at (-0.5, 2) {};
		\node [style=none] (16) at (-3, 0.75) {};
		\node [style=none] (17) at (-5, 1) {};
		\node [style=none] (18) at (-5, 1) {};
		\node [style=none] (19) at (-5, 2) {};
		\node [style=none] (20) at (-9, 3) {$k,a,i$};
		\node [style=none] (21) at (-0.5, 5) {$r,d,l$};
		\node [style=none] (22) at (-0.5, 4) {};
		\node [style=none] (23) at (-0.5, 4) {$p,b,j$};
		\node [style=none] (24) at (-0.5, 2) {$q,c,k$};
	\end{pgfonlayer}
	\begin{pgfonlayer}{edgelayer}
		\draw (0.center) to (1.center);
		\draw [style=gluon, bend right=15] (6.center) to (7.center);
		\draw [style=gluon] (2.center) to (3.center);
		\draw [style=gluon] (3.center) to (4.center);
		\draw [style=gluon] (3.center) to (5.center);
		\draw [style=interm] (8.center) to (9.center);
		\draw [style=interm] (10.center) to (11.center);
		\draw [style=interm] (19.center) to (18.center);
		\draw [style=interm] (11.center) to (16.center);
	\end{pgfonlayer}
\end{tikzpicture}
\end{figure}
\begin{align}
   \ket{(\Psi^{a}_{i}(k))^{1}_{ggg\,\rho}} & = \frac{1}{12}\, \int d^{3}p \, d^{3}q \, d^{3}r \, d^{3}u \, d^{3}v\, \ket{g^{b}_{j}(p)g^{c}_{k}(q)g^{d}_{l}(r)} \notag \\  & \frac{\bra{g^{b}_{j}(p)g^{c}_{k}(q)g^{d}_{l}(r)}H_{ggg}\ket{g^{f}_{n}(v)g^{e}_{m}(u)}}{E_{g}(k)-E_{ggg}(p,q,r)}\,\frac{\bra{g^{f}_{n}(v)g^{e}_{m}(u)}H_{g}\ket{g^{a}_{i}(k)}}{E_{g}(k)-E_{gg}(v,u)}, 
\end{align}
Using the matrix elements in eqs. \eqref{m2} and \eqref{m3}, we get
 \begin{align}
     \ket{(\Psi^{a}_{i}(k))^{1}_{ggg\,\rho}} & = -\frac{ig^{2}f^{abc}}{32\pi^{3}}\,\int d^{3}p \, d^{3}q \, d^{3}r \, \ket{g^{b}_{j}(p)g^{c}_{k}(q)g^{d}_{l}(r)}\, \frac{\rho^{d}(-\v{r})\delta^{(3)}(k-p-q)}{\sqrt{k^{+}p^{+}q^{+}r^{+}}}\times \frac{\v{r}^{l}}{\v{r}^{2}}\notag \\ &  \Bigg[\bigg(\v{p}^{i}-\v{q}^{i}+\frac{q^{+}-p^{+}}{k^{+}}\v{k}^{i}\bigg)\delta_{jk} +\bigg(\v{k}^{j}+\v{q}^{j}-\frac{k^{+}+q^{+}}{p^{+}}\v{p}^{j}\bigg)\delta_{ik} \notag \\ & +\bigg(\frac{k^{+}+p^{+}}{q^{+}}\v{q}^{k}-\v{p}^{k}-\v{k}^{k}\bigg)\delta_{ij}\Bigg]  \times \frac{1}{\bigg(\frac{\v{k}^{2}}{2k^{+}}-\frac{\v{p}^{2}}{2p^{+}}-\frac{\v{q}^{2}}{2q^{+}}-\frac{\v{r}^{2}}{2r^{+}}\bigg)}
 \end{align}
We consider the time-ordered complementary contribution to the previous diagram, in which the incoming gluon first splits into two-gluons, after which the valence charge emits a gluon. This corresponds to the alternative ordering of the same interaction vertices.

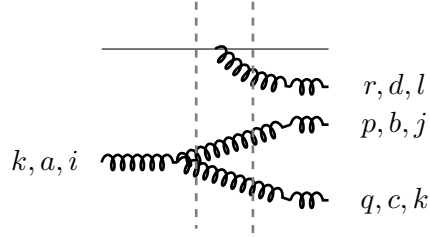
\begin{figure}[htbp]
    \centering
    \begin{tikzpicture}
	\begin{pgfonlayer}{nodelayer}
		\node [style=none] (0) at (-9, 6) {};
		\node [style=none] (1) at (-3, 6) {};
		\node [style=none] (2) at (-9, 3) {};
		\node [style=none] (3) at (-7, 3) {};
		\node [style=none] (4) at (-4, 4) {};
		\node [style=none] (5) at (-4, 2) {};
		\node [style=none] (6) at (-6, 6) {};
		\node [style=none] (7) at (-4, 5) {};
		\node [style=none] (8) at (-3, 5) {};
		\node [style=none] (9) at (-6.5, 7.25) {};
		\node [style=none] (10) at (-6.5, 1.25) {};
		\node [style=none] (11) at (-5, 7.25) {};
		\node [style=none] (12) at (-5, 1) {};
		\node [style=none] (13) at (-11, 3) {};
		\node [style=none] (14) at (-3, 4) {};
		\node [style=none] (15) at (-3, 2) {};
		\node [style=none] (16) at (-1.25, 5) {};
		\node [style=none] (17) at (-1.25, 4) {};
		\node [style=none] (18) at (-1.25, 2) {};
		\node [style=none] (19) at (-10.5, 3) {};
		\node [style=none] (20) at (-10.5, 3) {$k,a,i$};
		\node [style=none] (21) at (-1.25, 5) {$r,d,l$};
		\node [style=none] (22) at (-1.25, 4) {$p,b,j$};
		\node [style=none] (23) at (-1.25, 2) {$q,c,k$};
	\end{pgfonlayer}
	\begin{pgfonlayer}{edgelayer}
		\draw (0.center) to (1.center);
		\draw [style=gluon, bend right=15, looseness=1.25] (6.center) to (7.center);
		\draw [style=gluon] (2.center) to (3.center);
		\draw [style=gluon] (3.center) to (4.center);
		\draw [style=gluon] (3.center) to (5.center);
		\draw [style=gluon] (7.center) to (8.center);
		\draw [style=interm] (9.center) to (10.center);
		\draw [style=interm] (11.center) to (12.center);
		\draw [style=gluon] (4.center) to (14.center);
		\draw [style=gluon] (5.center) to (15.center);
	\end{pgfonlayer}
\end{tikzpicture}
\end{figure}
\begin{align}
  \ket{(\Psi^{a}_{i}(k))^{2}_{ggg\,\rho}} & = \frac{1}{12}\,\int d^{3}p \, d^{3}q \, d^{3}r \, d^{3}u \, d^{3}v\, \ket{g^{b}_{j}(p)g^{c}_{k}(q)g^{d}_{l}(r)}\,\notag \\ &    \frac{\bra{g^{b}_{j}(p)g^{c}_{k}(q)g^{d}_{l}(r)}H_{g}\ket{g^{e}_{m}(u)g^{f}_{n}(v)}}{E_{g}(k)-E_{ggg}(p,q,r)}\, \frac{\bra{g^{f}_{n}(v)g^{e}_{m}(u)}H_{ggg}\ket{g^{a}_{i}(k)}}{E_{g}(k)-E_{gg}(u,v)},
\end{align}
Using the matrix elements in eqs. \eqref{m2} and \eqref{m3}, we get
\begin{align}
   \ket{(\Psi^{a}_{i}(k))^{2}_{ggg\,\rho}}  & = - \frac{ig^{2}f^{abc}}{32\pi^{3}}\int d^{3}p \, d^{3}q \, d^{3}r \, \ket{g^{b}_{j}(p)g^{c}_{k}(q)g^{d}_{l}(r)}\,\frac{\delta^{(3)}(k-p-q)\rho^{d}(-\v{r})}{\sqrt{k^{+}p^{+}q^{+}r^{+}}}\notag \\ &  \times \frac{\v{r}^{l}}{r^{+}\,\left(\frac{\v{k}^{2}}{k^{+}}-\frac{\v{p}^{2}}{p^{+}}-\frac{\v{q}^{2}}{q^{+}}\right)} \Bigg[\bigg(\v{p}^{i}-\v{q}^{i}+\frac{q^{+}-p^{+}}{k^{+}}\v{k}^{i}\bigg)\delta_{jk}+\bigg(\v{k}^{j}+\v{q}^{j}-\frac{k^{+}+q^{+}}{p^{+}}\v{p}^{j}\bigg)\notag \\ & \delta_{ik} +\bigg(\frac{k^{+}+p^{+}}{q^{+}}\v{q}^{k}-\v{p}^{k}-\v{k}^{k}\bigg)\delta_{ij}\Bigg] \times \frac{1}{\bigg(\frac{\v{p}^{2}}{2p^{+}}+\frac{\v{q}^{2}}{2q^{+}}+\frac{\v{r}^{2}}{2r^{+}}-\frac{\v{k}^{2}}{2k^{+}}\bigg)}.
\end{align}

\subsubsection{Summary of results}\label{sec:norm}
We now assemble all contributions to the interacting single-gluon wave function up to order $g^{2}$. The complete result is obtained by summing all wave function components derived in the single-gluon incoming section. Organizing the result according to the number of soft gluons and powers of the valence color charge density, the interacting one gluon state takes the form
\begin{align}
 \notag 
  \ket{\Psi^{a}_{i}(k)}
 & = \mf{N} \ket{ g_i^a(k) }
 \\ \notag &+ 
 \ket{ \left(\Psi^{a}_{i}(k) \right)_{\rho} }
+\ket{ \left(\Psi^{a}_{i}(k) \right)_{gg} }
+\ket{ \left(\Psi^{a}_{i}(k) \right)_{gg\,\rho} }
\\ & +\ket{ \left(\Psi^{a}_{i}(k) \right)^{1}_{g\, \rho\ {\rm inst.} } }
+\ket{ \left(\Psi^{a}_{i}(k) \right)^{2}_{g \rho\ {\rm inst.} } }
+\ket{ \left(\Psi^{a}_{i}(k) \right)^{1}_{g \rho} }
+\ket{ \left(\Psi^{a}_{i}(k) \right)^{2}_{g \rho} }
\notag \\ & +\ket{ \left(\Psi^{a}_{i}(k) \right)_{g \{\rho\, \rho\}} }
+\ket{  \left(\Psi^{a}_{i}(k) \right)_{g\, [\rho ,\rho]}}\notag \\ &   +\ket{(\Psi^{a}_{i}(k))^{(1)}_{ggg}}+\ket{(\Psi^{a}_{i}(k))^{(2)}_{ggg}}+\ket{(\Psi^{a}_{i}(k))^{1}_{ggg\,\rho}}+\ket{(\Psi^{a}_{i}(k))^{2}_{ggg\,\rho}}
\label{Eq:intergluon}
\end{align}
The first line contains the non-interacting one gluon state multiplied by the normalization factor $\mf{N}$. The second and third lines contain all perturbative corrections contributing at orders $g$ and $g^{2}$. The various components correspond to transitions into sectors containing different numbers of soft gluons, as well as different color structures involving either single or multiple insertions of the color charge density operators $\rho$. The normalization factor $\mf{N}$ is determined by imposing the normalization condition on the interacting gluon state
\begin{align}
  \bra{\Psi^{a'}_{i'}(k')}
  {\Psi^{a}_{i}(k)} \rangle
  = \delta^{a a'} \delta_{i i'} \delta^{3}(k-k'). 
\end{align}
Since we work only up to order $g^{2}$, not all overlaps between the different wave function components contribute. In particular, products involving two terms already of order $g^{2}$, or products mixing $g$ and $g^{2}$ order components, contribute only beyond the perturbative accuracy considered here and can therefore be neglected. Among the nontrivial overlaps, one potentially contributing term is the overlap between the two-gluon components i.e., $\ket{ \left(\Psi^{a}_{i}(k) \right)_{gg} }$  and  $\ket{ \left(\Psi^{a}_{i}(k) \right)_{gg\rho} }$. However, due to the symmetry properties of the corresponding states, the combination
\begin{align}
    \bra{ (\Psi^{a'}_{i'}(k'))_{gg\rho} }
{ \left(\Psi^{a}_{i}(k) \right)_{gg} }\rangle + 
\bra{ (\Psi^{a'}_{i'}(k') )_{gg} }
{ \left(\Psi^{a}_{i}(k) \right)_{gg\,\rho} }\rangle  
 = 0  
\end{align}
vanishes identically. Another important contribution arises from the overlap between the bare one gluon state and the component proportional to the symmetric double source structure $\{\rho,\rho\}$:
\begin{align}
  \bra{(\Psi^{a^{\prime}}_{i^{\prime}}(k^{\prime}))_{g\, \{\rho,\rho\} } }  
 g^{a}_{i}(k) \rangle + 
\bra{ g^{a^{\prime}}_{i^{\prime}}(k^{\prime}) }  
   (\Psi^{a}_{i}(k))_{g\, \{\rho,\rho\}}  \rangle  = 
   - \frac{g^2}{4\pi^3 \sqrt{k^+ k^{\prime +}}} 
 \frac{k^{i} k^{\prime i^{\prime}}}{\v{k}^2 \v{k}^{\prime 2}} 
 \{\rho^{a}(\v{k}),\rho^{a^{\prime}}(-\v{k}^{\prime}) \}
\end{align}
Collecting all contributions relevant at order $g^{2}$, the normalization condition becomes
\begin{align}
  \bra{\Psi^{a'}_{i'}(k')}
  {\Psi^{a}_{i}(k)} \rangle
  & = \mf{N}^2 \delta^{a a'} \delta_{i i'} \delta^{3}(k-k') 
  \notag \\ & 
  + \mf{N}\,\left( \bra{ (\Psi^{a'}_{i'}(k'))_{g \{\rho,\rho\} } }  
    g^{a}_{i}(k) \rangle + 
    \bra{ g^{a'}_{i'}(k') }  
  (\Psi^{a}_{i}(k))_{g \{\rho,\rho\}}  \rangle \right) \\ \notag   
            & +  \bra{ (\Psi^{a'}_{i'}(k') )_{\rho} } (\Psi^{a}_{i}(k) )_{\rho}\rangle     
            +  \bra{ (\Psi^{a'}_{i'}(k') )_{gg\rho} } (\Psi^{a}_{i}(k) )_{gg\rho}\rangle     
           \notag \\ &  +  \bra{ (\Psi^{a'}_{i'}(k') )_{gg} } (\Psi^{a}_{i}(k) )_{gg}\rangle     
  \notag \\ &  = \delta^{a a'} \delta_{i i'} \delta^{3}(k-k') 
\end{align}
The normalization coefficient itself is expanded perturbatively, 
$$\mf{N} = 1 + g\, \mf{N}_{1} +  g^{2}\, \mf{N}_{2}.$$
and one immediately finds that $\mf{N}_{1} = 0$. Retaining terms only up to order $g^{2}$, the normalization condition reduces to
\begin{align}
\label{Eq:N2}
 & 2 g^2 {\cal N}_2  \delta^{a a'} \delta_{i i'} \delta^{3}(k-k') 
 +  \bra{ (\Psi^{a'}_{i'}(k') )_{gg} } (\Psi^{a}_{i}(k) )_{gg}\rangle     
  \\ \notag    & 
  + \bra{ (\Psi^{a'}_{i'}(k'))_{g \{\rho,\rho\} } }  
  g^{a}_{i}(k) \rangle \notag \\ & + 
  \bra{ g^{a'}_{i'}(k') }  
  (\Psi^{a}_{i}(k))_{g\, \{\rho,\rho\}}  \rangle 
  +  \bra{ (\Psi^{a'}_{i'}(k') )_{\rho} } (\Psi^{a}_{i}(k) )_{\rho}\rangle  
  +  \bra{ (\Psi^{a'}_{i'}(k') )_{gg\,\rho} } (\Psi^{a}_{i}(k) )_{gg\rho}\rangle 
  = 0
\end{align}
The overlaps involving the $\rho$ dependent terms can be evaluated explicitly and combine into a compact expression proportional to $\log{\left(\frac{\vee}{\Lambda}\right)}$. The only remaining contribution requiring special attention is the overlap of the purely two-gluon sector,
\begin{align}
  \bra{ (\Psi^{a^{\prime}}_{i^{\prime}}(k^{\prime}) )_{\rho} } (\Psi^{a}_{i}(k) )_{\rho}\rangle   = 
  \frac{g^2}{4 \pi^3 \sqrt{k^+ k^{\prime +}}} \frac{k^{i} k^{\prime i^{\prime}}}{\v{k}^2\v{k}^{'2'}} \rho^{a^{\prime}} (-\v{k}^\prime) \rho^a (\v{k})
\end{align}
and 
\begin{align}
  \bra{ (\Psi^{a^{\prime}}_{i^{\prime}}(k^{\prime}) )_{gg\rho} } (\Psi^{a}_{i}(k) )_{gg\rho}\rangle   &  = 
  \frac{g^2}{4 \pi^3 \sqrt{k^+ k^{\prime +}}} \frac{k^{i} k^{\prime i^{\prime}}}{\v{k}^2\v{k}^{'2'}}\rho^a (\v{k}) \rho^{a^{\prime}} (-\v{k}^{\prime})
 \notag \\ &  + 
 \delta^{a a^{\prime}} \delta_{i i^{\prime}}  \delta^3(k-k^{\prime})  
  \frac{g^2}{4 \pi^3 }\,\log{\left(\frac{\vee}{\Lambda}\right)}\,
  \int d^2p \frac{\rho^a(-\v{p}) \rho^a(\v{p})}{\v{p}^2}
  \end{align} 
Thus the contributions in the last line  of Eq.~\eqref{Eq:N2} sums up to 
\begin{align}
  &\bra{ (\Psi^{a^{\prime}}_{i^{\prime}}(k^{\prime}))_{g \{\rho,\rho\} } }  
  g^{a}_{i}(k) \rangle + 
  \bra{ g^{a^{\prime}}_{i^{\prime}}(k^{\prime}) }  
  (\Psi^{a}_{i}(k))_{g \{\rho,\rho\}}  \rangle 
  +  \bra{ (\Psi^{a^{\prime}}_{i^{\prime}}(k^{\prime}) )_{\rho} } (\Psi^{a}_{i}(k) )_{\rho}\rangle    
   \notag \\
  & +  \bra{ (\Psi^{a^{\prime}}_{i^{\prime}}(k^{\prime}) )_{gg\rho} } (\Psi^{a}_{i}(k) )_{gg\rho}\rangle   
 = 
 \delta^{a a^{\prime}} \delta_{i i^{\prime}}  \delta^3(k-k^{\prime})  
  \frac{g^2}{4 \pi^3 } \,\log{\left(\frac{\vee}{\Lambda}\right)}\,
  \int d^2p \frac{\rho^a(-\v{p}) \rho^a(\v{p})}{\v{p}^2} .
\end{align}
The remaining contribution in Eq.~\eqref{Eq:N2} requires special attention:
\begin{align}
 \bra{ (\Psi^{a^{\prime}}_{i^{\prime}}(k^{\prime}) )_{gg} } (\Psi^{a}_{i}(k) )_{gg}\rangle &  = 
 \frac{g^2N_c}{8\pi^3} \delta^{aa^{\prime}} \delta^{3}(k-k^{\prime}) \delta_{ii^{\prime}} 
  \int_{\frac{\Lambda}{k^{+}}}^{1-\frac{\Lambda}{k^{+}}}  d\xi \left(\xi(1-\xi) + \frac{1-\xi}{\xi}  + \frac{\xi}{1-\xi} \right)\notag \\ &  
 \int d^2 \tilde p \frac{1}{\tilde{\v{p}}^2}
\end{align}
The longitudinal fraction integral can be readily performed 
\begin{align}
  \int_{\frac{\Lambda}{k^{+}}}^{1-\frac{\Lambda}{k^{+}}}  d\xi \left(\xi(1-\xi) + \frac{1-\xi}{\xi} + \frac{\xi}{1-\xi} \right) = 
   -2 \log \bigg(\frac{\Lambda}{k^{+}} \bigg)- \frac{11}{6}\,.
\end{align}
The transverse integral is divergent both in UV and IR. Putting everything together, we obtain 
\begin{align}\label{onegrhonorm}
 \bra{ (\Psi^{a^{\prime}}_{i^{\prime}}(k^{\prime}) )_{gg} } (\Psi^{a}_{i}(k) )_{gg}\rangle  = 
 - \frac{g^2N_c}{8\pi^3} \delta_{aa^{\prime}} \delta^3(k-k^{\prime}) \delta_{ii^{\prime}} 
 \left(  2 \log \bigg(\frac{\Lambda}{k^{+}} \bigg) + \frac{11}{6}  \right)
 \int \frac{d^2\tilde p}{\tilde p^2} 
\end{align}
We need to note that the integral $\int \frac{d^2\tilde p}{\tilde p^2}$ vanishes because it is a scaleless integral ~\cite{peskin2018introduction}, because there is no intrinsic momentum scale in the integrand or integration measure. In dimensional regularization, such integrals are set to zero since they contain overlapping UV and IR divergences with no scale to distinguish them. Therefore the normalization coefficient is given by
\begin{align}
\mf{N} & =   1 - \frac{g^2}{8 \pi^3 }\,\log{\left(\frac{\vee}{\Lambda}\right)}
  \int d^2p \frac{\rho^a(-\v{p}) \rho^a(\v{p})}{\v{p}^2}.
\end{align}
\subsection{Calculation of the two-gluon incoming states}\label{sec:Twoglue}
We will now examine all contributions at order $g^{2}$ associated with an incoming state containing two-gluons. At this order in perturbation theory, the allowed final state configurations consist either of two outgoing gluons, corresponding to virtual corrections or four outgoing gluons, corresponding to real emission processes. We will focus only on those diagrams that contribute directly to the construction of the next-to-leading order evolution operator and therefore we will not discuss diagrams that do not enter the evolution kernel. Furthermore, since the relevant light-cone wave functions first appear at order $g^{2}$, their contribution to physical observables is already of the required perturbative order.

\subsubsection{Two outgoing gluons}
There are six diagrams which are required that corresponds to two outgoing gluons at order $g^{2}$. From this point onward, we will indicate all momentum-exchange symmetries (such as $p\leftrightarrow k$ and similar permutations) implicitly, as writing out every interchange explicitly would make the already lengthy wave function expressions unnecessarily cumbersome.

\subsubsection*{$\rho$ independent part}
In this contribution, the wave function is generated by the interaction of the two incoming gluons through the four gluon interaction vertex, in which the incoming gluon pair directly converts into two outgoing gluons. In addition to the ordinary four gluon interaction term, there is also an instantaneous contribution arising in the light-cone Hamiltonian formalism. Since both contributions possess the same external structure and enter at the same perturbative order, we will treat them together in the construction of the corresponding wave function. The entire process contributes at order $g^{2}$, as the four gluon interaction itself already carries two powers of the coupling constant. At this order, this constitutes the unique contribution of this type.

\begin{figure}[htbp]
    \centering
    \begin{tikzpicture}[baseline={([yshift=-.8ex] current bounding box.center)}]
	\begin{pgfonlayer}{nodelayer}
		\node [style=none] (4) at (-15, 12) {};
		\node [style=none] (5) at (-12, 10) {};
		\node [style=none] (6) at (-7, 12) {};
		\node [style=none] (7) at (-12, 10) {};
		\node [style=none] (11) at (-15, 8.75) {$k, a, i$};
		\node [style=none] (12) at (-15, 11.25) {$p, b, j$};
		\node [style=none] (17) at (-15, 8) {};
		\node [style=none] (18) at (-7, 8) {};
		\node [style=none] (19) at (-7.25, 11.25) {$q, c, k$};
		\node [style=none] (20) at (-7.25, 8.75) {$r, d, l$};
		\node [style=none] (21) at (-11, 14) {};
		\node [style=none] (22) at (-11, 6) {};
	\end{pgfonlayer}
	\begin{pgfonlayer}{edgelayer}
		\draw [style=gluon, bend left] (4.center) to (5.center);
		\draw [style=gluon, bend right=15] (6.center) to (7.center);
		\draw [style=gluon, bend left] (7.center) to (17.center);
		\draw [style=gluon, bend right=15] (7.center) to (18.center);
		\draw [style=interm] (21.center) to (22.center);
	\end{pgfonlayer}
\end{tikzpicture} 
\end{figure}
\begin{align}
     \ket{(\Psi^{ba}_{ji}(k,p))^{1}_{gg}} &  = \frac{1}{2}\,\int d^{3}r d^{3}q\, \ket{g^{d}_{l}(r)g^{c}_{k}(q)}\frac{\bra{g^{d}_{l}(r)g^{c}_{k}(q)}H_{gggg}\ket{g^{a}_{i}(k)g^{b}_{j}(p)}}{E_{gg}(k,p)-E_{gg}(q,r)} \notag \\ & + \frac{1}{2}\,\int d^{3}r d^{3}q\, \ket{g^{d}_{l}(r)g^{c}_{k}(q)}\frac{\bra{g^{d}_{l}(r)g^{c}_{k}(q)}H_{gg-gg\,\text{inst.}}\ket{g^{a}_{i}(k)g^{b}_{j}(p)}}{E_{gg}(k,p)-E_{gg}(q,r)}
\end{align}
Using the matrix element (Eq. \eqref{gg-gg}), we obtain
\begin{align}
 \ket{(\Psi^{ba}_{ji}(k,p))^{1}_{gg}}& =    \frac{g^{2}}{16(2\pi)^{3}}\int d^{3}r\, d^{3}q\, \ket{g^{d}_{l}(r)g^{c}_{k}(q)}\frac{\delta^{(3)}(-k+r-p+q)}{\sqrt{k^{+}q^{+}r^{+}p^{+}}}\notag \\ & \Bigg[f^{a'ad}f^{a'cb}\bigg(\delta_{ik}\delta_{jl}-\delta_{lk}\delta_{ij}+\frac{(q^{+}+p^{+})(r^{+}+k^{+})}{(k^{+}-r^{+})(p^{+}-q^{+})}\delta_{il}\delta_{jk}\bigg)\notag \\ & +f^{a'ac}f^{a'db}\bigg(\delta_{jk}\delta_{il}-\delta_{kl}\delta_{ij}+\frac{(p^{+}+r^{+})(q^{+}+k^{+})}{(k^{+}-q^{+})(p^{+}-r^{+})}\delta_{lj}\delta_{ik}\bigg)\notag \\ & + f^{a'ab}f^{a'dc}\bigg(\delta_{il}\delta_{jk}-\delta_{ik}\delta_{lj}+\frac{(q^{+}-r^{+})(k^{+}-p^{+})}{(k^{+}+p^{+})(r^{+}+q^{+})}\delta_{ij}\delta_{lk}\bigg) \Bigg]\times \notag \\ & \frac{1}{\left(\frac{\v{k}^{2}}{2k^{+}}+\frac{\v{p}^{2}}{2p^{+}}-\frac{\v{q}^{2}}{2q^{+}}-\frac{\v{r}^{2}}{2r^{+}}\right)} + p\leftrightarrow k 
\end{align}
Next, we consider two incoming gluons first combine through a three-gluon vertex to form an intermediate gluon, which then undergoes a second three-gluon interaction and splits into two outgoing gluons. This process therefore involves two distinct three-gluon vertices.

\begin{figure}[htbp]
    \centering
    \begin{tikzpicture}
	\begin{pgfonlayer}{nodelayer}
		\node [style=none] (0) at (-7, 5) {};
		\node [style=none] (1) at (-7, 2.25) {};
		\node [style=none] (2) at (-3, 3.75) {};
		\node [style=none] (3) at (0, 3.75) {};
		\node [style=none] (4) at (4, 5) {};
		\node [style=none] (5) at (4, 2) {};
		\node [style=none] (6) at (-2, 6) {};
		\node [style=none] (7) at (-2, 1) {};
		\node [style=none] (8) at (2, 6) {};
		\node [style=none] (9) at (2, 1) {};
		\node [style=none] (10) at (-9, 5) {};
		\node [style=none] (11) at (-9, 2.25) {};
		\node [style=none] (12) at (6, 5) {};
		\node [style=none] (13) at (6, 2) {};
		\node [style=none] (14) at (-9, 5) {$p,b,j$};
		\node [style=none] (15) at (-9, 2.25) {$q,c,k$};
		\node [style=none] (16) at (6, 5) {$r,d,l$};
		\node [style=none] (17) at (6, 2) {$s,e,m$};
	\end{pgfonlayer}
	\begin{pgfonlayer}{edgelayer}
		\draw [style=gluon] (0.center) to (2.center);
		\draw [style=gluon] (1.center) to (2.center);
		\draw [style=gluon] (2.center) to (3.center);
		\draw [style=gluon] (3.center) to (4.center);
		\draw [style=gluon] (3.center) to (5.center);
		\draw [style=interm] (6.center) to (7.center);
		\draw [style=interm] (8.center) to (9.center);
	\end{pgfonlayer}
\end{tikzpicture}
\end{figure}
\begin{align}
     \ket{(\Psi^{cb}_{kj}(p,q))^{2}_{gg}} & = \frac{1}{2}\, \int d^{3}r\, d^{3}s\, d^{3}r^{\prime}\, \ket{g^{d}_{l}(r)g^{e}_{m}(s)} \frac{\bra{g^{d}_{l}(r)g^{e}_{m}(s)}H_{ggg}\ket{g^{d^{\prime}}_{l^{\prime}}(r^{\prime})}}{E_{gg}(q,p)-E_{gg}(r,s)}\notag \\ & \frac{\bra{g^{d^{\prime}}_{l^{\prime}}(r^{\prime})}H_{ggg}\ket{g^{c}_{k}(q)g^{b}_{j}(p)}}{E_{gg}(q,p)-E_{g}(r^{\prime})}
\end{align}
By using the matrix element (Eq. \eqref{m3}), we obtain
\begin{align}
    \ket{(\Psi^{cb}_{kj}(p,q))^{2}_{gg}} & = \frac{g^{2}f^{d^{\prime}ed}f^{d^{\prime}cb}}{128\pi^{3}}\int d^{3}r\, d^{3}s\, d^{3}r^{\prime}\, \ket{g^{d}_{l}(r)g^{e}_{m}(s)}\, \frac{\delta^{(3)}(r^{\prime}-s-r)\delta^{(3)}(p+q-r^{\prime})}{\sqrt{s^{+}q^{+}r^{+}p^{+}}\bigg(\frac{\v{p}^{2}}{p^{+}}+\frac{\v{q}^{2}}{q^{+}}-\frac{\v{r}^{\prime 2}}{r^{\prime +}}\bigg)}\notag \\ & \Bigg[\bigg(\v{s}^{l^{\prime}}-\v{r}^{l^{\prime}}+\frac{r^{+}-s^{+}}{r^{\prime +}}\v{r}^{\prime l^{\prime}}\bigg)\delta_{ml} + \bigg(\v{r}^{\prime m}+\v{r}^{m}-\frac{r^{\prime +}+r^{+}}{s^{+}}\v{s}^{m}\bigg)\delta_{ll^{\prime}}\notag \\ & +\bigg(\frac{r^{\prime +}+s^{+}}{r^{+}}\v{r}^{l}-\v{s}^{l}-\v{r}^{\prime l}\bigg)\delta_{ml^{\prime}}\Bigg] \times \Bigg[\bigg(\v{p}^{l^{\prime}}-\v{q}^{l^{\prime}}+\frac{q^{+}-p^{+}}{r^{\prime +}}\v{r}^{\prime l^{\prime}}\bigg)\delta_{jk} \notag \\ & +\bigg(\v{r}^{\prime j}+\v{q}^{j}-\frac{r^{\prime +}+q^{+}}{p^{+}}\v{p}^{j}\bigg)\delta_{kl^{\prime}} +\bigg(\frac{r^{\prime +}+p^{+}}{q^{+}}\v{q}^{k}-\v{p}^{k}-\v{r}^{\prime k}\bigg)\delta_{ij}\Bigg]\notag \\ & \times \frac{1}{\bigg(\frac{\v{p}^{2}}{2p^{+}}+\frac{\v{q}^{2}}{2q^{+}}-\frac{\v{r}^{2}}{2r^{+}}-\frac{s^{2}}{2s^{+}}\bigg)} + p\leftrightarrow q.
\end{align}
    
\subsubsection*{One $\rho$ part}
One of the incoming gluons is first absorbed by the valence, while the other incoming gluon subsequently undergoes a splitting and produces two outgoing gluons. 

\begin{figure}[htbp]
    \centering
    \begin{tikzpicture}
	\begin{pgfonlayer}{nodelayer}
		\node [style=none] (0) at (-8, 5) {};
		\node [style=none] (1) at (-2, 5) {};
		\node [style=none] (2) at (-8, 3) {};
		\node [style=none] (3) at (-8, 1) {};
		\node [style=none] (4) at (-3, 1) {};
		\node [style=none] (5) at (0, 3) {};
		\node [style=none] (6) at (0, -1) {};
		\node [style=none] (7) at (-5, 5) {};
		\node [style=none] (8) at (-9.75, 3) {};
		\node [style=none] (9) at (-9.75, 1) {};
		\node [style=none] (10) at (2, 3) {};
		\node [style=none] (11) at (2, -1) {};
		\node [style=none] (12) at (-9.75, 3) {$s,e,m$};
		\node [style=none] (13) at (-9.75, 1) {$r,d,l$};
		\node [style=none] (14) at (2, 3) {};
		\node [style=none] (15) at (2, 3) {$p,b,j$};
		\node [style=none] (16) at (2, -1) {$q,c,k$};
		\node [style=none] (17) at (-4, 6.75) {};
		\node [style=none] (18) at (-4, -1.75) {};
		\node [style=none] (19) at (-1, 7) {};
		\node [style=none] (20) at (-1, -2) {};
		\node [style=none] (21) at (0, 5) {};
	\end{pgfonlayer}
	\begin{pgfonlayer}{edgelayer}
		\draw (0.center) to (1.center);
		\draw [style=gluon, bend right=45] (2.center) to (7.center);
		\draw [style=gluon] (3.center) to (4.center);
		\draw [style=gluon] (4.center) to (5.center);
		\draw [style=gluon] (4.center) to (6.center);
		\draw [style=interm] (17.center) to (18.center);
		\draw [style=interm] (19.center) to (20.center);
		\draw (1.center) to (21.center);
	\end{pgfonlayer}
\end{tikzpicture}
\end{figure}
\begin{align}
\ket{(\Psi^{ed}_{ml}(r,s))^{1}_{gg\,\rho}} & = \frac{1}{2}\int d^{3}p\, d^{3}q\, d^{3}r^{\prime} \ket{g^{b}_{j}(p)g^{c}_{k}(q)}\frac{\bra{g^{b}_{j}(p)g^{c}_{k}(q)}H_{ggg}\ket{g^{d^{\prime}}_{l^{\prime}}(r^{\prime})}}{E_{gg}(s,r)-E_{gg}(p,q)}\notag \\ & \frac{\bra{g^{d^{\prime}}_{l^{\prime}}(r^{\prime})}H_{g}\ket{g^{e}_{m}(s)g^{d}_{l}(r)}}{E_{gg}(s,r)-E_{g}(r^{\prime})}
\end{align}
Using the matrix elements (eqs. \eqref{m1} and \eqref{m3}), we obtain
\begin{align}
  \ket{(\Psi^{ed}_{ml}(r,s))^{1}_{gg\,\rho}} & =  -\frac{ig^{2}f^{dbe}}{32\pi^{3}}\int d^{3}p\, d^{3}q\,\ket{g^{b}_{j}(p)g^{c}_{k}(q)} \frac{\delta^{(3)}(p+q-r)\rho^{e}(-\v{s})}{\sqrt{p^{+}q^{+}r^{+}s^{+}}}\times \frac{\v{s}^{m}}{\v{s}^{2}}\notag \\ & \Bigg[\bigg(\v{p}^{l}-\v{q}^{l}+\frac{q^{+}-p^{+}}{r^{+}}\v{r}^{l}\bigg)\delta_{jk} +\bigg(\v{r}^{j}+\v{q}^{j}-\frac{r^{+}+q^{+}}{p^{+}}\v{p}^{j}\bigg)\delta_{lk}\notag \\ & +\bigg(\frac{r^{+}+p^{+}}{q^{+}}\v{q}^{k}-\v{p}^{k}-\v{r}^{k}\bigg)\delta_{lj}\Bigg]\times \frac{1}{\bigg(\frac{\v{s}^2}{2s^+}+\frac{\v{r}^2}{2r^+}-\frac{\v{p}^2}{2p^+}-\frac{\v{q}^2}{2q^+}\bigg)} + s\leftrightarrow r.
\end{align}
In the complementary time ordering, one of the incoming gluons is absorbed by the valence at a later stage, while the other gluon splits into two beforehand.

\begin{figure}[htbp]
    \hspace{5cm}
    \begin{tikzpicture}
	\begin{pgfonlayer}{nodelayer}
		\node [style=none] (0) at (-7, 5) {};
		\node [style=none] (1) at (-1, 5) {};
		\node [style=none] (2) at (-7, 3.25) {};
		\node [style=none] (3) at (-7, 1) {};
		\node [style=none] (4) at (-4, 1) {};
		\node [style=none] (5) at (-1, 3) {};
		\node [style=none] (6) at (-1, 0) {};
		\node [style=none] (7) at (-2, 5) {};
		\node [style=none] (8) at (-1, -1) {};
		\node [style=none] (9) at (-3, 7) {};
		\node [style=none] (10) at (-3, -2) {};
		\node [style=none] (11) at (-1.5, 7) {};
		\node [style=none] (12) at (-1.5, -2) {};
		\node [style=none] (13) at (-8.75, 3.25) {};
		\node [style=none] (14) at (-8.75, 1) {};
		\node [style=none] (15) at (0.5, 3) {};
		\node [style=none] (16) at (0.5, -1) {};
		\node [style=none] (17) at (-8.75, 3.25) {$s,e,m$};
		\node [style=none] (18) at (-8.75, 1) {$r,d,l$};
		\node [style=none] (19) at (0.5, 3) {$p,b,j$};
		\node [style=none] (20) at (0.5, -1) {$q,c,k$};
		\node [style=none] (21) at (16, 5.75) {};
	\end{pgfonlayer}
	\begin{pgfonlayer}{edgelayer}
		\draw (0.center) to (1.center);
		\draw [style=gluon, bend right, looseness=1.25] (2.center) to (7.center);
		\draw [style=gluon] (3.center) to (4.center);
		\draw [style=gluon] (4.center) to (5.center);
		\draw [style=gluon] (4.center) to (8.center);
		\draw [style=interm] (9.center) to (10.center);
		\draw [style=interm] (11.center) to (12.center);
	\end{pgfonlayer}
\end{tikzpicture}
\end{figure}
\begin{align}
    \ket{(\Psi^{ed}_{ml}(r,s))^{2}_{gg\,\rho}} & = \frac{1}{12}\int d^{3}p\, d^{3}q\, d^{3}r^{\prime}\,d^{3}s^{\prime} \,d^{3}q^{\prime}\,\ket{g^{b}_{j}(p)g^{c}_{k}(q)}\notag \\ & \frac{\bra{g^{b}_{j}(p)g^{c}_{k}(q)}H_{g}\ket{g^{e^{\prime}}_{m^{\prime}}(s^{\prime})g^{b^{\prime}}_{j^{\prime}}(p^{\prime})g^{c^{\prime}}_{k^{\prime}}(q^{\prime})}}{E_{gg}(s,r)-E_{gg}(p,q)}\, \frac{\bra{g^{e^{\prime}}_{m^{\prime}}(s^{\prime})g^{b^{\prime}}_{j^{\prime}}(p^{\prime})g^{c^{\prime}}_{k^{\prime}}(q^{\prime})}H_{ggg}\ket{g^{e}_{m}(s)g^{d}_{l}(r)}}{E_{gg}(s,r)-E_{gg}(s^{\prime},p^{\prime},q^{\prime})} 
\end{align}
Using relevant matrix elements, we obtain
\begin{align}
    \ket{(\Psi^{ed}_{ml}(r,s))^{2}_{gg\,\rho}} & = -\frac{ig^{2}f^{dbc}}{32\pi^{3}}\int\, d^{3}p\, d^{3}q\, \ket{g^{b}_{j}(p)g^{c}_{k}(q)} \frac{\delta^{(3)}(r-p-q)\rho^{e}(\v{s})\v{s}^{m}}{s^{+}\sqrt{r^{+}p^{+}q^{+}s^{+}}\bigg(\frac{\v{r}^{2}}{r^{+}}-\frac{\v{p}^{2}}{p^{+}}-\frac{\v{q}^{2}}{q^{+}}\bigg)}\times\notag \\ &   \Bigg[\bigg(\v{p}^{l}-\v{q}^{l} +\frac{q^{+}-p^{+}}{r^{+}}\v{r}^{l}\bigg)\delta_{jk} +\bigg(\v{r}^{j}+\v{q}^{j}-\frac{q^{+}+r^{+}}{p^{+}}\v{p}^{j}\bigg)\delta_{lk}\notag\\ & +\bigg(\frac{r^{+}+p^{+}}{q^{+}}\v{q}^{k}-\v{p}^{k}-\v{r}^{k}\bigg)\delta_{lj}\Bigg] \times \frac{1}{\bigg(\frac{\v{s}^2}{2s^+}+\frac{\v{r}^2}{2r^+}-\frac{\v{p}^2}{2p^+}-\frac{\v{q}^2}{2q^+}\bigg)} + s\leftrightarrow r.
\end{align}
After the two incoming gluons merge into a single intermediate gluon, the valence emits an additional gluon in the subsequent time ordering.

\begin{figure}[htbp]
    \centering
    \begin{tikzpicture}
	\begin{pgfonlayer}{nodelayer}
		\node [style=none] (0) at (-9, 7) {};
		\node [style=none] (1) at (-3, 7) {};
		\node [style=none] (2) at (-6, 7) {};
		\node [style=none] (3) at (-3, 5.75) {};
		\node [style=none] (4) at (-9, 5) {};
		\node [style=none] (5) at (-9, 3) {};
		\node [style=none] (6) at (-7, 4) {};
		\node [style=none] (7) at (-3, 4) {};
		\node [style=none] (8) at (-10, 5) {$r,d,l$};
		\node [style=none] (9) at (-10.25, 3) {$s,e,m$};
		\node [style=none] (10) at (-2, 4) {$p,b,j$};
		\node [style=none] (11) at (-2, 5.75) {$q,c,k$};
		\node [style=none] (12) at (-6.75, 8) {};
		\node [style=none] (13) at (-6.75, 8) {};
		\node [style=none] (14) at (-6.75, 2.5) {};
		\node [style=none] (15) at (-5, 8) {};
		\node [style=none] (16) at (-5, 2.5) {};
	\end{pgfonlayer}
	\begin{pgfonlayer}{edgelayer}
		\draw (0.center) to (1.center);
		\draw [style=gluon, bend right=15] (2.center) to (3.center);
		\draw [style=gluon] (4.center) to (6.center);
		\draw [style=gluon] (5.center) to (6.center);
		\draw [style=gluon] (6.center) to (7.center);
		\draw [style=interm] (13.center) to (14.center);
		\draw [style=interm] (15.center) to (16.center);
	\end{pgfonlayer}
\end{tikzpicture}
\end{figure}
\begin{align}
     \ket{(\Psi^{ed}_{ml}(r,s))^{3}_{gg\rho}} & = \frac{1}{2} \int d^{3}p\, d^{3}q\, d^{3}p^{\prime}\ket{g^{c}_{k}(q)g^{b}_{j}(p)}\frac{\bra{g^{c}_{k}(q)g^{b}_{j}(p)}H_{g}\ket{g^{b^{\prime}}_{j^{\prime}}(p^{\prime})}}{E_{gg}(s,r)-E_{gg}(p,q)}\notag \\ & \frac{\bra{g^{b^{\prime}}_{j^{\prime}}(p^{\prime})}H_{ggg}\ket{g^{d}_{l}(r)g^{e}_{m}(s)}}{E_{gg}(s,r)-E_{g}(p^{\prime})}
\end{align}
Using the relevant matrix elements (eqs. \eqref{m1} and \eqref{m3}), we get
\begin{align}
    \ket{(\Psi^{ed}_{ml}(r,s))^{3}_{gg\rho}} & = -\frac{ig^{2}f^{bed}}{32\pi^{3}}\int\, d^{3}p\, d^{3}q\, \ket{g^{b}_{j}(p)g^{c}_{k}(q)}\, \frac{\delta^{(3)}(-p+s+r)\rho^{c}(-q)}{\sqrt{p^{+}s^{+}r^{+}q^{+}}\bigg(\frac{s^{2}}{s^{+}}-\frac{r^{2}}{r^{+}}-\frac{p^{2}}{p^{+}}\bigg)}\times \frac{\v{q}^{k}}{q^{+}}\notag \\ & \Bigg[\bigg(\v{s}^{j}-\v{r}^{j}+\frac{r^{+}-s^{+}}{p^{+}}\v{p}^{j}\bigg)\delta_{lm} +\bigg(\v{p}^{m}+\v{r}^{m}-\frac{p^{+}+r^{+}}{s^{+}}\v{s}^{m}\bigg)\delta_{lj}\notag \\ & +\bigg(\frac{p^{+}+s^{+}}{r^{+}}\v{r}^{l}-\v{s}^{l}-\v{p}^{l}\bigg)\delta_{mj}\Bigg] \times \frac{1}{\bigg(\frac{\v{s}^{2}}{2s^{+}}+\frac{\v{r}^{2}}{2r^{+}}-\frac{\v{p}^{2}}{2p^{+}}-\frac{\v{q}^{2}}{2q^{+}}\bigg)} + r\leftrightarrow s.
\end{align}
Next, we consider the case in which a gluon is emitted from the valence before the two incoming gluons merge into a single-gluon. 

\begin{figure}[htbp]
    \centering
   \begin{tikzpicture}
	\begin{pgfonlayer}{nodelayer}
		\node [style=none] (0) at (-10, 7) {};
		\node [style=none] (1) at (-4, 7) {};
		\node [style=none] (2) at (-8.75, 7) {};
		\node [style=none] (3) at (-4, 5.75) {};
		\node [style=none] (4) at (-11, 5) {};
		\node [style=none] (5) at (-11, 3) {};
		\node [style=none] (6) at (-7, 4) {};
		\node [style=none] (7) at (-4, 4) {};
		\node [style=none] (8) at (-8, 8) {};
		\node [style=none] (9) at (-8, 2) {};
		\node [style=none] (10) at (-6, 8) {};
		\node [style=none] (11) at (-6, 2) {};
		\node [style=none] (12) at (-3.25, 5.75) {$p,b,j$};
		\node [style=none] (13) at (-3, 4) {$q,c,k$};
		\node [style=none] (14) at (-12, 5) {$r,d,l$};
		\node [style=none] (15) at (-12.25, 3) {$s,e,m$};
	\end{pgfonlayer}
	\begin{pgfonlayer}{edgelayer}
		\draw (0.center) to (1.center);
		\draw [style=gluon, bend right=15] (2.center) to (3.center);
		\draw [style=gluon] (4.center) to (6.center);
		\draw [style=gluon] (5.center) to (6.center);
		\draw [style=gluon] (6.center) to (7.center);
		\draw [style=interm] (8.center) to (9.center);
		\draw [style=interm] (10.center) to (11.center);
	\end{pgfonlayer}
\end{tikzpicture}
    \label{fig:enter-label}
\end{figure}
\begin{align}
    \ket{(\Psi^{ed}_{ml}(r,s))^{4}_{gg\rho}} & = \frac{1}{12}\int d^{3}p\, d^{3}q\, d^{3}r^{\prime}\,d^{3}s^{\prime} \,d^{3}q^{\prime}\,\ket{g^{b}_{j}(p)g^{c}_{k}(q)}\notag \\  & \frac{\bra{g^{b}_{j}(p)g^{c}_{k}(q)}H_{ggg}\ket{g^{e^{\prime}}_{m^{\prime}}(s^{\prime})g^{d^{\prime}}_{l^{\prime}}(r^{\prime})g^{c^{\prime}}_{k^{\prime}}(q^{\prime})}}{E_{gg}(s,r)-E_{gg}(p,q)}\,\frac{\bra{g^{e^{\prime}}_{m^{\prime}}(s^{\prime})g^{d^{\prime}}_{l^{\prime}}(r^{\prime})g^{c^{\prime}}_{k^{\prime}}(q^{\prime})}H_{g}\ket{g^{e}_{m}(s)g^{d}_{l}(r)}}{E_{gg}(s,r)-E_{gg}(s^{\prime},r^{\prime},q^{\prime})}  
\end{align}
Using the relevant matrix elements, we get
\begin{align}
    \ket{(\Psi^{ed}_{ml}(r,s))^{4}_{gg\rho}} & =  -\frac{ig^{2}f^{cde}}{32\pi^{3}}\int\, d^{3}p \, d^{3}q\, \ket{g^{b}_{j}(p)g^{c}_{k}(q)}\, \frac{\delta^{(3)}(q-r-s)\rho^{b}(\v{-p})}{\sqrt{r^{+}s^{+}p^{+}q^{+}}}\times \frac{\v{p}^{j}}{\v{p}^{2}}\notag \\ & \Bigg[\bigg(\v{r}^{k}-\v{s}^{k}+\frac{s^{+}-r^{+}}{q^{+}}  \v{q}^{k}\bigg)\delta_{ml} +\bigg(\v{q}^{l}+\v{s}^{l}-\frac{q^{+}+s^{+}}{r^{+}}\v{r}^{l}\bigg)\delta_{mk}\notag \\ & +\bigg(\frac{q^{+}+r^{+}}{s^{+}}\v{s}^{m}-\v{r}^{m}-\v{q}^{m}\bigg)\delta_{lk}\Bigg] \times \frac{1}{\bigg(\frac{\v{s}^{2}}{2s^{+}}+\frac{\v{r}^{2}}{2r^{+}}-\frac{\v{p}^{2}}{2p^{+}}-\frac{\v{q}^{2}}{2q^{+}}\bigg)} + s\leftrightarrow r.
\end{align}

\subsubsection{Four outgoing gluons}

\subsubsection*{$\rho$ independent part}
At order $g^{2}$, there is a single diagram in which two incoming gluons produce four outgoing gluons. Since our goal is to construct the next to leading order evolution operator, we will restrict our analysis to the portion of the wave function that directly contributes to this operator, and omit any additional structures that do not enter its determination.

\begin{figure}[htbp]
    \centering
    \begin{tikzpicture}
	\begin{pgfonlayer}{nodelayer}
		\node [style=none] (0) at (-5, 4) {};
		\node [style=none] (1) at (-2, 4) {};
		\node [style=none] (2) at (1, 6) {};
		\node [style=none] (3) at (1, 2) {};
		\node [style=none] (4) at (-5, -2) {};
		\node [style=none] (5) at (-3, -2) {};
		\node [style=none] (6) at (1, 0) {};
		\node [style=none] (7) at (1, -3) {};
		\node [style=none] (8) at (1, -4.75) {};
		\node [style=none] (9) at (-2.5, 7) {};
		\node [style=none] (10) at (-2.5, -5) {};
		\node [style=none] (11) at (-1, 7) {};
		\node [style=none] (12) at (-1, -5) {};
		\node [style=none] (13) at (-7, 4) {};
		\node [style=none] (14) at (-7, -2) {};
		\node [style=none] (15) at (3, 6) {};
		\node [style=none] (16) at (3, 2) {};
		\node [style=none] (17) at (3, 0) {};
		\node [style=none] (18) at (3, -5) {};
		\node [style=none] (19) at (-7, 4) {$k,a,i$};
		\node [style=none] (20) at (-7, -2) {$p,b,j$};
		\node [style=none] (21) at (3, 6) {$r,d,l$};
		\node [style=none] (22) at (3, 2) {$s,e,m$};
		\node [style=none] (23) at (3, 0) {$u,f,n$};
		\node [style=none] (24) at (3, -5) {$v,h,o$};
	\end{pgfonlayer}
	\begin{pgfonlayer}{edgelayer}
		\draw [style=gluon] (0.center) to (1.center);
		\draw [style=gluon] (1.center) to (2.center);
		\draw [style=gluon] (1.center) to (3.center);
		\draw [style=gluon] (4.center) to (5.center);
		\draw [style=gluon] (5.center) to (6.center);
		\draw [style=gluon] (5.center) to (8.center);
		\draw [style=interm] (9.center) to (10.center);
		\draw [style=interm] (11.center) to (12.center);
	\end{pgfonlayer}
\end{tikzpicture}
\end{figure}
\begin{align}
     \ket{(\Psi^{ba}_{ji}(k,p))^{3}_{gg}} & =  \frac{1}{144}\, \int d^{3}r\, d^{3}s\, d^{3}u\, d^{3}v \, d^{3}r^{\prime}\, d^{3}s^{\prime} \, d^{3}q^{\prime}\, \ket{g^{d}_{l}(r)g^{e}_{m}(s)g^{f}_{n}(u)g^{h}_{o}(v)}\notag \\ & \frac{\bra{g^{d}_{l}(r)g^{e}_{m}(s)g^{f}_{n}(u)g^{h}_{o}(v)}H_{ggg}\ket{g^{c^{\prime}}_{k^{\prime}}(q^{\prime})g^{e^{\prime}}_{m^{\prime}}(s^{\prime})g^{d^{\prime}}_{l^{\prime}}(r^{\prime})}}{E_{gg}(k,p)-E_{gggg}(r,s,u,v)}\notag \\ &  \frac{\bra{g^{c^{\prime}}_{k^{\prime}}(q^{\prime})g^{e^{\prime}}_{m^{\prime}}(s^{\prime})g^{d^{\prime}}_{l^{\prime}}(r^{\prime})}H_{ggg}\ket{g^{a}_{i}(k)g^{b}_{j}(p)}}{E_{gg}(k,p)-E_{ggg}(q^{\prime},s^{\prime},r^{\prime})}
\end{align}
Using relevant matrix elements that are derived from Eq. \eqref{m3}, we get
\begin{align}
     \ket{(\Psi^{ba}_{ji}(k,p))_{gg}} & =  -\frac{g^{4}f^{ade}f^{bfh}}{128\pi^{3}} \int d^{3}r\, d^{3}s\, d^{3}u\,d^{3}v\, \frac{\ket{g^{d}_{l}(r)g^{e}_{m}(s)g^{f}_{n}(u)g^{h}_{o}(v)}}{\bigg(\frac{p^{2}}{2p^{+}}+\frac{k^{2}}{2k^{+}}-\frac{r^{2}}{2r^{+}}-\frac{s^{2}}{2s^{+}}-\frac{u^{2}}{2u^{+}}-\frac{v^{2}}{2v^{+}}\bigg)} \notag \\ & \frac{\delta^{(3)}(k-r-s)\delta^{(3)}(p-u-v)}{\sqrt{k^{+}r^{+}s^{+}}\sqrt{p^{+}u^{+}v^{+}}}\Bigg[\bigg(r^{i}-s^{i}+\frac{s^{+}-r^{+}}{k^{+}}k^{i}\bigg)\delta_{lm}\notag \\ & +\bigg(k^{l}+s^{l}-\frac{k^{+}+s^{+}}{r^{+}}r^{l}\bigg)\delta_{im} +\bigg(\frac{k^{+}+r^{+}}{s^{+}}s^{m}-r^{m}-k^{m}\bigg)\delta_{il}\Bigg]\times \notag \\ & \Bigg[\bigg(u^{j}-v^{j}+\frac{v^{+}-u^{+}}{p^{+}}p^{j}\bigg)\delta_{on}+\bigg(p^{n}+v^{n} -\frac{p^{+}+v^{+}}{u^{+}}u^{n}\bigg)\delta_{oj}\notag \\ & +\bigg(\frac{p^{+}+u^{+}}{v^{+}}v^{o}-u^{o}-p^{o}\bigg)\delta_{jn}\Bigg]\times \frac{1}{\bigg(\frac{p^{2}}{p^{+}}-\frac{u^{2}}{u^{+}}-\frac{v^{2}}{v^{+}}\bigg)}\,+ p\leftrightarrow k
\end{align}

\subsubsection{Summary of the results}
The complete incoming two-gluon wave function up to order $g^{2}$ is obtained by summing all contributions derived in the previous subsection. These include both the purely gluonic components as well as the terms involving the color charge density insertion. The resulting wave function can therefore be written as
\begin{align}
    \ket{(\Psi^{ba}_{ji}(k,p))_{gg}} & = \mf{N} \ket{g^{a}_{i}(k)g^{b}_{j}(p)} \notag \\ & + \ket{(\Psi^{ba}_{ji}(k,p))^{1}_{gg}} + \ket{(\Psi^{ba}_{ji}(k,p))^{2}_{gg}} + \ket{(\Psi^{ba}_{ji}(k,p))^{3}_{gg}}\notag \\ & + \ket{(\Psi^{ba}_{ji}(k,p))^{1}_{gg\,\rho}}+ \ket{(\Psi^{ba}_{ji}(k,p))^{2}_{gg\,\rho}} + \ket{(\Psi^{ba}_{ji}(k,p))^{3}_{gg\,\rho}}+ \ket{(\Psi^{ba}_{ji}(k,p))^{4}_{gg\,\rho}}
\end{align}
The first term represents the bare two-gluon Fock state multiplied by the normalization constant $\mf{N}$, while the remaining terms correspond to the various interaction terms computed previously. The normalization factor $\mf{N}$ is determined from the normalization condition imposed on the full wave function. However, an important simplification occurs at this order in perturbation theory. Since the interacting part of the wave function itself starts at order $g^{2}$, corrections to the normalization factor generated by these interactions contribute only at higher perturbative order. Consequently, when the calculation is consistently truncated at order $g^{2}$, the higher order normalization contributions lie beyond the required accuracy and can therefore be neglected.

\subsection{Calculation of the three-gluon incoming states} \label{sec:Threeglue}
We restrict the calculation to the diagrams that are actually needed to obtain the evolution operator at next to leading order. In practice, this means that only two diagrams contribute to the result and these are the ones we evaluate.

\subsubsection*{$\rho$ independent part}
After two of the incoming gluons merge into a single outgoing gluon, then an another incoming gluon splits into two outgoing gluons.

\begin{figure}[htbp]
    \centering
    \begin{tikzpicture}
	\begin{pgfonlayer}{nodelayer}
		\node [style=none] (0) at (-6, 6) {};
		\node [style=none] (1) at (-3, 6) {};
		\node [style=none] (2) at (0, 7) {};
		\node [style=none] (3) at (0, 5) {};
		\node [style=none] (4) at (-6, 4) {};
		\node [style=none] (5) at (-6, 2) {};
		\node [style=none] (6) at (-1.5, 3) {};
		\node [style=none] (7) at (0, 3) {};
		\node [style=none] (8) at (-4, 3) {};
		\node [style=none] (9) at (-6, 4) {};
		\node [style=none] (10) at (-3.75, 8) {};
		\node [style=none] (11) at (-3.75, 1) {};
		\node [style=none] (12) at (-2, 8) {};
		\node [style=none] (13) at (-2, 1) {};
		\node [style=none] (14) at (-7.5, 6) {};
		\node [style=none] (15) at (-7.5, 4) {};
		\node [style=none] (16) at (-7.5, 2) {};
		\node [style=none] (17) at (1.75, 7) {};
		\node [style=none] (18) at (1.75, 5) {};
		\node [style=none] (19) at (1.75, 3) {};
		\node [style=none] (20) at (-7.5, 6) {$k,a,i$};
		\node [style=none] (21) at (-7.5, 4) {$r,d,l$};
		\node [style=none] (22) at (-7.5, 2) {$s,e,m$};
		\node [style=none] (23) at (1.75, 7) {p,b,j};
		\node [style=none] (24) at (1.75, 5) {$q,c,k$};
		\node [style=none] (25) at (1.75, 3) {};
		\node [style=none] (26) at (1.75, 3) {$u,f,n$};
		\node [style=none] (27) at (-10.75, 8.25) {};
	\end{pgfonlayer}
	\begin{pgfonlayer}{edgelayer}
		\draw [style=gluon] (0.center) to (1.center);
		\draw [style=gluon] (1.center) to (2.center);
		\draw [style=gluon] (1.center) to (3.center);
		\draw [style=gluon] (9.center) to (8.center);
		\draw [style=gluon] (5.center) to (8.center);
		\draw [style=gluon] (8.center) to (7.center);
		\draw [style=interm] (10.center) to (11.center);
		\draw [style=interm] (12.center) to (13.center);
	\end{pgfonlayer}
\end{tikzpicture}
\end{figure}
\begin{align}
    \ket{(\Psi^{eda}_{mli}(k,r,s))^{1}_{ggg}} & =  \frac{1}{12}\, \int\, d^{3}p\, d^{3}q\, d^{3}u\, d^{3}r^{\prime}\, d^{3}s^{\prime}\ket{g^{b}_{j}(p)g^{c}_{k}(q)g^{f}_{n}(u)}\notag \\ &   \frac{\bra{g^{b}_{j}(p)g^{c}_{k}(q)g^{f}_{n}(u)}H_{ggg}\ket{g^{d^{\prime}}_{l^{\prime}}(r^{\prime})g^{e^{\prime}}_{m^{\prime}}(s^{\prime})}}{E_{ggg}(k,r,s)-E_{ggg}(p,q,u)}\notag \\ & \frac{\bra{g^{d^{\prime}}_{l^{\prime}}(r^{\prime})g^{e^{\prime}}_{m^{\prime}}(s^{\prime})}H_{ggg}\ket{g^{a}_{i}(k)g^{d}_{l}(r)g^{e}_{m}(s)}}{E_{ggg}(k,r,s)-E_{gg}(r^{\prime},s^{\prime})}
\end{align}
Using the relevant matrix element derived from Eq. \eqref{m3}, we obtain
\begin{align}
     \ket{(\Psi^{eda}_{mli}(k,r,s))^{1}_{ggg}} & = \frac{g^{2}\,f^{abc}f^{fed}}{64\pi^{3}}\int\, d^{3}p\, d^{3}q \,d^{3}u\,\frac{\ket{g^{b}_{j}(p)g^{c}_{k}(q)g^{f}_{n}(u)}}{\bigg(\frac{k^{2}}{2k^{+}}+\frac{r^{2}}{2r^{+}}+\frac{s^{2}}{2s^{+}}-\frac{p^{2}}{2p^{+}}-\frac{q^{2}}{2q^{+}}-\frac{u^{2}}{2u^{+}}\bigg)}\,\notag \\ & \Bigg[\frac{\delta^{(3)}(k-p-q)\delta^{(3)}(u-s-r)}{\sqrt{k^{+}p^{+}q^{+}}\sqrt{u^{+}s^{+}r^{+}}}\times \frac{1}{\bigg(\frac{u^{2}}{u^{+}}-\frac{s^{2}}{s^{+}}-\frac{r^{2}}{r^{+}}\bigg)}\times\notag \\ &   \bigg\{\bigg(p^{i}-q^{i} +\frac{q^{+}-p^{+}}{k^{+}}k^{i}\bigg)\delta_{jk} +\bigg(k^{j}+q^{j}-\frac{k^{+}+q^{+}}{p^{+}}p^{j}\bigg)\delta_{ik}\notag \\ & +\bigg(\frac{k^{+}+p^{+}}{q^{+}}q^{k}-p^{k}-k^{k}\bigg)\delta_{ij}\bigg\} \times \bigg\{\bigg(s^{n}-r^{n}+\frac{r^{+}-s^{+}}{u^{+}}u^{n}\bigg)\delta_{lm}\notag \\ & +\bigg(u^{m}+r^{m}-\frac{u^{+}+r^{+}}{s^{+}}s^{m}\bigg)\delta_{ln} +\bigg(\frac{u^{+}+s^{+}}{r^{+}}r^{l}-s^{l}-u^{l}\bigg)\bigg\} \notag \\ & + (r\leftrightarrow s) + (k\leftrightarrow s)  + (k\rightarrow s),\,(s\rightarrow r)\,, (r\rightarrow k) \notag \\  &+ (k \rightarrow r)\, , (s\rightarrow k)\,, (r\rightarrow s)+(k\leftrightarrow r) \Bigg]
\end{align}

Finally, we consider the case in which one of the incoming gluons splits into two, after which two of the incoming gluons merge into a single-gluon.

\begin{figure}[htbp]
    \centering
    \begin{tikzpicture}
	\begin{pgfonlayer}{nodelayer}
		\node [style=none] (0) at (-7, 4) {};
		\node [style=none] (1) at (-7, 2) {};
		\node [style=none] (2) at (-4, 3) {};
		\node [style=none] (3) at (-1.25, 3) {};
		\node [style=none] (4) at (-7, 7) {};
		\node [style=none] (5) at (-5, 7) {};
		\node [style=none] (6) at (-2, 8.25) {};
		\node [style=none] (7) at (-2, 6) {};
		\node [style=none] (8) at (-3.25, 3) {};
		\node [style=none] (9) at (-4.25, 8.75) {};
		\node [style=none] (10) at (-4, 1.25) {};
		\node [style=none] (11) at (-2.75, 8.75) {};
		\node [style=none] (12) at (-2.5, 1.25) {};
		\node [style=none] (13) at (-8.75, 7) {};
		\node [style=none] (14) at (-8.75, 2) {};
		\node [style=none] (15) at (-8.75, 4) {};
		\node [style=none] (16) at (0.5, 3) {};
		\node [style=none] (17) at (0.5, 6) {};
		\node [style=none] (18) at (0.5, 8.5) {};
		\node [style=none] (19) at (-8.75, 7) {$k,a,i$};
		\node [style=none] (20) at (-8.75, 4) {$r,d,l$};
		\node [style=none] (21) at (-8.75, 2) {};
		\node [style=none] (22) at (-8.75, 2) {$s,e,m$};
		\node [style=none] (23) at (0.5, 8.5) {$p,b,j$};
		\node [style=none] (24) at (0.5, 6) {$q,c,k$};
		\node [style=none] (25) at (0.5, 3) {$u,f,n$};
	\end{pgfonlayer}
	\begin{pgfonlayer}{edgelayer}
		\draw [style=gluon] (0.center) to (8.center);
		\draw [style=gluon] (1.center) to (8.center);
		\draw [style=gluon] (8.center) to (3.center);
		\draw [style=gluon] (4.center) to (5.center);
		\draw [style=gluon] (5.center) to (6.center);
		\draw [style=gluon] (5.center) to (7.center);
		\draw [style=interm] (9.center) to (10.center);
		\draw [style=interm] (11.center) to (12.center);
	\end{pgfonlayer}
\end{tikzpicture}
\end{figure}
\begin{align}
    \ket{(\Psi^{eda}_{mli}(k,r,s))^{2}_{ggg}} & =  \frac{1}{144}\, \int\, d^{3}p\, d^{3}q\, d^{3}u\, d^{3}r^{\prime}\, d^{3}s^{\prime}\, d^{3}p^{\prime}\,d^{3}q^{\prime}\ket{g^{b}_{j}(p)g^{c}_{k}(q)g^{f}_{n}(u)} \notag \\ &  \frac{\bra{g^{b}_{j}(p)g^{c}_{k}(q)g^{f}_{n}(u)}H_{ggg}\ket{g^{d^{\prime}}_{l^{\prime}}(r^{\prime})g^{e^{\prime}}_{m^{\prime}}(s^{\prime})g^{b^{\prime}}_{j^{\prime}}(p^{\prime})g^{c^{\prime}}_{k^{\prime}}(q^{\prime})}}{E_{ggg}(k,r,s)-E_{ggg}(p,q,u)}\notag \\ & \frac{\bra{g^{d^{\prime}}_{l^{\prime}}(r^{\prime})g^{e^{\prime}}_{m^{\prime}}(s^{\prime})g^{b^{\prime}}_{j^{\prime}}(p^{\prime})g^{c^{\prime}}_{k^{\prime}}(q^{\prime})}H_{ggg}\ket{g^{a}_{i}(k)g^{d}_{l}(r)g^{e}_{m}(s)}}{E_{ggg}(k,r,s)-E_{gggg}(r^{\prime},s^{\prime},p^{\prime},q^{\prime})}
\end{align}
Using the relevant matrix element derived from Eq. \eqref{m3}, we obtain
\begin{align}
    \ket{(\Psi^{eda}_{mli}(k,r,s))^{2}_{ggg}} & = - \frac{g^{2}\,f^{abc}f^{fed}}{64\pi^{3}}\,\int\, \frac{d^{3}p}{(2\pi)^{3}}\, \frac{d^{3}q}{(2\pi)^{3}}\, \frac{d^{3}u}{(2\pi)^{3}}\,\ket{g^{b}_{j}(p)g^{c}_{k}(q)g^{f}_{n}(u)}\notag \\ & \frac{1}{\bigg(\frac{k^{2}}{2k^{+}}+\frac{r^{2}}{2r^{+}}+\frac{s^{2}}{2s^{+}}-\frac{p^{2}}{2p^{+}}-\frac{q^{2}}{2q^{+}}-\frac{u^{2}}{2u^{+}}\bigg)}\, \Bigg[\frac{\delta^{(3)}(k-p-q)\delta^{(3)}(u-s-r)}{\sqrt{k^{+}p^{+}q^{+}}\sqrt{u^{+}s^{+}r^{+}}} \times\notag \\ &  \frac{1}{\bigg(\frac{k^{2}}{k^{+}}-\frac{p^{2}}{p^{+}}-\frac{q^{2}}{q^{+}}\bigg)}\times\bigg\{\bigg(p^{i}-q^{i} +\frac{q^{+}-p^{+}}{k^{+}}k^{i}\bigg)\delta_{jk} +\bigg(k^{j}+q^{j}-\frac{k^{+}+q^{+}}{p^{+}}p^{j}\bigg)\notag \\ & \delta_{ik} +\bigg(\frac{k^{+}+p^{+}}{q^{+}}q^{k}-p^{k}-k^{k}\bigg)\delta_{ij}\bigg\} \times \bigg\{\bigg(s^{n}-r^{n}+\frac{r^{+}-s^{+}}{u^{+}}u^{n}\bigg)\delta_{lm}\notag \\ & +\bigg(u^{m}+r^{m}-\frac{u^{+}+r^{+}}{s^{+}}s^{m}\bigg)\delta_{ln}+\bigg(\frac{u^{+}+s^{+}}{r^{+}}r^{l}-s^{l}-u^{l}\bigg)\bigg\}\notag \\ &  + (r\leftrightarrow s) + (k\leftrightarrow s)  + (k\rightarrow s),\,(s\rightarrow r)\,, (r\rightarrow k)  \notag \\ & + (k \rightarrow r)\, , (s\rightarrow k)\,, (r\rightarrow s)+(k\leftrightarrow r) \Bigg]
\end{align}
    
\subsubsection{Summary of results}
The final result for the three-gluon incoming wave function at order $g^{2}$ is obtained by summing the contributions obtained in this subsection. As in the earlier case, the wave function is already at order $g^{2}$, so the normalization factor would contribute only at higher order and is therefore ignored.
\begin{align}
    \ket{(\Psi^{cba}_{kji}(k,p,q))_{ggg}} & =  \ket{(\Psi^{cba}_{kji}(k,p,q))^{1}_{ggg}} + \ket{(\Psi^{cba}_{kji}(k,p,q))^{2}_{ggg}}.
\end{align}

\section{Coefficients of the evolution operator up to $\mathcal{O}(g^2)$ through LCPT} \label{sec:coeff}
To determine the coefficients appearing in the evolution operator, we compare its action on simple Fock space states with the corresponding light-cone wave functions computed directly within light-cone perturbation theory. The basic idea underlying this procedure is that the evolution operator $\Omega$ must reproduce the perturbative structure generated by the pure Yang-Mills light-cone Hamiltonian, $H_{\text{LCYM}}$, when acting on physical states. In other words, the operator $\Omega$ is constructed such that its perturbative expansion generates exactly the same multi particle components as those obtained from the explicit LCPT calculation of the light-cone wave function. Operationally, this matching is implemented by evaluating the action of $\Omega$ on a set of elementary basis states in Fock space and comparing the resulting expressions with the perturbative wave functions derived from the Hamiltonian formalism. In particular, we match matrix elements of the form $\Omega\ket{0}, \Omega\ket{g^{a}_{i}(k)}, \Omega\ket{g^{a}_{i}(k),g^{b}_{j}(p)}$ with the corresponding vacuum, one gluon and two-gluon light-cone wave functions obtained by perturbatively expanding $H_{\text{LCYM}}$. Since these states probe different sectors of Fock space, they constrain different structures appearing in the evolution operator and thereby allow the coefficients to be fixed systematically order by order in the coupling.

We begin with the vacuum sector, which provides the simplest nontrivial consistency condition. By comparing the coefficients appearing in Eq.~\eqref{eq:Omegaonvacmom1} with the explicit vacuum wave functions derived in Sec.~\ref{sec:vac}, we obtain the following relations among the coefficients appearing in the evolution operator:
\begin{equation}\label{Ncoeffmom}
    \mf{N} = 1- g^{2} \int \frac{d^3p} {(2\pi)^3} \frac{\rho^{b}(-p)\rho^{b}(p)}{p^{+}p^{2}}  
\end{equation}
\begin{align}
\label{A1coeffM}
    \mf{A}^{b}_{j}(p) & = -\sqrt{2} \frac{gp^{j}\rho^{b}(-p)}{ \sqrt{p^{+}}p^{2}} 
\end{align}
\begin{align}
     \mf{B}^{cb}_{2\,kj}(p,q) & = (2\pi)^{3}\Bigg\{\frac{1}{\frac{\mathbf{p}^{2}}{2p^{+}}\left(\frac{\mathbf{p}^{2}}{2p^{+}}+\frac{\mathbf{q}^{2}}{2q^{+}}\right)} \times \frac{g^{2}\,q^{j}p^{k}\rho^{b}(-q)\rho^{c}(-p)}{16\pi^{3}|q^{+}|^{3/2}|p^{+}|^{3/2}}+\frac{g^{2}p^{k}(q)^{j}\rho^{c}(-p)\rho^{b}(-q)}{16\pi^{3}|q^{+}|^{3/2}|p^{+}|^{3/2}}\notag \\ & \frac{1}{\frac{\mathbf{q}^{2}}{2q^{+}}\left(\frac{\mathbf{p}^{2}}{2p^{+}}+\frac{\mathbf{q}^{2}}{2q^{+}}\right)}  + \frac{igf^{acb}}{16\pi^{3/2}\left(\frac{\mathbf{p}^{2}}{2p^{+}}+\frac{\mathbf{q}^{2}}{2q^{+}}\right)\left(\frac{(\mathbf{p}+\mathbf{q})^{2}}{2\,(p^{+}+q^{+})}\right)\sqrt{(p^{+}+q^{+})\,p^{+}q^{+}}}\notag \\ & \Bigg(\left[2\mathbf{p}^{i}-\frac{2p^{+}}{(p^{+}+q^{+})}\,(\mathbf{p}^{i}+\mathbf{k}^{i})\right]\delta_{jk} +\left[\frac{q^{+}+2\,p^{+}}{q^{+}}\mathbf{q}^{j}-\mathbf{q}^{j}-2\,\mathbf{p}^{j}\right]\delta_{ik} \notag \\ & +\left[2(\mathbf{p}^{k}+\mathbf{q}^{k})-\frac{2(p^{+}+q^{+})}{p^{+}}\mathbf{p}^{k}\right]\delta_{ij}\Bigg)\,\times \left(\frac{g\rho^{a}(-\mathbf{p}-\mathbf{q})\,(\mathbf{p}^{i}+\mathbf{q}^{i})}{4\pi^{3/2}|p^{+}+q^{+}|^{3/2}}\right) \notag \\ & -\frac{ig^{2}f^{abc}(p^{+}-q^{+})\rho^{a}(-\v{p}-\v{q})\delta_{jk}}{2(2\pi)^{3}\sqrt{p^{+}q^{+}}(p^{+}+q^{+})^{2}\,\left(\frac{q^{2}}{q^{+}}+\frac{p^{2}}{p^{+}}\right)}\Bigg\}.
\end{align}
Having fixed the coefficients associated with the vacuum sector, we now proceed to the next nontrivial sector of the theory, namely the light-cone wave function corresponding to a single incoming gluon state. This provides us additional constraints on the structure of the evolution operator, particularly on those terms responsible for generating one to two and one to three particle transitions in Fock space. As in the vacuum case, the coefficients are determined through a direct matching procedure. The relevant expressions are obtained by comparing the coefficients in Eq.~\eqref{eq:Omegaongmom} with the wave functions presented in sec.~\ref{sec:singleglue}.
\begin{align} \label{eq:B3}
    \mf{B}^{ba}_{3ji}(k, p) & = (2\pi)^{3}\Bigg\{\frac{ig^2 f^{abc}}{(2\pi)^3} \frac{\sqrt{p^{+}k^{+}} (k^++p^+) } {(p^+-k^+)^2} \frac{\rho^c(-\v{p}+\v{k})}{p^{+}\v{k}^2 - k^{+}\v{p}^2} \delta_{ij}\bigg[\Theta\bigg(p^{+} - k^{+} - \Lambda\bigg)\notag \\ & +\Theta\bigg(k^{+}-\Lambda-p^{+}\bigg)\bigg]\notag \\ & -\frac{g^{2}}{8\pi^{3}}\frac{k^{i}p^{j}}{k^{2}p^{2}}\bigg[\frac{1}{\sqrt{k^{+}p^{+}}}\{\rho^{a}(k),\rho^{b}(-p)\}+f^{abc}\rho^{c}(k-p)\frac{p^{2}k^{+}+k^{2}p^{+}}{p^{2}k^{+}-k^{2}p^{+}}\bigg]\notag \\ & + \frac{ig^{2}f^{abc}\rho^{c}(k-p)(k-p)^{l}}{8\pi^{3}\sqrt{k^{+}p^{+}}(k^{+}-p^{+})^{2}}\times \frac{1}{\bigg(\frac{k^{2}}{k^{+}}-\frac{p^{2}}{p^{+}}\bigg)} \Bigg[\bigg(2k^{j}-p^{j}+\frac{p^{+}-2k^{+}}{p^{+}}p^{j}\bigg)\delta_{il}\notag \\ & +\bigg(2p^{i}-k^{i}-\frac{2p^{+}-k^{+}}{k^{+}}k^{i}\bigg)\delta_{jl} +\bigg(\frac{p^{+}+k^{+}}{p^{+}-k^{+}}(p-k)^{l}-k^{l}-p^{l}\bigg)\delta_{ij}\Bigg]\notag \\ & \times \Bigg[\frac{(k-p)^{+}}{(k-p)^{2}}\Theta(p^{+}-k^{+}-\Lambda)+\frac{1}{\frac{k^{2}}{k^{+}}-\frac{(k-p)^{2}}{(k-p)^{+}}-\frac{p^{2}}{p^{+}}}\Theta(k^{+}-\Lambda-p^{+})\Bigg] \Bigg\}
\end{align}
\begin{align} \label{eq:C}
  \mf{C}^{cba}_{kji}(k,p,q) & = \frac{igf^{abc}\delta^{(3)}(k-p-q)}{8\pi^{3/2}\sqrt{k^{+}p^{+}q^{+}}\bigg(\frac{k^{2}}{k^{+}}-\frac{p^{2}}{p^{+}}-\frac{q^{2}}{q^{+}}\bigg)}\Bigg[\bigg(p^{i}-q^{i}+\frac{q^{+}-p^{+}}{k^{+}}k^{i}\bigg)\delta_{jk}\notag \\ & +\bigg(k^{j}+q^{j}-\frac{k^{+}+q^{+}}{p^{+}}p^{j}\bigg)\delta_{ik} +\bigg(\frac{k^{+}+p^{+}}{q^{+}}q^{k}-p^{k}-k^{k}\bigg)\delta_{ij}\Bigg]\times (2\pi)^{9/2}
\end{align}
The above equation Eq.~\eqref{eq:C} represents the LCWF component generated by the three-gluon interaction vertex. For the sake of simplicity, we will express the coefficients $\mf{D}_{1}, \mf{D}_{2}, \mf{E}_{1}, \mf{E}_{2}$ in terms of $\mf{A}$ and $\mf{C}$ wherever possible.
\begin{align}
    \mf{D}^{dcba}_{1\,lkji}(k,p,q,r) & = 
     \frac{g^{2}}{24} \frac{(2\pi)^{3}\,\delta^{(3)}(-k+p+q+r)}{\sqrt{k^{+}p^{+}q^{+}r^{+}}}\Bigg[f^{a'ab}f^{a'cd}\bigg(\delta_{lj}\delta_{ik}-\delta_{il}\delta_{kj} \notag \\ & +\frac{(p^{+}+k^{+})(r^{+}-q^{+})}{(r^{+}+q^{+})(k^{+}-p^{+})}\delta_{lk}\delta_{ij}\bigg) +f^{a'ac}f^{a'bd}\bigg(\delta_{lk}\delta_{ij}-\delta_{jk}\delta_{il} \notag \\ & +\frac{(q^{+}+k^{+})(r^{+}-p^{+})}{(r^{+}+p^{+})(k^{+}-q^{+})}\delta_{ik}\delta_{lj}\bigg)+f^{a'ad}f^{a'bc}\bigg(\delta_{lk}\delta_{ij}-\delta_{lj}\delta_{ik} \notag \\ & +\frac{(r^{+}+k^{+})(q^{+}-p^{+})}{(p^{+}+q^{+})(k^{+}-r^{+})}\delta_{jk}\delta_{il}\bigg)\Bigg]\times \frac{1}{\left(\frac{\v{k}^{2}}{2k^{+}}-\frac{\v{p}^{2}}{2p^{+}}-\frac{\v{q}^{2}}{2q^{+}}-\frac{\v{r}^{2}}{2r^{+}}\right)}\notag \\ & +\mf{C}^{cba}_{kji}(k,p,q) \mf{A}^{d}_{l}(r)\bigg(\frac{\v{k}^{2}}{2k^{+}}-\frac{\v{p}^{2}}{2p^{+}}-\frac{\v{q}^{2}}{2q^{+}}\bigg) \frac{1}{\bigg(\frac{\v{k}^{2}}{2k^{+}}-\frac{\v{p}^{2}}{2p^{+}}-\frac{\v{q}^{2}}{2q^{+}}-\frac{\v{r}^{2}}{2r^{+}}\bigg)}\notag \\ &  + \mf{A}^{d}_{l}(r)\mf{C}^{bca}_{jki}(k,q,p)\left(\frac{\v{r}^{2}}{2r^{+}}\right) \frac{1}{\bigg(\frac{\v{p}^{2}}{2p^{+}}+\frac{\v{q}^{2}}{2p^{+}}+\frac{\v{r}^{2}}{2r^{+}}-\frac{\v{k}^{2}}{2k^{+}}\bigg)}\notag \\ & +\int\, \frac{d^{3}v}{(2\pi)^{3}}\, \mf{C}^{a'da}_{i'li}(k,r,v)\,\mf{C}^{cba'}_{kji'}(v,p,q)\,\left(\frac{\v{v}^{2}}{2v^{+}}-\frac{\v{p}^{2}}{2p^{+}}-\frac{\v{q}^{2}}{2q^{+}}\right)\notag \\ & \frac{1}{\bigg(\frac{\v{k}^{2}}{2k^{+}}-\frac{\v{p}^{2}}{2p^{+}}-\frac{\v{q}^{2}}{2q^{+}}-\frac{\v{r}^{2}}{2r^{+}}\bigg)}.
\end{align}
The coefficients $\mf{D}_{2},\mf{E}_{1}$ can be obtained by comparing the coefficients derived in Eq. \eqref{eq:Omegaongg} with the explicit two-gluon incoming wave function components obtained in sec. \ref{sec:Twoglue}. 
\begin{align}
    \mf{D}^{dcba}_{2\,lkji}(k,p,q,r) & = 
\frac{g^{2}(2\pi)^{3}}{16}\frac{\delta^{(3)}(-k+r-p+q)}{\sqrt{k^{+}q^{+}r^{+}p^{+}}}\Bigg[f^{a'ad}f^{a'cb}\bigg(\delta_{ik}\delta_{jl}-\delta_{lk}\delta_{ij}\notag \\ & +\frac{(q^{+}+p^{+})(r^{+}+k^{+})}{(k^{+}-r^{+})(p^{+}-q^{+})}\delta_{il}\delta_{jk}\bigg) +f^{a'ac}f^{a'db}\bigg(\delta_{jk}\delta_{il}-\delta_{kl}\delta_{ij}\notag \\ & +\frac{(p^{+}+r^{+})(q^{+}+k^{+})}{(k^{+}-q^{+})(p^{+}-r^{+})}\delta_{lj}\delta_{ik}\bigg) + f^{a'ab}f^{a'dc}\bigg(\delta_{il}\delta_{jk}-\delta_{ik}\delta_{lj}\notag \\ & +\frac{(q^{+}-r^{+})(k^{+}-p^{+})}{(k^{+}+p^{+})(r^{+}+q^{+})}\delta_{ij}\delta_{lk}\bigg) \Bigg]\times  \frac{1}{\left(\frac{\v{k}^{2}}{2k^{+}}+\frac{\v{p}^{2}}{2p^{+}}-\frac{\v{q}^{2}}{2q^{+}}-\frac{\v{r}^{2}}{2r^{+}}\right)}\notag \\ &  + \int\,\frac{d^{3}r^{\prime}}{(2\pi)^{3}}\,\mf{C}(r^{\prime},k,p)\mf{C}^{\dagger}(r^{\prime},r,q)  \times \frac{\left(\frac{\v{r}^{\prime\, 2}}{2r^{\prime\, +}}-\frac{\v{k}^{2}}{2k^{+}}-\frac{\v{p}^{2}}{2p^{+}}\right)}{\bigg(\frac{\v{q}^{2}}{2q^{+}}+\frac{\v{r}^{2}}{2r^{+}}-\frac{\v{p}^{2}}{2p^{+}}-\frac{k^{2}}{2k^{+}}\bigg)} \notag \\ & -\mf{C}(p,r,q)\mf{A}^{\dagger}(k)\bigg(\frac{\v{p}^2}{2p^+}-\frac{\v{q}^2}{2q^+}-\frac{\v{r}^2}{r^+}\bigg) \times \frac{1}{\bigg(\frac{\v{k}^2}{2k^+}+\frac{\v{p}^2}{2p^+}-\frac{\v{q}^2}{2q^+}-\frac{\v{r}^2}{2r^+}\bigg)}\notag \\ & - \mf{A}^{\dagger}(k)\mf{C}(p,q,r)\bigg(\frac{\v{k}^{2}}{2k^{+}}\bigg) \times \frac{1}{\bigg(\frac{\v{k}^2}{2k^+}+\frac{\v{p}^2}{2p^+}-\frac{\v{q}^2}{2q^+}-\frac{\v{r}^2}{2r^+}\bigg)}\notag \\ & +\mf{C}^{\dagger}(q,k,p)\mf{A}(r) \bigg(\frac{\v{r}^{2}}{2r^{+}}\bigg)\frac{1}{\bigg(\frac{\v{k}^{2}}{2k^{+}}+\frac{\v{p}^{2}}{2p^{+}}-\frac{\v{q}^{2}}{2q^{+}}-\frac{\v{r}^{2}}{2r^{+}}\bigg)}\notag \\ & + \mf{C}^{\dagger}(r,p,k)\mf{A}(q)\bigg(\frac{\v{r}^{2}}{2r^{+}}-\frac{\v{p}^{2}}{2p^{+}}-\frac{\v{k}^{2}}{2k^{+}}\bigg)\frac{1}{\bigg(\frac{\v{k}^{2}}{2k^{+}}+\frac{\v{p}^{2}}{2p^{+}}-\frac{\v{q}^{2}}{2q^{+}}-\frac{\v{r}^{2}}{2r^{+}}\bigg)}
\end{align}
\begin{align}
    \mf{E}^{fedcba}_{1\,nmlkji}(k,p,q,r,s,u) & =  \frac{\mf{C}(k,r,s)\mf{C}(p,u,q)}{\bigg(\frac{p^{2}}{2p^{+}}+\frac{k^{2}}{2k^{+}}-\frac{r^{2}}{2r^{+}}-\frac{s^{2}}{2s^{+}}-\frac{u^{2}}{2u^{+}}-\frac{q^{2}}{2q^{+}}\bigg)} \times \bigg(\frac{k^{2}}{2k^{+}}-\frac{r^{2}}{2r^{+}}-\frac{s^{2}}{2s^{+}}\bigg)
\end{align}
Similarly, the coefficient $\mf{E}_{2}$ is obtained by comparing Eq. \eqref{eq:Omegaonggg} with the corresponding wave functions derived in sec. \ref{sec:Threeglue}.
\begin{align}
    \mf{E}^{fedcba}_{2nmlkji}(k,p,q,r,s,u) & =  \frac{\mf{C}(k,r,s) \mf{C}^{\dagger\,}(u,q,p)}{\bigg(\frac{k^{2}}{2k^{+}}+\frac{p^{2}}{2p^{+}}+\frac{q^{2}}{2q^{+}}-\frac{r^{2}}{2r^{+}}-\frac{s^{2}}{2s^{+}}-\frac{u^{2}}{2u^{+}}\bigg)}\ \times \bigg(\frac{k^{2}}{2k^{+}}-\frac{r^{2}}{2r^{+}}-\frac{s^{2}}{2s^{+}}\bigg) \notag \\ & + \frac{\mf{C}(k,r,s)\mf{C}^{\dagger}(u,p,q)}{\bigg(\frac{k^{2}}{2k^{+}}+\frac{p^{2}}{2p^{+}}+\frac{q^{2}}{2q^{+}}-\frac{r^{2}}{2r^{+}}-\frac{s^{2}}{2s^{+}}-\frac{u^{2}}{2u^{+}}\bigg)} \times \bigg(\frac{p^{2}}{2p^{+}}+\frac{q^{2}}{2q^{+}}-\frac{u^{2}}{2u^{+}}\bigg).
\end{align}
The remaining coefficients appearing in the perturbative expansion of the evolution operator $\Omega$ are determined from the constraint equations that follow from the unitarity condition discussed in Sec.~\ref{sec:Omega}. While the matching to the light-cone wave functions fixes the coefficients associated with the physical emission and virtual processes, unitarity imposes additional nontrivial relations among them. These relations ensure that probability is conserved order by order in perturbation theory and that the resulting evolution operator generates a consistent quantum mechanical evolution in Fock space.

Having determined all coefficients entering the perturbative expansion of $\Omega$, the final step is to reconstruct the complete evolution operator. The operator is organized as a perturbative expansion in the coupling constant,
\begin{align}
\Omega & = 1+ g\, \Omega^{(1)} + g^{2}\,\Omega^{(2)}+\cdots
\end{align}
The perturbative construction therefore determines the infinitesimal structure of the light-cone boost order by order in $g$. To extract the connected contributions, one may introduce $\mf{G}$ through
\begin{align}
    \Omega = \text{exp}\left(i\,\mf{G}\right).
\end{align}
Physically, this exponentiation promotes the infinitesimal evolution generated by $\mf{G}$ into a finite light-cone boost acting on the hadronic state, thereby resumming repeated gluon emissions and virtual corrections into a single unitary operator acting on the full Fock space. Explicitly, the generating function $\mf{G}$ takes the form\footnote{$\int_{p}\equiv \int\, \frac{d^{3}p}{(2\pi)^{3}}$}
\begin{align} \label{eq:expon}
    \mf{G} & = -i\int_{p}\left[\mf{A}^{b}_{j}(p)\,a^{\dagger\, b}_{j}(p) - \mf{A}^{\dagger\, b}_{j}(p)\, a^{b}_{j}(p) \right]-\frac{i}{2}\,\int_{p,q}\, \left[\mf{B}^{cb}_{1kj}(p,q)-\mf{B}^{\dagger\, cb}_{2kj}(p,q)\right]\,a^{b}_{j}(p)\,a^{c}_{k}(q) \notag \\ & -\frac{i}{2}\,\int_{p,q}\, \left[\mf{B}^{cb}_{2kj}(p,q)  - \mf{B}^{\dagger\, cb}_{1kj}(p,q)\right]\,a^{\dagger b}_{j}(p)a^{\dagger c}_{k}(q) -\frac{i}{2}\, \int_{p,q}\,\left[\mf{B}^{cb}_{3kj}(p,q)-\mf{B}^{\dagger\,cb}_{3kj}(p,q)\right]a^{\dagger\, b}_{j}(p) a^{c}_{k}(q) \notag \\ & -i\, \int_{p,q,r}\, \left[\mf{C}^{ dcb}_{1lkj}(p,q,r) a^{\dagger d}_{l}(r)  a^{\dagger c}_{k}(q)  a^{b}_{j}(p) -\mf{C}^{\dagger\, bcd}_{1jkl}(r,q,p) a^{\dagger d}_{l}(r) a^{c}_{k}(q) a^{b}_{j}(p)\right]\notag \\ & -\frac{i}{2}\, \int_{p,q,r,s}\, \left[\mf{D}^{edcb}_{1mlkj}(p,q,r,s)-\mf{D}^{\dagger\,edcb}_{3mlkj}(p,q,r,s)\right]a^{\dagger e}_{m}(s)a^{\dagger d}_{l}(r)a^{\dagger c}_{k}(q)a^{b}_{j}(p) \notag \\ & -\frac{i}{2}\,\int_{p,q,r,s}\,\left[\mf{D}^{edcb}_{3mlkj}(p,q,r,s)-\mf{D}^{\dagger\,edcb}_{1mlkj}(p,q,r,s)  \right]\,a^{\dagger e}_{m}(s)a^{d}_{l}(r)a^{c}_{k}(q)a^{b}_{j}(p)\notag \\ & -\frac{i}{2}\, \int_{p,q,r,s}\, \left[\mf{D}^{edcb}_{2mlkj}(p,q,r,s)-\mf{D}^{\dagger\, edcb}_{2mlkj}(p,q,r,s)\right]\,a^{e\dagger}_{m}(s)a^{d\dagger}_{l}(r)a^{c}_{k}(q)a^{b}_{j}(p)\notag \\ & -\, \frac{i}{2}\,\int_{p,q,r,s,u,v}\,\left[\mf{E}^{hfedcb}_{1\,onmlkj}(p,q,r,s,u,v)-\mf{E}^{\dagger\, hfedcb}_{3\,onmlkj}(p,q,r,s,u,v)\right]\notag \\ & a^{\dagger h}_{o}(v)a^{\dagger f}_{n}(u)a^{\dagger e}_{m}(s)a^{\dagger d}_{l}(r)a^{c}_{k}(q)a^{b}_{j}(p) -\,\frac{i}{2}\,\int_{p,q,r,s,u,v}\, \bigg[\mf{E}^{hfedcb}_{3\,onmlkj}(p,q,r,s,u,v)\notag \\ & -\,\mf{E}^{\dagger\,hfedcb}_{1\,onmlkj}(p,q,r,s,u,v)\bigg]\,a^{\dagger h}_{o}(v)a^{\dagger f}_{n}(u)a^{e}_{m}(s)a^{d}_{l}(r)a^{c}_{k}(q)a^{b}_{j}(p) -\frac{i}{2}\, \int_{p,q,r,s,u,v}\,\notag \\ &  \left[\mf{E}^{hfedcb}_{2\,onmlkj}(p,q,r,s,u,v)-\mf{E}^{\dagger\, hfedcb}_{2\,onmlkj}(p,q,r,s,u,v)\right] a^{\dagger h}_{o}(v)a^{\dagger f}_{n}(u)a^{\dagger e}_{m}(s)a^{d}_{l}(r)a^{c}_{k}(q)a^{b}_{j}(p),
\end{align}
where $\mf{A}$ and $\mf{C}$ are at order $g$, while  $\mf{B}_{1}$, $\mf{B}_{2}$, $\mf{B}_{3}$, $\mf{D}_{1}$, $\mf{D}_{2}$, $\mf{D}_{3}$, $\mf{E}_{1}$, $\mf{E}_{2}$, $\mf{E}_{3}$ are at order $g^{2}$. $\mf{B}_{1}, \mf{D}_{3},\mf{E}_{3}$ are fixed by the constraint equation. Note that $\mf{G}$ is Hermitian by construction.
 
\section{Diagonalization of the pure Yang-Mills Hamiltonian at $\mathcal{O}(g)$ and $\mathcal{O}(g^{2})$}\label{sec:diagonal}
The evolution operator provides a systematic framework for diagonalizing the pure Yang-Mills Hamiltonian governing the soft gluon sector of a fast-moving hadron. We introduce the unitary similarity transformation
\begin{equation}
    H_{\text{diag}} = \Omega^{\dagger}\cdot H \cdot \Omega,
\end{equation}
constructed perturbatively in $g$ with coefficients fixed by matching to the LCWF and by unitarity. The role of $\Omega$ is to cancel the off-diagonal matrix elements of the Hamiltonian that connect different Fock sectors perturbatively up to the desired accuracy,
\begin{equation}
    H_{\text{diag}} = \sum_{n}\,E_{n}\ket{n}\bra{n}.
\end{equation}
In Ref.~\cite{kovner2019entanglement}, the diagonalization was carried out using a coherent operator $\mathcal{C}$ that reproduces the classical Weizsäcker-Williams field. However, the non-Abelian three-gluon interaction generates additional operator structures at the same order that must be included for a complete diagonalization. The present work extends the coherent state construction by incorporating all operators at order $g$, including those from the three-gluon vertex. The Hamiltonian at order $g$ is
\begin{align}
    H_{1} & = H_{0} + H_{g} + H_{ggg},
\end{align}
where $H_{0}$ is the free light-cone Hamiltonian, $H_{g}$ is the soft gluon--valence charge interaction and $H_{ggg}$ is the non-Abelian three-gluon term. Substituting the expansion~\eqref{Eq:OmegaExp} into the transformed Hamiltonian gives
\begin{align}
    \Omega^{\dagger}\cdot H_{1} \cdot \Omega & = H_{0}+H_{g}+ H_{ggg}
    + \int \frac{d^{3}p}{(2\pi)^{3}}\Bigg[- \mf{A}^{b}_{j}(p)\frac{p^{2}}{2p^{+}} a^{\dagger\, b}_{j}(p) + \mf{A}^{b\, \dagger}_{j}(p)\frac{p^{2}}{2p^{+}} a^{b}_{j}(p)\Bigg]\notag \\
    & + \int \frac{d^{3}p}{(2\pi)^{3}} \frac{d^{3}q}{(2\pi)^{3}} \Bigg[\mf{C}^{dcb}_{lkj}(p,q,r) a^{\dagger d}_{l}(r)  a^{\dagger c}_{k}(q)  a^{b}_{j}(p)\left(\frac{r^{2}}{2r^{+}}+\frac{q^{2}}{2q^{+}}-\frac{p^{2}}{2p^{+}}\right) \notag \\
    & \qquad\qquad -  \mf{C}^{\dagger\, bcd}_{jkl}(r,q,p) a^{\dagger d}_{l}(r) a^{c}_{k}(q) a^{b}_{j}(p)\left(\frac{r^{2}}{2r^{+}}-\frac{q^{2}}{2q^{+}}-\frac{p^{2}}{2p^{+}}\right)\Bigg].
\end{align}
Upon substituting the explicit values of $H_{g}$, $H_{ggg}$, $\mf{A}(p)$ and $\mf{C}(p,q,r)$, all off-diagonal contributions cancel identically:
\begin{align}\label{eq:diag1}
     \Omega^{\dagger}\cdot H_{1} \cdot \Omega  = H_{0}.
\end{align}
The transformed Hamiltonian is thus fully diagonal within the complete operator basis at order $g$. The remaining free Hamiltonian term reflects the fact that the classical Weizsäcker-Williams energy, which is proportional to the square of the valence color charge density, is already described in the choice of $\mf{A}(p)$.

We now proceed to diagonalization at order $g^{2}$. The full Hamiltonian including all terms through $\mathcal{O}(g^{2})$ is
\begin{align}
    H = H_{0} + H_{g} + H_{ggg} + H_{gggg} + H_{gg\text{-inst.}} + H_{gggg\text{-inst.}},
\end{align}
where $H_{gggg}$ denotes the four-gluon interaction and the remaining terms are instantaneous interactions arising in light-cone quantization after eliminating the non-dynamical gauge field components. Substituting all previously derived coefficients gives
\begin{align}\label{eq:diag2}
     \Omega^{\dagger}\cdot H \cdot \Omega & = H_{0}\notag \\ & 
     - \int \frac{d^{3}p}{(2\pi)^{3}} \mf{A}^{\dagger\, b}_{j}(p)\mf{A}^{b}_{j}(p)\frac{p^{2}}{2p^{+}} \notag \\
     & + 2\int\frac{d^{3}p}{(2\pi)^{3}}\frac{d^{3}q}{(2\pi)^{3}} \frac{d^{3}r}{(2\pi)^{3}}\frac{d^{3}s}{(2\pi)^{3}} \left(\frac{r^{2}}{2r^{+}}+\frac{s^{2}}{2s^{+}}-\frac{p^{2}}{2p^{+}}\right)\notag \\ &
     \mf{C}^{\dagger\,dec}_{lmk}(q,s,r)\mf{C}^{edb}_{mlj}(p,r,s)\,a^{\dagger\, c}_{k}(q)a^{b}_{j}(p).
\end{align}
The first correction is the coherent background-field energy associated with the Weizsäcker-Williams field. The second term arises from the product of two, three-gluon operators and describes higher order contributions from successive soft gluon interactions. Evaluating the second term explicitly, one finds
\begin{align}
     2\int\frac{d^{3}p}{(2\pi)^{3}}\frac{d^{3}q}{(2\pi)^{3}} \frac{d^{3}r}{(2\pi)^{3}}\frac{d^{3}s}{(2\pi)^{3}}\left(\frac{r^{2}}{2r^{+}}+\frac{s^{2}}{2s^{+}}-\frac{p^{2}}{2p^{+}}\right) \mf{C}^{\dagger\,dec}_{lmk}(q,s,r)\mf{C}^{edb}_{mlj}(p,r,s)\,a^{\dagger\, c}_{k}(q)a^{b}_{j}(p) = 0.
\end{align}
This confirms that the three-gluon corrections produce no residual off-diagonal contribution at this order. The remaining coherent term evaluates to
\begin{align}
    \int\frac{d^{3}p}{(2\pi)^{3}} \mf{A}^{\dagger\, b}_{j}(p)\mf{A}^{b}_{j}(p)\frac{p^{2}}{2p^{+}} = \int\frac{d^{3}p}{(2\pi)^{3}} \frac{g^{2}\rho^{b}(p)\rho^{b}(-p)}{p^{+\, 2}},
\end{align}
so the final diagonalized Hamiltonian through $\mathcal{O}(g^{2})$ is
\begin{align}\label{eq:final}
     \Omega^{\dagger}\cdot H \cdot \Omega  = H_{0} -\int \frac{d^{3}p}{(2\pi)^{3}} \frac{g^{2}\rho^{b}(p)\rho^{b}(-p)}{p^{+\, 2}}.
\end{align}
The only surviving correction is the $\rho^{2}$ background field energy, demonstrating explicitly that the perturbatively constructed evolution operator $\Omega$ diagonalizes the pure Yang-Mills Hamiltonian through $\mathcal{O}(g^{2})$ within the CGC framework.

\section{Concluding remarks}\label{sec:conclusion}
We have constructed the soft gluon light-cone wave function and the corresponding unitary evolution operator $\Omega$ of a fast-moving hadron explicitly through $\mathcal{O}(g^{2})$ in pure Yang-Mills theory within the CGC framework. Starting from the light-cone pure Yang-Mills Hamiltonian in the eikonal and Born-Oppenheimer approximations, $\Omega$ was organized as a fully normal ordered series in soft gluon creation and annihilation operators with operator valued coefficients that are functionals of the valence color charge density $\rho$. The non-Abelian nature of these color charges means the coefficients are themselves non commuting operators whose ordering was tracked consistently throughout the construction.

The coefficients were determined through two complementary procedures. The unitarity condition $\Omega^{\dagger}\Omega = 1$, imposed order by order in the coupling, which provided relations among the coefficients and the remaining coefficients were fixed by matching the action of $\Omega$ on the vacuum, one, two and three-gluon Fock states to the LCWF computed using LCPT. The resulting expressions for all coefficients through $\mathcal{O}(g^{2})$ complete and extend the partial results of Ref.~\cite{lublinsky2017high}.

As an application, we used $\Omega$ to diagonalize the soft pure Yang-Mills Hamiltonian. At order $g$, a complete diagonalization requires both the coherent operator and the operator corresponding to the three-gluon vertex. It diagonalizes the Hamiltonian exactly to its free part. At order $g^{2}$, the only diagonal correction is the coherent background field energy proportional to $\rho^{2}$. The evolution operator thus plays a dual role: it generates the perturbative evolution of the light-cone wave function and simultaneously diagonalizes the pure Yang-Mills Hamiltonian order by order in the coupling.

The construction of $\Omega$ presented in this paper constitutes a necessary step towards the computation of single inclusive gluon production at next-to-leading order at mid-rapidity~\cite{rr}. The single inclusive gluon spectrum is obtained as the expectation value of the soft gluon number operator in the projectile state that has been evolved with $\Omega$ and scattered off the target color field; its evaluation at NLO therefore requires the explicit coefficients of the evolution operator derived here~\cite{kovner2006one}. A complete computation of single-gluon production at NLO also involves certain coefficients of $\Omega$ at $\mathcal{O}(g^{3})$, which were derived in Ref.~\cite{lublinsky2017high}. We also leave for future work the computation of double and triple gluon production.

\acknowledgments
The author is grateful to Michael Lublinsky and Vladimir Skokov for constant encouragement to continue this work and to Tiyasa Kar for cross checking critical parts of the calculation. The author also thanks the Department of Physics at North Carolina State University for providing research assistantship support during the course of this work. This work is supported by the U.S. Department of Energy, Office of Science, Office of Nuclear Physics through Contract No. DE-SC0020081.
\appendix
\section{From YM Lagrangian to Light-cone YM Hamiltonian}\label{sec:app0}
The pure Yang-Mills Lagrangian:
\begin{equation}\label{eq:QCD}
\mathcal{L}_{YM}=-\frac{1}{4}F^{a\mu\nu}F_{\mu\nu}^{a}.
 \end{equation}
The tensor $F_{\mu\nu}$ defined as $F_{\mu\nu}\equiv t^{a}F_{\mu\nu}^{a}=-\frac{i}{g}\left[D_{\mu},\, D_{v}\right]$, 
satisfies the following equations of motion:
\begin{equation}\label{glueq}
D_{\mu}F^{a\mu\nu}=\partial_{\mu}F^{a\mu\nu}-gf^{abc}A_{\mu}^{b}F^{c\mu\nu}=-gJ^{a\nu}.
 \end{equation}
The canonical momenta:
\begin{equation}(\Pi_{A})^{a\mu}(x)\,\equiv\,\frac{\delta\mathcal{L}_{YM}}{\delta(\partial^{-}A_{\mu}^{a})}=\frac{1}{2}F^{a\mu+}.
\end{equation}
The following constraints are obtained from the $+$ component of (\ref{glueq}):\begin{equation}D_{\mu}F^{a\mu+}=-gJ^{a+},\qquad\qquad\qquad\qquad\partial^{+}A_{+}^{a}=-\frac{1}{\partial^{+}}(D_{i}^{ab}\partial^{+}A_{i}^{b}-gJ^{a+}).
\end{equation}
Eliminating $A^{-}$, the LC Hamiltonian is:
\begin{eqnarray}
H_{LC\; YM}&&=P_{+}=\int dx^{-}\, d^{2}\mathbf{x}\,\mathcal{H}_{YM}=\int dx^{-}\, d^{2}\mathbf{x}\,\left((\Pi_{A})^{\mu}\partial^{-}A_{\mu}-\mathcal{L}_{YM}\right)\nonumber\\
&&=\int dx^{-}\, d^{2}\mathbf{x}\,\left(-F^{\mu+}F_{\mu+}+\frac{1}{4}F^{\mu\nu}F_{\mu\nu}\right).
\end{eqnarray}
where
\begin{equation}F^{ij}F_{ij}\,=\,2\left((\partial_{i}A_{j}^{a})(\partial_{i}A_{j}^{a})\,-\,(\partial_{i}A_{j}^{a})(\partial_{j}A_{i}^{a})\right)\,-\,4gf^{abc}A_{i}^{b}A_{j}^{c}\partial_{i}A_{j}^{a}\,+\, g^{2}f^{abc}f^{ade}A_{i}^{b}A_{j}^{c}A_{i}^{d}A_{j}^{e}.
\end{equation}
\begin{eqnarray}
&&\Pi^{a}(x^{-},\, \mathbf{x})\Pi^{a}(x^{-},\, \mathbf{x})\\
&&\quad=\,\left(\frac{1}{\partial^{+}}D_{i}\partial^{+}A_{i}\right)^{a}\left(\frac{1}{\partial^{+}}D_{j}\partial^{+}A_{j}\right)^{a}\,+\,2g\left(\frac{1}{\partial^{+}}D_{i}\partial^{+}A_{i}\right)^{a}\frac{1}{\partial^{+}}J^{a+}\,+\, g^{2}\frac{1}{\partial^{+}}J^{a+}\frac{1}{\partial^{+}}J^{a+}.\nonumber
\end{eqnarray}

\section{State definitions and normalization}\label{sec:app1}
Following conventional notation, $\left|0\right\rangle$ denotes the vacuum state of the free Hamiltonian on the soft Hilbert space.
The energy of this state will be set to zero, $E_{0}=0$. 
We define the vacuum state as 
\begin{equation}
 a_{i}^{a}(k^{+},\, \mathbf{k})\left|0\right\rangle =0,
\end{equation}
for any $\Lambda<k^{+}< \vee$. The soft vacuum $\ket{0}$  should be understood as $\ket{0}\otimes\ket{v}$  and, because of its valence component, it  differs from the QCD vacuum. Additionally, we will use the following three eigen states of free Hamiltonian:\\
\begin{itemize}
  \item One gluon state:  
    \begin{equation}
      \label{gsta}
  \left|g_{i}^{a}(k)\right\rangle \, = \,\frac{a_{i}^{a\dagger}(k)}{(2\pi)^{3/2}}\left|0\right\rangle 
\end{equation}
The normalization of the state is 
$$
\left\langle g_{j}^{b}(p)\left|g_{i}^{a}(k)\right.\right\rangle
=\delta^{ab}\delta_{ij}\delta^{(2)}(\mathbf{k}-\mathbf{p})\delta(k^{+}-p^{+})\;$$ 
and 
\begin{equation}
 a_{i}^{a}(k^{+},\, \mathbf{k})\left|g^b_j(q)\right\rangle = (2\pi )^{3/2}\delta^{ab} \delta_{ij}\delta^{(3)}(k-q)  \left| 0 \right\rangle
\end{equation}
For completeness we  note $E_{g}(k)=\frac{\mathbf{k}^{2}}{2k^{+}}.$ 
\item Two-gluon state: 
\begin{equation}
\begin{split}\label{ggst}
&\ket{g_{i}^{a}(k)\, g_{j}^{b}(p)} \,= \,\frac{a_{i}^{a\dagger}(k)\, a_{j}^{b\dagger}(p)}{(2\pi)^{3}}\left|0\right\rangle.
\end{split}
\end{equation}\\
  with the energy $E_{gg}(k,\, p)\,\equiv\,\frac{\mathbf{k}^{2}}{2k^{+}}+\frac{\mathbf{p}^{2}}{2p^{+}}$.\\
\item Three-gluon state: 
  \begin{equation}\begin{split}\label{gggst}
&\ket{g_{i}^{a}(k)\, g_{j}^{b}(p)\, g^{c}_{k}(q)} \,= \,\frac{a_{i}^{a\dagger}(k)\, a_{j}^{b\dagger}(p)a_{k}^{c\dagger}(q)}{(2\pi)^{9/2}}\left|0\right\rangle.
  \end{split}\end{equation}
with the energy $E_{gg}(k,\, p\,q)\,\equiv\,\frac{\mathbf{k}^{2}}{2k^{+}}+\frac{\mathbf{p}^{2}}{2p^{+}}+\frac{\mathbf{q}^{2}}{2q^{+}}$.  \\
The creation and annihilation operators obey the bosonic algebra:
\begin{equation}
\left[a_{i}^{a}(k^{+},\mathbf{k}),\, a_{j}^{b\dagger}(p^{+},\mathbf{p})\right]=(2\pi)^{3}\delta^{ab}\delta_{ij}\delta(k^{+}-p^{+})\delta^{(2)}(\mathbf{k}-\mathbf{p}).
 \end{equation}
\end{itemize}
Here, we describe the notation used in the representation of the wave function. As an example, we consider the leading order diagram:
\begin{figure}[htbp]
\centering
    \begin{tikzpicture}[baseline={([yshift=-.8ex] current bounding box.center)}]
	\begin{pgfonlayer}{nodelayer}
		\node [style=none] (4) at (-12, 12) {};
		\node [style=none] (5) at (-7, 10) {};
		\node [style=none] (15) at (-9, 13) {};
		\node [style=none] (16) at (-9, 8) {};
        \node [style=none] (11) at (-5.5, 9.5) {$k, a, i$};
		\node [style=none] (18) at (-15, 12) {};
		\node [style=none] (19) at (-7, 12) {};
	\end{pgfonlayer}
	\begin{pgfonlayer}{edgelayer}
		\draw [style=gluon, bend right=15] (4.center) to (5.center);
		\draw [style=interm] (15.center) to (16.center);
		\draw [style=basic] (18.center) to (19.center);
	\end{pgfonlayer}
\end{tikzpicture}
\end{figure}
The LCWF is given by
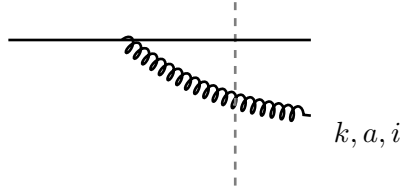
\begin{align}
    \left|\Psi_{g\rho}\right\rangle
    = - \int_{\Lambda}^{\vee} dk^{+} \int d^{2}\mathbf{k}\,
    \ket{g^{a}_{i}(k)}
    \frac{\left\langle g_{i}^{a}(k)\left| H_{g}\right|0\right\rangle}
    {E_{g}(k)}.
\end{align}
The dashed lines in the diagram denote intermediate states. In the subscript of the wave function, $g$ denotes the outgoing gluon, while $\rho$ indicates that the corresponding contribution contains one insertion of the color charge density $\rho$. The wave function is written as a function of the incoming momentum, whereas all outgoing and intermediate state momenta are integrated over.
\section{The CGC Hamiltonian: Eikonal Approximation for pure Yang Mills Hamiltonian} \label{sec:App2}
In order to introduce the eikonal approximation for the LC YM Hamiltonian, we split the gluon fields into  the soft and valence modes: 
\begin{equation}\label{Am}
\underline{A}_{i}^{a}(x)=\int_{\Lambda}^{\vee}\frac{dk^{+}}{2\pi}\int\frac{d^{2}\mathbf{k}}{(2\pi)^{2}}\frac{1}{\sqrt{2k^{+}}}\left(a_{i}^{a}(k^{+},\mathbf{k})e^{-ik\cdot x}+a_{i}^{a\dagger}(k^{+},\mathbf{k})e^{ik\cdot x}\right),
 \end{equation}
and
\begin{equation}\label{Ap}
\overline{A}_{i}^{a}(x)=\int_{\vee}^{\infty}\frac{dk^{+}}{2\pi}\int\frac{d^{2}\mathbf{k}}{(2\pi)^{2}}\frac{1}{\sqrt{2k^{+}}}\left(a_{i}^{a}(k^{+},\mathbf{k})e^{-ik\cdot x}+a_{i}^{a\dagger}(k^{+},\mathbf{k})e^{ik\cdot x}\right).
 \end{equation}
After inserting $A_{i}^{a}(x)=\underline{A}_{i}^{a}(x)+\overline{A}_{i}^{a}(x)$  to $H_{\text{int.}}$, we define the following components of the interaction Hamiltonian:
 \begin{equation}\label{Hg}
H_{g}\,\equiv\,-g\int dx^{-}\, d^{2}\mathbf{x}\,(\partial_{i}\underline{A}_{i}^{a})\left(f^{abc}\frac{1}{\partial^{+}}(\overline{A}_{j}^{b}\partial^{+}\overline{A}_{j}^{c})\right),
  \end{equation}
\begin{equation}\label{hggg}
H_{ggg}\,\equiv\,-gf^{abc}\int dx^{-}\, d^{2}\mathbf{x}\,\left((\partial_{i}\underline{A}_{j}^{a})\underline{A}_{i}^{b}\underline{A}_{j}^{c}+(\partial_{i}\underline{A}_{i}^{a})\frac{1}{\partial^{+}}(\underline{A}_{j}^{b}\partial^{+}\underline{A}_{j}^{c})\right),
 \end{equation}
  \begin{equation}\begin{split}\label{ggi}
&H_{gg-inst}\,\equiv\,g^{2}f^{abc}\int dx^{-}\, d^{2}\mathbf{x}\,\frac{1}{\partial^{+}}(\underline{A}_{i}^{b}\partial^{+}\underline{A}_{i}^{c})\left(f^{ade}\frac{1}{\partial^{+}}(\overline{A}_{j}^{d}\partial^{+}\overline{A}_{j}^{e})\right),
 \end{split}\end{equation}
 \begin{equation}H_{gggg}+H_{gggg\,\text{inst.}}\,\equiv\,\frac{g^{2}}{4}f^{abc}f^{ade}\int dx^{-}\, d^{2}\mathbf{x}\,\left(\underline{A}_{i}^{b}\underline{A}_{j}^{c}\underline{A}_{i}^{d}\underline{A}_{j}^{e}-2\frac{1}{\partial^{+}}(\underline{A}_{i}^{b}\partial^{+}\underline{A}_{i}^{c})\frac{1}{\partial^{+}}(\underline{A}_{j}^{d}\partial^{+}\underline{A}_{j}^{e})\right).
 \end{equation}
The eikonal approximation means that  all  contributions  involving division by large (valence) longitudinal momenta are neglected, such as $\frac{1}{\partial^{+}}(\underline{A}_{j}^{b}\partial^{+}\overline{A}_{j}^{c})\sim0$. 
Furthermore, we retain only the terms needed to compute the wave functions at next-to-leading order, namely those describing interactions among soft modes or interactions between soft and valence modes.

\section{Matrix elements} \label{sec:App3}
We write down all the matrix elements that are relevant for us to compute all the wave functions at the next to leading order. 

\begin{itemize}
  
  \item   
    Gluon emission from valence $\rho$
    \begin{align}
      \label{m1}
      \left\langle g_{i}^{a}(k)\left|\, H_{g}\,\right|0\right\rangle &= 
      \frac{g{k}^{i}\rho^{a}(-\v{k})}{4\pi^{3/2}|k^{+}|^{3/2}} 
    \end{align}
  
  \item Valence adds an extra gluon to an existing one 
    \begin{align}
      \left\langle g_{l}^{c}(q)\, g_{j}^{b}(p)\left|H_{g}\right|g_{i}^{a}(k)\right\rangle &\,=\,
      \delta^{ac}\delta_{il}\delta^{(3)}(k-q)\,\frac{g {p}^{j}\rho^{b}(-\mathbf{p})}{4\pi^{3/2}|p^{+}|^{3/2}}
      \,\notag \\ & +\,\delta^{ab}\delta_{ij}\delta^{(3)}(k-p)\,\frac{g {q}^{l}\rho^{c}(-\mathbf{q})}{4\pi^{3/2}|q^{+}|^{3/2}}
      \label{m2}
    \end{align}

  \item Three-gluon interaction 
    \begin{align}
      \label{m3}
      &\left\langle g_{l}^{c}(q)\, g_{j}^{b}(p)\left|H_{ggg}\right|g_{i}^{a}(k)\right\rangle = \frac{igf^{abc}\delta^{(3)}(k-p-q)}{8\pi^{3/2}\sqrt{k^{+}p^{+}q^{+}}}
      \left[\left({p}^{i}-{q}^{i}+\frac{q^{+}-p^{+}}{k^{+}}{k}^{i}\right)\delta_{jl}\right.\nonumber\\
      &\quad\quad\quad \left.+\left({k}^{j}+{q}^{j}-\frac{k^{+}+q^{+}}{p^{+}}{p}^{j}\right)
      \delta_{il}+\left(\frac{k^{+}+p^{+}}{q^{+}}{q}^{l}-{p}^{l}-{k}^{l}\right)\delta_{ij}\right].
    \end{align}
    \item Valence instantaneously creates two-gluons
    \begin{align}\label{gg1}
\left\langle g_{j}^{c}(q)\, g_{i}^{b}(p)\left|H_{gg-inst}\right|0\right\rangle \,& =\,\frac{ig^{2}f^{abc}(q^{+}-p^{+})\delta_{ij}\rho^{a}(-\mathbf{p}-\mathbf{q})}{2(2\pi)^{3}\sqrt{p^{+}q^{+}}(p^{+}+q^{+})^{2}}.
 \end{align}
  \item Instantaneous  interaction between valence and gluon  
    \begin{align}
      \label{gg2}
      \left\langle g_{j}^{c}(q)\left|H_{gg-inst}\right|g_{i}^{b}(p)\right\rangle \,
      =\,
      \frac{ig^{2}f^{abc}(p^{+}+q^{+})\delta_{ij}\rho^{a}(\mathbf{p}-\mathbf{q})}{2(2\pi)^{3}\sqrt{p^{+}q^{+}}(p^{+}-q^{+})^{2}}.
    \end{align}
    \item Four gluon interaction
      \begin{align}
    \bra{g^{b}_{j}(p)g^{c}_{k}(q)g^{d}_{l}(r)}& H_{gggg}\ket{g^{a}_{i}(k)}  = \frac{g^{2}}{4(2\pi)^{3}}\frac{\delta^{(3)}(-k+p+q+r)}{\sqrt{k^{+}p^{+}q^{+}r^{+}}}\Bigg[f^{a'ab}f^{a'cd}\bigg(\delta_{lj}\delta_{ik}\notag \\ & -\delta_{il}\delta_{kj} +\frac{(p^{+}+k^{+})(r^{+}-q^{+})}{(r^{+}+q^{+})(k^{+}-p^{+})}\delta_{lk}\delta_{ij}\bigg) +f^{a'ac}f^{a'bd}\bigg(\delta_{lk}\delta_{ij}-\delta_{jk}\delta_{il} \notag \\ & +\frac{(q^{+}+k^{+})(r^{+}-p^{+})}{(r^{+}+p^{+})(k^{+}-q^{+})}\delta_{ik}\delta_{lj}\bigg)+f^{a'ad}f^{a'bc}\bigg(\delta_{lk}\delta_{ij}-\delta_{lj}\delta_{ik} \notag \\ & +\frac{(r^{+}+k^{+})(q^{+}-p^{+})}{(p^{+}+q^{+})(k^{+}-r^{+})}\delta_{jk}\delta_{il}\bigg)\Bigg]
    \label{ggg-g}
\end{align}
\begin{align}\label{gg-gg}
   \bra{g^{b}_{j}(p)g^{c}_{k}(q)}&\left[H_{gggg}+H_{gg-gg\,\text{inst.}}\right] \ket{g^{a}_{i}(k)g^{d}_{l}(r)} \notag \\ &  = \frac{g^{2}}{4(2\pi)^{3}}\frac{\delta^{(3)}(-k+p-r+q)}{\sqrt{k^{+}q^{+}r^{+}p^{+}}}\notag \\ & \Bigg[f^{a'ab}f^{a'cd}\bigg(\delta_{ik}\delta_{jl}-\delta_{jk}\delta_{il}+\frac{(q^{+}+r^{+})(p^{+}+k^{+})}{(k^{+}-p^{+})(r^{+}-q^{+})}\delta_{ij}\delta_{lk}\bigg)\notag \\ & +f^{a'ac}f^{a'bd}\bigg(\delta_{lk}\delta_{ij}-\delta_{kj}\delta_{il}+\frac{(p^{+}+r^{+})(q^{+}+k^{+})}{(k^{+}-q^{+})(r^{+}-p^{+})}\delta_{lj}\delta_{ik}\bigg)\notag \\ & + f^{a'ad}f^{a'bc}\bigg(\delta_{ij}\delta_{lk}-\delta_{ik}\delta_{lj}+\frac{(q^{+}-p^{+})(k^{+}-r^{+})}{(k^{+}+r^{+})(p^{+}+q^{+})}\delta_{il}\delta_{jk}\bigg) \Bigg]
\end{align}
  \end{itemize}
  \section{Perturbation Theory with non-commutative Matrix Elements} \label{sec:App5}
In this Appendix we outline the derivation of the wave function up to
order $g^2$. The derivation follows standard Rayleigh--Schr\"odinger
(time independent) perturbation theory, as presented in textbooks on
quantum mechanics~\cite{Landau:1991wop,shankar2012principles}; our aim here is to present it in a form that remains valid when the relevant matrix elements are operator valued and non-commuting. We start from the unperturbed (free) Hamiltonian $H_{0}$, whose eigenvalues and eigenstates are denoted by $E_{n}^{(0)}$ and $\left|n^{(0)}\right\rangle $, respectively. We then introduce a weak interaction Hamiltonian $H_{\text{int.}}$, and denote the eigenvalues and eigenstates of the full Hamiltonian by $E_{n}$, and $\left|n\right\rangle$:
\begin{equation}\label{dwf1}
H_{0}\,\left|n^{(0)}\right\rangle \,=\, E_{n}^{(0)}\,\left|n^{(0)}\right\rangle \:;\qquad\qquad\qquad\left(H_{0}\,+\, H_{int}\right)\,\left|n\right\rangle \,=\, E_{n}\,\left|n\right\rangle .
  \end{equation}
  \begin{equation}\begin{split}\label{poas}
  &E_{n}\,=\, E_{n}^{(0)}\,+\, E_{n}^{(1)}\,+\, E_{n}^{(2)}\,+\,\ldots\qquad\qquad\left|n\right\rangle \,=\,\left|n^{(0)}\right\rangle \,+\,\left|n^{(1)}\right\rangle \,+\,\left|n^{(2)}\right\rangle \,+\,\ldots
 \end{split}\end{equation}
 where $\left|n^{(i)}\right\rangle $ are orthogonal states, corresponding to $i$-th order in perturbation theory. By inserting (\ref{poas}) in (\ref{dwf1}), expanding and equating terms in the same order of $g$, we obtain
\begin{eqnarray}\label{waex}
\left|n\right\rangle \,&& =\,\left|n^{(0)}\right\rangle \,-\,\sum_{i\neq0}\frac{1}{E_{i}^{(0)}\,-\, E_{0}^{(0)}\,-\, E_{i}^{(1)}\,-\, E_{i}^{(2)}}\,\left|n^{(i)}\right\rangle \left\langle n^{(i)}\left|H_{int}\right|n^{(0)}\right\rangle\nonumber\\ 
&&+\,\sum_{i,j\neq0}\,\frac{1}{E_{i}^{(0)}-E_{0}^{(0)}-E_{i}^{(1)}}\,\left|n^{(i)}\right\rangle \,\left\langle n^{(i)}\left|H_{int}\right|n^{(j)}\right\rangle \,\frac{1}{E_{j}^{(0)}-E_{0}^{(0)}-E_{j}^{(1)}}\,\left\langle n^{(j)}\left|H_{int}\right|n^{(0)}\right\rangle ,
\end{eqnarray}
where $$E_{i}^{(1)}\,\equiv\,\left\langle n^{(i)}\right|H_{int}\left|n^{(i)}\right\rangle$$
Let the vacuum state to be denoted $\left|\psi\right\rangle _{N}$ and we set  $E_{0}^{(0)}=0$. Substituting this in Eq.  (\ref{waex}), we  get:
\begin{equation}\begin{split}\left|\psi\right\rangle _{N}&=\,\left|0\right\rangle \,-\,\left|n^{(i)}\right\rangle \frac{\left\langle n^{(i)}\left|H_{int}\right|0\right\rangle }{E_{i}^{(0)}}\,+\,\left|n^{(i)}\right\rangle \frac{\left\langle n^{(i)}\left|H_{int}\right|n^{(j)}\right\rangle \,\left\langle n^{(j)}\left|H_{int}\right|0\right\rangle }{E_{i}^{(0)}E_{j}^{(0)}}.
\end{split}
\end{equation}

\bibliography{ref}

\providecommand{\href}[2]{#2}\begingroup\raggedright\begin{thebibliography}{10}

\bibitem{morreale2021mining}
A.~Morreale and F.~Salazar, \emph{Mining for \textsc{G}luon \textsc{S}aturation at \textsc{C}olliders}, \href{https://doi.org/10.3390/universe7080312}{\emph{Universe} {\bfseries 7} (2021) 312}.

\bibitem{albacete2014gluon}
J.~Albacete and C.~Marquet, \emph{Gluon saturation and initial conditions for relativistic heavy ion collisions}, \href{https://doi.org/10.1016/j.ppnp.2014.01.004}{\emph{Progress in Particle and Nuclear Physics} {\bfseries 76} (2014) 1–42}.

\bibitem{blaizot2017high}
J.-P.~Blaizot, \emph{High gluon densities in heavy ion collisions}, \href{https://doi.org/10.1088/1361-6633/aa5435}{\emph{Reports on Progress in Physics} {\bfseries 80} (2017) 032301}.

\bibitem{mueller1986gluon}
A.~Mueller and J.~Qiu, \emph{Gluon recombination and shadowing at small values of $x$}, \href{https://doi.org/https://doi.org/10.1016/0550-3213(86)90164-1}{\emph{Nuclear Physics B} {\bfseries 268} (1986) 427}.

\bibitem{gelis2010color}
F.~Gelis, E.~Iancu, J.~Jalilian-Marian and R.~Venugopalan, \emph{\textsc{T}he \textsc{C}olor \textsc{G}lass \textsc{C}ondensate}, {\emph{Annual Review of Nuclear and Particle Science} {\bfseries 60} (2010) 463}.

\bibitem{weigert2005evolution}
H.~Weigert, \emph{Evolution at small-$x$ : \textsc{T}he \textsc{C}olor \textsc{G}lass \textsc{C}ondensate}, \href{https://doi.org/10.1016/j.ppnp.2005.01.029}{\emph{Progress in Particle and Nuclear Physics} {\bfseries 55} (2005) 461–565}.

\bibitem{McLerran:1993ka}
L.D.~McLerran and R.~Venugopalan, \emph{{Gluon distribution functions for very large nuclei at small transverse momentum}}, \href{https://doi.org/10.1103/PhysRevD.49.3352}{\emph{Phys. Rev. D} {\bfseries 49} (1994) 3352} [\href{https://arxiv.org/abs/hep-ph/9311205}{{\ttfamily hep-ph/9311205}}].

\bibitem{McLerran:1993ni}
L.D.~McLerran and R.~Venugopalan, \emph{{Computing quark and gluon distribution functions for very large nuclei}}, \href{https://doi.org/10.1103/PhysRevD.49.2233}{\emph{Phys. Rev. D} {\bfseries 49} (1994) 2233} [\href{https://arxiv.org/abs/hep-ph/9309289}{{\ttfamily hep-ph/9309289}}].

\bibitem{iancu2001renormalization}
E.~Iancu, A.~Leonidov and L.~McLerran, \emph{The renormalization group equation for the \textsc{T}he \textsc{C}olor \textsc{G}lass \textsc{C}ondensate}, \href{https://doi.org/10.1016/s0370-2693(01)00524-x}{\emph{Physics Letters B} {\bfseries 510} (2001) 133–144}.

\bibitem{iancu2001nonlinear}
E.~Iancu, A.~Leonidov and L.~McLerran, \emph{Nonlinear gluon evolution in the \textsc{T}he \textsc{C}olor \textsc{G}lass \textsc{C}ondensate: I}, \href{https://doi.org/10.1016/s0375-9474(01)00642-x}{\emph{Nuclear Physics A} {\bfseries 692} (2001) 583–645}.

\bibitem{weigert2002unitarity}
H.~Weigert, \emph{Unitarity at small \textsc{B}jorken\, $x$}, \href{https://doi.org/10.1016/s0375-9474(01)01668-2}{\emph{Nuclear Physics A} {\bfseries 703} (2002) 823–860}.

\bibitem{kovner2000relating}
A.~Kovner, J.G.~Milhano and H.~Weigert, \emph{Relating different approaches to nonlinear \textsc{QCD} evolution at finite gluon density}, \href{https://doi.org/10.1103/physrevd.62.114005}{\emph{Physical Review D} {\bfseries 62} (2000) }.

\bibitem{jalilian1998wilson}
J.~Jalilian-Marian, A.~Kovner, A.~Leonidov and H.~Weigert, \emph{Wilson renormalization group for low - $x$ physics: \textsc{T}owards the high density regime}, \href{https://doi.org/10.1103/physrevd.59.014014}{\emph{Physical Review D} {\bfseries 59} (1998) }.

\bibitem{jalilian19981wilson}
J.~Jalilian-Marian, A.~Kovner and H.~Weigert, \emph{Wilson renormalization group for low $x$ physics: \textsc{G}luon evolution at finite parton density}, \href{https://doi.org/10.1103/physrevd.59.014015}{\emph{Physical Review D} {\bfseries 59} (1998) }.

\bibitem{jalilian1997bfkl}
J.~Jalilian-Marian, A.~Kovner, A.~Leonidov and H.~Weigert, \emph{The \textsc{BFKL} equation from the \textsc{W}ilson renormalization group}, \href{https://doi.org/10.1016/s0550-3213(97)00440-9}{\emph{Nuclear Physics B} {\bfseries 504} (1997) 415–431}.

\bibitem{jalilian1997intrinsic}
J.~Jalilian-Marian, A.~Kovner, L.~McLerran and H.~Weigert, \emph{Intrinsic glue distribution at very small-$x$}, \href{https://doi.org/10.1103/physrevd.55.5414}{\emph{Physical Review D} {\bfseries 55} (1997) 5414–5428}.

\bibitem{kovchegov2000unitarization}
Y.V.~Kovchegov, \emph{Unitarization of the \textsc{BFKL} \textsc{P}omeron on a nucleus}, \href{https://doi.org/10.1103/physrevd.61.074018}{\emph{Physical Review D} {\bfseries 61} (2000) }.

\bibitem{kovchegov1999small}
Y.V.~Kovchegov, \emph{Small-$x$ structure function of a nucleus including multiple \textsc{P}omeron exchanges}, \href{https://doi.org/10.1103/physrevd.60.034008}{\emph{Physical Review D} {\bfseries 60} (1999) }.

\bibitem{lublinsky2017high}
M.~Lublinsky and Y.~Mulian, \emph{High energy \textsc{QCD} at \textsc{NLO}: from light-cone wave function to \textsc{JIMWLK} evolution}, \href{https://doi.org/10.1007/JHEP05%282017%29097}{\emph{Journal of High Energy Physics} {\bfseries 2017} (2017) 1}.

\bibitem{kovner2011particle}
A.~Kovner, \emph{Particle production and angular correlations at high energy.}, {\emph{Acta Physica Polonica B} {\bfseries 42} (2011) }.

\bibitem{altinoluk2018double}
T.~Altinoluk, N.~Armesto, A.~Kovner and M.~Lublinsky, \emph{Double and triple inclusive gluon production at mid rapidity: quantum interference in $p-a$ scattering}, \href{https://doi.org/10.1140/epjc/s10052-018-6186-1}{\emph{The European Physical Journal C} {\bfseries 78} (2018) }.

\bibitem{altinoluk2011particle}
T.~Altinoluk and A.~Kovner, \emph{Particle production at high energy and large transverse momentum: “\textsc{T}he hybrid formalism” revisited}, \href{https://doi.org/10.1103/physrevd.83.105004}{\emph{Physical Review D} {\bfseries 83} (2011) }.

\bibitem{iancu2021forward}
E.~Iancu and Y.~Mulian, \emph{Forward dijets in proton-nucleus collisions at next-to-leading order: the real corrections}, \href{https://doi.org/10.1007/jhep03(2021)005}{\emph{Journal of High Energy Physics} {\bfseries 2021} (2021) }.

\bibitem{dominguez2013universality}
F.~Dominguez, C.~Marquet, A.M.~Stasto and B.-W.~Xiao, \emph{Universality of multiparticle production in \textsc{QCD} at high energies}, \href{https://doi.org/10.1103/physrevd.87.034007}{\emph{Physical Review D} {\bfseries 87} (2013) }.

\bibitem{iancu2023dihadron}
E.~Iancu and Y.~Mulian, \emph{Dihadron production in \textsc{DIS} at \textsc{NLO}: the real corrections}, \href{https://doi.org/10.1007/jhep07(2023)121}{\emph{Journal of High Energy Physics} {\bfseries 2023} (2023) }.

\bibitem{levin2010gluon}
E.~Levin and A.H.~Rezaeian, \emph{Gluon saturation and inclusive hadron production at \textsc{LHC}}, \href{https://doi.org/10.1103/physrevd.82.014022}{\emph{Physical Review D} {\bfseries 82} (2010) }.

\bibitem{dumitru2010two}
A.~Dumitru and J.~Jalilian-Marian, \emph{Two-particle correlations in high-energy collisions and the gluon four-point function}, \href{https://doi.org/10.1103/physrevd.81.094015}{\emph{Physical Review D} {\bfseries 81} (2010) }.

\bibitem{iancu2019forward}
E.~Iancu and Y.~Mulian, \emph{Forward trijet production in proton–nucleus collisions}, \href{https://doi.org/10.1016/j.nuclphysa.2019.02.003}{\emph{Nuclear Physics A} {\bfseries 985} (2019) 66–127}.

\bibitem{iancu2013jimwlk}
E.~Iancu and D.~Triantafyllopoulos, \emph{\textsc{JIMWLK} evolution for multi-particle production in \textsc{L}angevin form}, \href{https://doi.org/10.1007/jhep11(2013)067}{\emph{Journal of High Energy Physics} {\bfseries 2013} (2013) }.

\bibitem{levin2019j}
E.~Levin and M.~Siddikov, \emph{J / $\psi$ production in hadron scattering: three-pomeron contribution}, \href{https://doi.org/10.1140/epjc/s10052-019-6894-1}{\emph{The European Physical Journal C} {\bfseries 79} (2019) }.

\bibitem{gotsman2015cgc}
E.~Gotsman, E.~Levin and U.~Maor, \emph{\textsc{CGC}/saturation approach for soft interactions at high energy: long range rapidity correlations},  2015.

\bibitem{bartels2008inclusive}
J.~Bartels, M.~Salvadore and G.~Vacca, \emph{Inclusive $1$-jet production cross section at small - $x$ in \textsc{QCD}: multiple interactions}, \href{https://doi.org/10.1088/1126-6708/2008/06/032}{\emph{Journal of High Energy Physics} {\bfseries 2008} (2008) 032–032}.

\bibitem{altinoluk2009inclusive}
T.~Altinoluk, A.~Kovner and M.~Lublinsky, \emph{Inclusive gluon production in the \textsc{QCD} \textsc{R}eggeon field theory: \textsc{P}omeron loops included}, \href{https://doi.org/10.1088/1126-6708/2009/03/110}{\emph{Journal of High Energy Physics} {\bfseries 2009} (2009) 110–110}.

\bibitem{altinoluk2012particle}
T.~Altinoluk and A.~Kovner, \emph{Particle production at high energy and large transverse momentum: “\textsc{T}he hybrid formalism” revisited}, \href{https://doi.org/10.1103/physrevd.83.105004}{\emph{Physical Review D} {\bfseries 83} (2011) }.

\bibitem{gribov1983semihard}
L.~Gribov, E.~Levin and M.~Ryskin, \emph{Semihard processes in \textsc{QCD}}, \href{https://doi.org/https://doi.org/10.1016/0370-1573(83)90022-4}{\emph{Physics Reports} {\bfseries 100} (1983) 1}.

\bibitem{salzaar1}
P.~Caucal, M.G.~Morales and F.~Salazar, \emph{Forward trijet production in proton-nucleus collisions: gluon initiated channel},  \href{https://arxiv.org/abs/2604.07509}{{\ttfamily 2604.07509}}.

\bibitem{venugopalan2007introduction}
R.~Venugopalan, \emph{Introduction to light cone field theory and high energy scattering},  in \emph{Hadrons in Dense Matter and Hadrosynthesis}, p.~89–112, Springer Berlin Heidelberg (2007), \href{https://doi.org/10.1007/bfb0107312}{DOI}.

\bibitem{Li:2021zmf}
M.~Li and V.V.~Skokov, \emph{{First saturation correction in high energy proton-nucleus collisions. Part I. Time evolution of classical Yang-Mills fields beyond leading order}}, \href{https://doi.org/10.1007/JHEP06(2021)140}{\emph{JHEP} {\bfseries 06} (2021) 140} [\href{https://arxiv.org/abs/2102.01594}{{\ttfamily 2102.01594}}].

\bibitem{Li:2021yiv}
M.~Li and V.V.~Skokov, \emph{{First saturation correction in high energy proton-nucleus collisions. Part II. Single inclusive semi-hard gluon production}}, \href{https://doi.org/10.1007/JHEP06(2021)141}{\emph{JHEP} {\bfseries 06} (2021) 141} [\href{https://arxiv.org/abs/2104.01879}{{\ttfamily 2104.01879}}].

\bibitem{Li:2021ntt}
M.~Li and V.V.~Skokov, \emph{{First saturation correction in high energy proton-nucleus collisions. Part III. Ensemble averaging}}, \href{https://doi.org/10.1007/JHEP01(2022)160}{\emph{JHEP} {\bfseries 01} (2022) 160} [\href{https://arxiv.org/abs/2111.05304}{{\ttfamily 2111.05304}}].

\bibitem{Kovchegov:2024aus}
Y.V.~Kovchegov and M.~Li, \emph{{Gluon double-spin asymmetry in the longitudinally polarized p + p collisions}}, \href{https://doi.org/10.1007/JHEP05(2024)177}{\emph{JHEP} {\bfseries 05} (2024) 177} [\href{https://arxiv.org/abs/2403.06959}{{\ttfamily 2403.06959}}].

\bibitem{kovner2005pursuit}
A.~Kovner and M.~Lublinsky, \emph{In pursuit of \textsc{P}omeron loops: The \textsc{JIMWLK} equation and the \textsc{W}ess-\textsc{Z}umino term}, \href{https://doi.org/10.1103/physrevd.71.085004}{\emph{Physical Review D} {\bfseries 71} (2005) }.

\bibitem{kovner2017exploring}
A.~Kovner, M.~Lublinsky and V.~Skokov, \emph{Exploring correlations in the \textsc{CGC} wave function: \textsc{O}dd azimuthal anisotropy}, \href{https://doi.org/10.1103/physrevd.96.016010}{\emph{Physical Review D} {\bfseries 96} (2017) }.

\bibitem{McLerran:2014uka}
L.~McLerran and V.V.~Skokov, \emph{{The Eccentric Collective BFKL Pomeron}}, \href{https://doi.org/10.5506/APhysPolB.46.1513}{\emph{Acta Phys. Polon. B} {\bfseries 46} (2015) 1513} [\href{https://arxiv.org/abs/1407.2651}{{\ttfamily 1407.2651}}].

\bibitem{Dumitru:2014yza}
A.~Dumitru, L.~McLerran and V.~Skokov, \emph{{Azimuthal asymmetries and the emergence of {\textquotedblleft}collectivity{\textquotedblright} from multi-particle correlations in high-energy pA collisions}}, \href{https://doi.org/10.1016/j.physletb.2015.02.046}{\emph{Phys. Lett. B} {\bfseries 743} (2015) 134} [\href{https://arxiv.org/abs/1410.4844}{{\ttfamily 1410.4844}}].

\bibitem{McLerran:2015sva}
L.~McLerran and V.~Skokov, \emph{{Finite Numbers of Sources, Particle Correlations and the Color Glass Condensate}}, \href{https://doi.org/10.1016/j.nuclphysa.2015.12.005}{\emph{Nucl. Phys. A} {\bfseries 947} (2016) 142} [\href{https://arxiv.org/abs/1510.08072}{{\ttfamily 1510.08072}}].

\bibitem{kovchegov2013quantum}
Y.V.~Kovchegov and E.~Levin, \emph{Quantum chromodynamics at high energy}, Cambridge University Press (2013).

\bibitem{munier2025unitary}
S.~Munier, \emph{Unitary perturbation theory on the light cone using adiabatic switching},  2025.

\bibitem{munier2025extracting}
S.~Munier, \emph{Extracting light-cone wave functions from covariant amplitudes: a detailed study in scalar field theory},  \href{https://arxiv.org/abs/2512.22345}{{\ttfamily 2512.22345}}.

\bibitem{lappi2026two}
T.~Lappi, R.~Paatelainen and M.~Seppälä, \emph{Two-\textsc{L}oop \textsc{DGLAP} splitting functions from \textsc{L}ight \textsc{C}one \textsc{P}erturbation \textsc{T}heory},  \href{https://arxiv.org/abs/2601.10374}{{\ttfamily 2601.10374}}.

\bibitem{munier2026unitary}
S.~Munier, \emph{Unitary perturbation theory on the light cone using adiabatic switching},  \href{https://arxiv.org/abs/2510.05256}{{\ttfamily 2510.05256}}.

\bibitem{hanninen2018one}
H.~Hänninen, T.~Lappi and R.~Paatelainen, \emph{One-loop corrections to light cone wave functions: \textsc{T}he dipole picture \textsc{DIS} cross section}, \href{https://doi.org/10.1016/j.aop.2018.04.015}{\emph{Annals of Physics} {\bfseries 393} (2018) 358–412}.

\bibitem{lappi2017one}
T.~Lappi and R.~Paatelainen, \emph{The one loop gluon emission light cone wave function}, \href{https://doi.org/10.1016/j.aop.2017.02.002}{\emph{Annals of Physics} {\bfseries 379} (2017) 34–66}.

\bibitem{rr}
M.~Lublinsky, R.~Radhakrishnan and V.~Skokov, \emph{Single gluon inclusive production in the mid rapidity regime at the next-to-leading-order}, {\emph{In preparation} }.

\bibitem{harindranath1996introduction}
A.~Harindranath, \emph{An \textsc{I}ntroduction to \textsc{L}ight-\textsc{F}ront \textsc{D}ynamics for \textsc{P}edestrians},  \href{https://arxiv.org/abs/hep-ph/9612244}{{\ttfamily hep-ph/9612244}}.

\bibitem{burkardt1996light}
M.~Burkardt, \emph{Light \textsc{F}ront \textsc{Q}uantization},  in \emph{Advances in Nuclear Physics}, p.~1–74, Kluwer Academic Publishers (1996), \href{https://doi.org/10.1007/0-306-47067-5_1}{DOI}.

\bibitem{pritchard1980qcd}
D.~Pritchard and W.~Stirling, \emph{\textsc{QCD} calculations in the light-cone gauge}, \href{https://doi.org/https://doi.org/10.1016/0550-3213(80)90086-3}{\emph{Nuclear Physics B} {\bfseries 165} (1980) 237}.

\bibitem{kovner2005high}
A.~Kovner, \emph{High \textsc{E}nergy \textsc{E}volution - \textsc{T}he \textsc{W}ave \textsc{F}unction \textsc{P}oint of \textsc{V}iew},  \href{https://arxiv.org/abs/hep-ph/0508232}{{\ttfamily hep-ph/0508232}}.

\bibitem{kovner2005remarks}
A.~Kovner and M.~Lublinsky, \emph{Remarks on high energy evolution}, \href{https://doi.org/10.1088/1126-6708/2005/03/001}{\emph{Journal of High Energy Physics} {\bfseries 2005} (2005) 001–001}.

\bibitem{Landau:1991wop}
L.D.~Landau and E.M.~Lifshits, \emph{{Quantum Mechanics}: {Non-Relativistic Theory}}, vol.~v.3 of \emph{Course of Theoretical Physics}, Butterworth-Heinemann, Oxford (1991), \href{https://doi.org/10.1016/C2013-0-02793-4}{10.1016/C2013-0-02793-4}.

\bibitem{shankar2012principles}
R.~Shankar, \emph{Principles of quantum mechanics}, Springer Science \& Business Media (2012).

\bibitem{kovner2019entanglement}
A.~Kovner, M.~Lublinsky and M.~Serino, \emph{Entanglement entropy, entropy production and time evolution in high energy \textsc{QCD}}, \href{https://doi.org/10.1016/j.physletb.2018.10.043}{\emph{Physics Letters B} {\bfseries 792} (2019) 4–15}.

\bibitem{peskin2018introduction}
M.E.~Peskin, \emph{An Introduction to quantum field theory}, CRC press (2018).

\bibitem{kovner2006one}
A.~Kovner and M.~Lublinsky, \emph{One gluon, two gluon: multigluon production via high energy evolution}, \href{https://doi.org/10.1088/1126-6708/2006/11/083}{\emph{Journal of High Energy Physics} {\bfseries 2006} (2006) 083–083}.

\end{thebibliography}\endgroup
\end{document}